\documentclass[aps,preprint,nofootinbib,preprintnumbers,eqsecnum]{revtex4-1}

\usepackage[usenames, dvipsnames]{color}

\newcommand{\nb}[1]{\color{blue}}

\newcommand{\hl}[1]{\color{magenta}}

\usepackage{geometry,tikz}

\usepackage[
	  pagebackref=false,
	  colorlinks=true,
      linkcolor=blue,
      urlcolor=blue,
      filecolor=black,
      citecolor=red,
      pdfstartview=FitV,
      pdftitle={},
        pdfauthor={},
        pdfsubject={},
        pdfkeywords={},
        pdfpagemode=None,
        bookmarksopen=true
      ]{hyperref}

\usepackage[normalem]{ulem}
\usepackage{amsmath}
\usepackage{enumerate}
\usepackage{amsfonts}
\usepackage{epsfig}
\usepackage{amssymb}

\usepackage{setspace}

\def\Tr{\mathop{\rm Tr}}

\def\Im{\mathop{\rm Im} }
\def\Re{\mathop{\rm Re} }

\newcommand\half{{\ensuremath{\frac{1}{2}}}}
\newcommand\p{\ensuremath{\partial}}

\newcommand\field[1]{{\ensuremath{\mathbb{{#1}}}}}

\newcommand\vev[1]{{\ensuremath{\left\langle{#1}\right\rangle}}}

\newcommand\ket[1]{\ensuremath{\lvert{#1}\rangle}}

\newcommand{\CC}{\field{C}}

\newcommand{\FF}{\field{F}}
\newcommand{\OO}{\field{O}}

\newcommand{\NN}{\field{N}}

\newcommand{\RR}{\field{R}}

\newcommand{\ZZ}{\field{Z}}

\newcommand{\be}{\begin{equation}}
\newcommand{\ee}{\end{equation}}
\newcommand{\bea}{\begin{eqnarray}}
\newcommand{\eea}{\end{eqnarray}}
\newcommand{\bega}{\begin{gather}}
\newcommand{\eega}{\end{gather}}

\newcommand{\bi}{\begin{itemize}}
\newcommand{\ei}{\end{itemize}}
\newcommand{\ben}{\begin{enumerate}}
\newcommand{\een}{\end{enumerate}}
\newcommand{\bca}{\begin{cases}}
\newcommand{\eca}{\end{cases}}
\newcommand{\bln}{\begin{align}}
\newcommand{\eln}{\end{align}}
\newcommand{\bst}{\begin{split}}
\newcommand{\est}{\end{split}}
\def\ie{\begin{equation}\begin{aligned}}
\def\fe{\end{aligned}\end{equation}}
\newcommand{\bma}{\le(\begin{matrix}}
\newcommand{\ema}{\end{matrix}\ri)}

\newcommand\al{{\alpha}}
\def\b{{\beta}}
\newcommand\ep{\epsilon}

\newcommand\Sig{\Sigma}
\newcommand\lam{\lambda}
\newcommand\Lam{\Lambda}
\newcommand\om{\omega}
\newcommand\Om{\Omega}

\newcommand\ga{{\ensuremath{{\gamma}}}}
\newcommand\Ga{{\ensuremath{{\Gamma}}}}
\newcommand\de{{\ensuremath{{\delta}}}}
\newcommand\De{{\ensuremath{{\Delta}}}}

\newcommand\ze{\zeta}

\newcommand\ov{\over}
\newcommand\ha{{\half}}

\def\le{\left}
\def\ri{\right}

\newcommand\sB{{\ensuremath{{\mathcal B}}}}

\newcommand\sE{{\ensuremath{{\mathcal E}}}}
\newcommand\sF{{\ensuremath{{\mathcal F}}}}
\newcommand\sI{{\ensuremath{{\mathcal I}}}}

\newcommand\sH{{\ensuremath{{\mathcal H}}}}

\newcommand\sM{{\ensuremath{{\mathcal M}}}}
\newcommand\sN{{\ensuremath{{\mathcal N}}}}

\newcommand\sS{{\mathcal S}}

\newcommand\sX{{\mathcal X}}

\newcommand\sZ{{\mathcal Z}}

\newcommand\bA{{\boldsymbol  A}}

\newcommand\bD{{\boldsymbol  D}}
\newcommand\bR{{\boldsymbol R}}
\newcommand\bX{{\boldsymbol  X}}

\newcommand\bZ{{\boldsymbol  Z}}

\newcommand\brho{{\boldsymbol \rho}}

\newcommand\bp{{\mathbf p}}
\newcommand\bx{{\mathbf x}}

\newcommand{\fn}{{\mathfrak n}}

\newcommand{\bid}{\mathbf{1}}

\newcommand\Fil[1]{{\ensuremath{\FF\{#1\}}}}

\begin{document}

\title{
Ramp, Plateau, and Wormholes without Averaging, and Hyper-non-perturbative Structures in Gravity}

\preprint{MIT-CTP/6074}

\author{Hong Liu}
\vspace{0.5cm}
\affiliation{MIT Center for Theoretical Physics---a Leinweber Institute,\\ 
Massachusetts Institute of Technology, \\
77 Massachusetts Ave.,  Cambridge, MA 02139\\}


\begin{abstract}
 
 \vspace{0.5cm}

Universal hallmarks of quantum chaos---such as the ramp and plateau in the spectral form factor---and the ramp's gravitational duals involving wormholes are widely interpreted as consequences of spectral or ensemble averaging. In this paper, following an earlier proposal of~\cite{Liu25c}, we develop an alternative approach: these phenomena arise as macroscopic smooth structures hidden within erratic microscopic data, which can be isolated through a smooth filter projection.  Using the semiclassical Gutzwiller trace formula as a paradigmatic example, we illustrate how many features characteristic of random matrix models---including the ramp, the plateau, the spectral curve, and single-eigenvalue instantons---can be derived in the semiclassical limit without invoking ensemble or explicit spectral averages.

We postulate the existence of a minimal Gutzwiller-like structure in the large-$N$ limit of holographic systems and explore its consequences. Beyond deriving the ramp and the plateau, this Gutzwiller-like structure predicts universal rapid macroscopic oscillations in the density of states and the possible existence of hyper-instantons, both of which involve double exponentials in $1/N^2$.

On the gravity side, we demonstrate how spacetime wormholes enable the construction of emergent hyper-non-perturbative objects---such as baby-universe and wormhole condensates---which yield double exponential effects in $G_N$. This mirrors the postulated boundary Gutzwiller-like structure and provides a dual gravitational derivation of the universal rapid macroscopic oscillations in the density of states and the spectral plateau.

\end{abstract}

\today

\maketitle

\tableofcontents


\section{Introduction}

\subsection{Motivation}

Quantum chaotic systems exhibit universal random matrix-like spectral correlations that are largely insensitive to microscopic details, including the 
rigid level repulsion between nearby energy levels and the associated ramp and plateau behavior of the spectral form factor. In holographic systems, these random matrix-like features are closely connected to Euclidean wormholes and related semiclassical gravitational saddles~\cite{SaaShe18,SaaShe19,CotJen20}. This connection sharpens a long-standing puzzle regarding the interpretation of Euclidean wormholes in AdS/CFT~\cite{WitYau99,MalMao04,ArkOrg07}: wormholes appear to imply nontrivial ``correlations'' between partition functions of CFTs defined on disconnected manifolds, even though these quantities are expected to factorize exactly in an individual quantum theory.\footnote{Precursors of the present factorization puzzle and related interpretations of Euclidean wormholes can be found in the seminal early literature~\cite{Haw87,LavRub87,Col88,GidStr88a,GidStr88,BanKle89}.} A natural interpretation is that semiclassical gravitational path integrals compute certain ensemble-averaged quantities, which stands in significant tension with the standard formulation of holography in string theory, where the duality is expected to apply to individual quantum systems rather than ensembles.

Motivated by earlier discussions in~\cite{SchWit22,Liu25,KudWit25}, in~\cite{Liu25c} we proposed a new interpretation of Euclidean wormholes that preserves the successes of the ensemble-average picture while avoiding the introduction of an actual ensemble of theories. This proposal rests on two central elements:

\begin{enumerate}
    \item \textbf{Macroscopic decomposition and the smooth filter.} 
    
    We challenge the standard assumption that a typical observable in a large-$N$ system, such as the Euclidean partition function on a compact manifold, {\it always} has a smooth limit in terms of an asymptotic transseries expansion in $1/N^2$ (see also~\cite{SchWit22}). Instead, the limit can be generically ``singular.'' It is accordingly postulated that, in the large-$N$ limit, an observable $A$ of a boundary CFT admits a decomposition of the form
    \be\label{Decom}
    A = A^{(\rm sm)} + A^{(\rm err)} \ ,
    \ee
    where $A^{(\rm sm)}$ denotes the smooth, ``macroscopic'' component expressed as an asymptotic transseries in $1/N^2$, while $A^{(\rm err)}$ denotes the erratic, ``microscopic'' component capturing the singular nature of the limit.\footnote{In particular, two classes of microscopic features are expected to play especially important roles: (a) the erratic behavior of OPE coefficients among heavy operators of dimension $O(N^2)$ (and above), and (b) the fine-grained structure of the density of states at energies of order $O(N^2)$ (and above). See e.g~\cite{Moo98a,Moo98b,MilMoo99,DijMal00,SchWit22,Liu25,KudWit25,Liu25c,KudWit26,Per26} for discussions.}

    
 It is convenient to introduce a projection operation (a smooth ``filter'') $\FF$ to isolate the smooth component, $\Fil{A} \equiv A^{(\rm sm)}$. By definition, the filter satisfies the properties 
    \be\label{0CC0}
    \Fil{A^{(\rm err)}} = 0, \quad \FF^2 = \FF , \quad \Fil{a A + b B} = a \Fil{A} + b \Fil{B} \ ,
    \ee
    where $a$ and $b$ are smooth objects (i.e., transseries terms in $1/N^2$). The filter $\FF$ is also assumed to be positive,\footnote{Note that~\eqref{CC0} implies that $\Fil{A^*} = (\Fil{A})^*$. \label{ft:real}} meaning
    \be \label{CC0}
    \Fil{A^* A} \geq 0 \ .
    \ee
  
    \item \textbf{Holographic dictionary: the gravitational path integral as a filter.}

The second element posits that the gravitational path integral captures only the smooth component of boundary observables, with the bulk-boundary dictionary taking the form
\be \label{gid}
\Fil{A}\equiv A^{(\rm sm)}=A_{\rm GPI}\,,
\ee
where $A_{\rm GPI}$ denotes the result obtained from the gravitational path integral (GPI). Equation~\eqref{gid} therefore implies that, in the semiclassical regime, the gravitational path integral itself acts as a filter: it is the bulk counterpart of the smooth boundary projection $\FF$.

A qualification is important here. In general dimensions, gravitational path integrals are not well defined nonperturbatively and are not expected, by themselves, to provide a fundamental definition of quantum gravity. Throughout our discussion, we therefore use ``gravitational path integral'' only in a semiclassical sense. More precisely, we treat it as a formal semiclassical device for extracting the following smooth gravitational data:
\ben
\item a discrete, possibly infinite, set of semiclassical saddle points, including perturbative fluctuations around each saddle; and
\item constrained saddles, characterized by a finite number of collective parameters that must subsequently be integrated over.
\een
Gravitational path integrals in JT gravity, and perhaps in pure AdS$_3$ gravity, may be viewed as falling into the second category. In higher dimensions, examples of constrained saddles include those discussed in~\cite{CotJen20b,CotJen21,CotJen22}.

\end{enumerate}

Let $Z_1$ and $Z_2$ denote the partition functions of the boundary CFT on manifolds $M_1$ and $M_2$, respectively. Although the partition function factorizes in the large-$N$ limit,
\be
Z[M_1\cup M_2]=Z_1 Z_2 \ ,
\ee
its smooth component need not factorize. This is because the product of two erratic components, $Z_{1}^{(\rm err)} Z_{2}^{(\rm err)}$, may itself contain a smooth contribution. More precisely,
\be
\Fil{Z_1 Z_2}_c
\equiv
\Fil{Z_1 Z_2} - \Fil{Z_1}\Fil{Z_2} = \Fil{Z_1^{(\rm err)} Z_2^{(\rm err)}} \neq 0 \ ,
\ee
and equation~\eqref{gid} then identifies this connected component with its counterpart on the gravity side, namely the wormhole amplitude. This yields~\cite{Liu25c}
\be\label{worm}
\Fil{Z_1 Z_2}_c = \Fil{Z_1^{(\rm err)} Z_2^{(\rm err)}} = Z^{(\mathrm{wormhole})}[M_1,M_2] \ ,
\ee
where the right-hand side denotes the gravitational wormhole amplitude connecting the two boundary manifolds.

In this framework, wormhole ``correlations'' arise not from ensemble averaging, but from projecting onto the macroscopic component of a single theory. The projection isolates smooth contributions hidden within products of erratic microscopic quantities, thereby generating non-factorizing macroscopic observables even though the underlying microscopic theory factorizes exactly.

At present, explicitly formulating the decomposition~\eqref{Decom} or the projection operation $\Fil{\cdot}$ for a general holographic system remains an outstanding challenge.\footnote{See~\cite{Kli26,KudWit26,Wu26} for recent discussions in this direction.} There is, however, a well-studied physical context where the singular nature of the limit and the macroscopic/microscopic decomposition are understood: the familiar semiclassical $\hbar \to 0$ limit of a quantum chaotic system. 

For a physical observable $A$, the $\hbar \to 0$ limit can be singular: $A$ need not admit a smooth asymptotic expansion in $\hbar$.  In a quantum chaotic system, the celebrated Gutzwiller trace formula~\cite{Gut71,BalBlo72,Gut90} precisely captures this singular structure via a decomposition of the  form~\eqref{Decom}. Schematically, the Gutzwiller formula expresses the resolvent $R_+ (E) = \Tr {1 \ov E- H + i0}$ of a quantum chaotic system as\footnote{See Sec.~\ref{sec:Gutz} for further details.}
\be \label{00poR}
R_+ (E)= \bar R (E) - {i \ov \hbar} \sum_a  A_a e^{{i \ov \hbar} S_a(E)} \ ,
\ee
where $\bar R (E)$ is a smooth component that admits a well-defined asymptotic expansion in $\hbar$ starting with the classical Weyl approximation. The second term is an infinite sum over all periodic orbits $a$ of the system at energy $E$, where $S_a (E)$ and $A_a$ denote the classical action and amplitude of the orbit, respectively. 

At first glance, the phase factors $e^{\frac{i}{\hbar}S_a(E)}$ have the familiar form of semiclassical saddle-point contributions. Indeed, each individual term in the periodic-orbit sum ultimately arises from a saddle-point approximation to the coordinate-space propagator $\vev{\bx|e^{-iHt}|\bx}$. It is important to note, however, that $\vev{\bx|e^{-iHt}|\bx}$ is inherently a microscopic quantity, and the resulting spectral information is likewise microscopic: periodic orbits encode fine-grained dynamical data that are absent from the smooth semiclassical density of states. In a chaotic system, periodic orbits are isolated and proliferate rapidly with increasing period, while their classical actions $S_a(E)$ vary ``randomly'' from orbit to orbit. The periodic-orbit contribution therefore involves an infinite sum of oscillatory terms with effectively random phases $e^{iS_a(E)/\hbar}$. Although the contribution from each individual orbit admits the usual semiclassical asymptotic expansion, their infinite, erratic superposition does not: the full periodic-orbit sum has an irregular dependence on $\hbar$ and cannot in general be represented by a smooth asymptotic series in $\hbar$.

The dependence on the energy $E$ is similarly fine-grained. A given periodic orbit generally varies smoothly with $E$, but its phase varies according to $dS_a(E)/dE=T_a(E)$ where $T_a$ is the period of the orbit. Consequently, a change $\delta E$ produces a phase variation of order $T_a\delta E/\hbar$, so that long periodic orbits generate oscillations on increasingly fine energy scales. This fine-grained $E$-dependence is precisely what is needed to reproduce the highly irregular microscopic density of states, which cannot be captured by the smooth component $\bar R(E)$.

Thus, the periodic-orbit sum makes manifest the singular nature of the $\hbar\to0$ limit: while each individual saddle contribution has a conventional semiclassical expansion, the infinite sum over them exhibits an erratic dependence on $\hbar$. By contrast, the smooth component $\bar R(E)$ reflects only coarse-grained, macroscopic properties of the underlying phase space.

Acting the smooth filer projection on~\eqref{00poR} then gives 
\be
\Fil{R_+ (E)} = \bar R(E) \ .
\ee


\subsection{Three-tier description of a large-$N$ system and quantum gravity}

\begin{figure}
\begin{center}
\includegraphics[width=12cm]{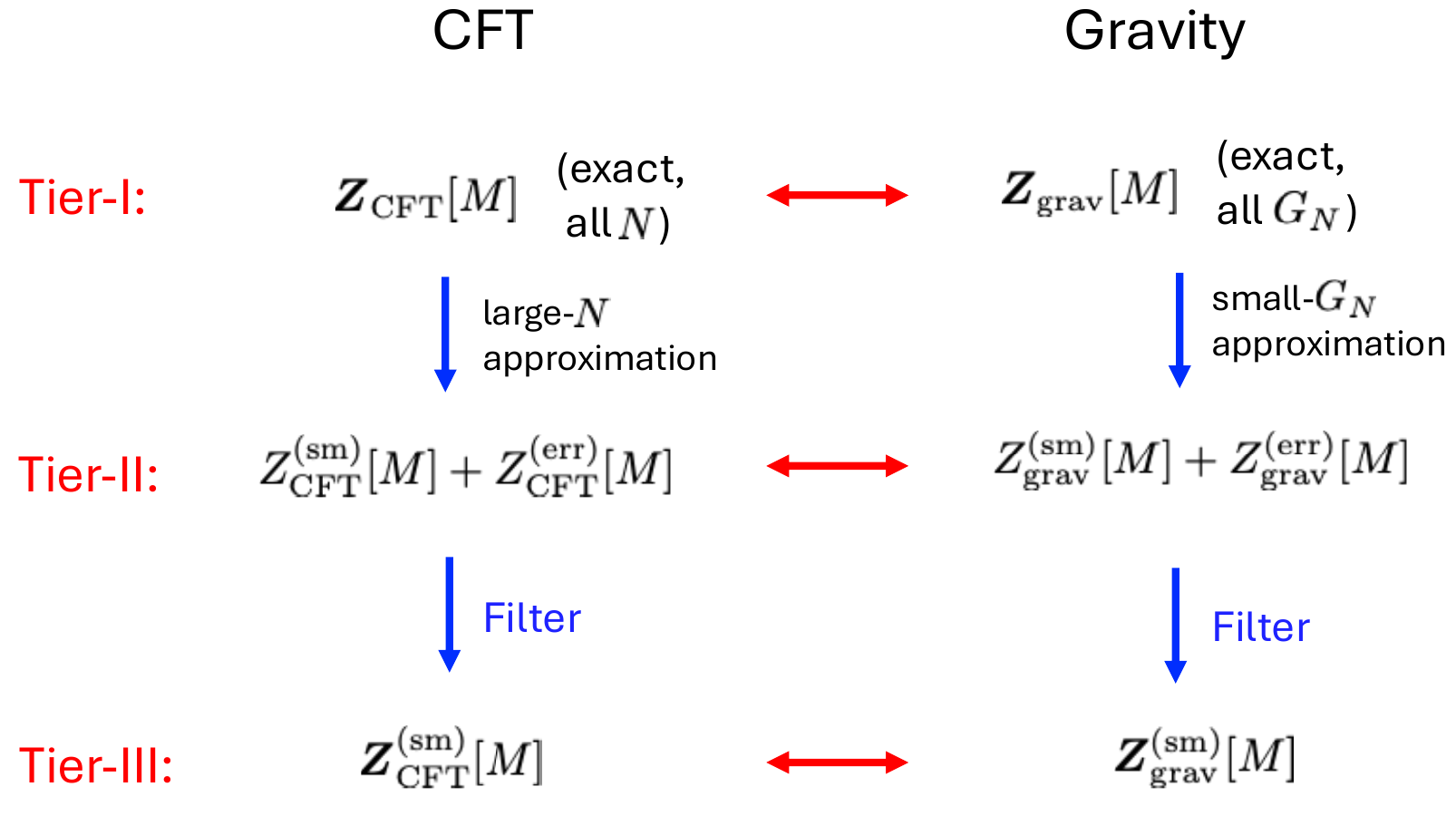}
\caption{\small Three-tier structure on the CFT and gravity sides. Corresponding quantities on the two sides are identified under holographic duality. In particular, at Tier II, $Z_{\rm CFT}^{(\rm sm)}[M]\leftrightarrow Z_{\rm grav}^{(\rm sm)}[M]$ and $Z_{\rm CFT}^{(\rm err)}[M]\leftrightarrow Z_{\rm grav}^{(\rm err)}[M]$. As discussed in the main text, a Tier-II description need not be unique, and filtering a given Tier-II representation generally extracts only a subset of the Tier-III asymptotic data. Specifically, under filtering $Z_{\rm CFT}^{(\rm err)}[M]$ is removed, while $Z_{\rm CFT}^{(\rm sm)}[M]$ yields a subset of the macroscopic asymptotic data contained in $\bZ_{\rm CFT}^{(\rm sm)}[M]$, and similarly on the gravity side. Results obtained from gravitational path integrals belong to the Tier-III description, but need not exhaust the full Tier-III asymptotic data.
}
\label{fig:3tiers}
\end{center}
\end{figure}

While we do not yet have an explicit demonstration of the decomposition~\eqref{Decom} for large-$N$ systems, the Gutzwiller description of the $\hbar\to0$ limit provides a compelling proof of concept and a paradigmatic example. If~\eqref{Decom} is correct, it suggests that the usual dichotomy between an exact microscopic description and a smooth asymptotic description should be refined by introducing a third, intermediate tier: a large-$N$ description that remains microscopic in the sense that it retains erratic ``fluctuations.'' This leads to the three-tier structure illustrated in Fig.~\ref{fig:3tiers}. Holography then suggests a corresponding three-tier structure in gravity, including a microscopic $G_N\to0$ description beyond the gravitational path integral.

It is useful at this point to establish precise terminology and notation for the three tiers. More explicitly, for a given observable $A$ (such as the partition function), we have the following tiers of description:
\bi
\item {\bf Tier I:} The exact microscopic description, valid at any $N$, which we denote by $\bA$.

\item {\bf Tier II:} A ``microscopic'' large-$N$ description, denoted by $A$.\footnote{In~\cite{Liu25c}, objects at this tier were heuristically denoted by $\lim_{N\to\infty}\bA$ (translated into the notation used here for the exact quantity), although, as emphasized there, this ``limit'' should not be understood in the conventional sense.} A Tier-II representation describes $\bA$ from a large-$N$ perspective while retaining a nonvanishing component of erratic microscopic contributions.

\item {\bf Tier III:} The smooth, ``macroscopic'' large-$N$ asymptotic description, denoted by $\bA^{(\rm sm)}$. It is organized as a transseries in $1/N^2$ and, according to~\eqref{gid}, it is quantities at this tier that are directly captured by the semiclassical gravitational path integral.
\ei


The filter $\Fil{\cdot}$ maps a Tier-II representation to smooth Tier-III data. For a given microscopic large-$N$ representation $A$ of $\bA$, the filtered quantity $A^{\rm sm}=\Fil{A}$ extracts a corresponding subset of the macroscopic asymptotic information contained in $\bA^{(\rm sm)}$.\footnote{It is also possible to pass directly from a Tier-I to a Tier-III description without explicitly introducing an intermediate Tier-II representation.} As we will demonstrate explicitly in the Gutzwiller description, the Tier-II representation of a given microscopic quantity need not be unique. Different choices of $A$ may expose different, and possibly overlapping, subsets of the Tier-III asymptotic data.

The Tier-III asymptotic expansion can itself be organized into distinct transseries {\it levels} according to its ``exponential depth,''
\be
\bA^{(\rm sm)}
=
\bA^{(\rm sm)}_0
+
\bA^{(\rm sm)}_1
+
\bA^{(\rm sm)}_2
+\cdots\,.
\ee
The level-0 contribution $\bA^{(\rm sm)}_0$ consists of a formal perturbative power series in $1/N^2$. The level-1 contribution $\bA^{(\rm sm)}_1$ includes ordinary nonperturbative exponentials of the form
\be \label{cftE}
\sum_i e^{-N^2V_0^{(i)}}V_1^{(i)}
\left(1+\frac{1}{N^2}V_2^{(i)}+\cdots\right)\,.
\ee
The level-2 contribution $\bA^{(\rm sm)}_2$ includes double-exponential structures such as
\be\label{1dex}
e^{f e^{O(N^2)}}\,,
\qquad
f\in\CC\,,
\ee
and so on. In~\cite{Liu25c}, it was implicitly assumed that filtering a Tier-II description produces contributions of the conventional level-1 form~\eqref{cftE}. We will see that suitable Tier-II representations can also reveal higher-level contributions, including level-2 terms of the form~\eqref{1dex}, which we refer to as hyper-non-perturbative. Different Tier-II representations need not correspond to successive truncations of the Tier-III transseries; rather, they may expose different sets of contributions within one or more transseries levels.

We stress that the Tier-III description strictly refers to a smooth, macroscopic representation of a quantity. While it captures nonperturbative and hyper-non-perturbative information, it fundamentally remains a macroscopic asymptotic description---distinct from the exact microscopic data of Tier I and the erratic contributions of Tier II. For example, each individual periodic-orbit contribution in~\eqref{00poR} admits a conventional semiclassical expansion around its saddle, but it should not be classified as Tier III: it remains a microscopic contribution.

Within the present classification, a conventional weak-coupling theory such as QED is not expected to require a distinct Tier-II description, since the small-coupling expansion in QED is not believed to exhibit erratic behavior. (Unlike the large-$N$ limit, there is no indication that this conventional wisdom should be challenged.) In contrast, in the Marolf--Maxfield topological model~\cite{MarMax20}, where the genus expansion yields a convergent series (i.e., one with a finite radius of convergence) rather than an asymptotic one, there is only a Tier-I description.\footnote{In this context, the issue of ``null states'' arises because truncating this convergent series at small couplings---even where it might seem mathematically justified---destroys delicate exact relations and yields misleading answers to certain physical questions. Separately, the complete absence of Tier-II and Tier-III descriptions highlights an inherent limitation of such topological models when used to capture the rich chaotic dynamics expected in higher-dimensional quantum gravity.}

\subsection{Summary of main results} 

In this paper, using the Gutzwiller representation~\eqref{00poR} as a paradigmatic example, we explicitly illustrate the mechanics of the general framework proposed in~\cite{Liu25c}. Furthermore, we postulate the existence of a minimal Gutzwiller-like structure in the large-$N$ limit of holographic systems and discuss its implications. Chief among these is the emergence of hyper-non-perturbative structures that give rise to double exponentials in $1/N^2$. On the gravity side, we explore how spacetime wormholes enable the construction of emergent hyper-non-perturbative objects, which correspondingly yield double exponentials in $G_N$. More explicitly: 

\ben

\item We show that a smooth filter projection can be explicitly defined in the context of Gutzwiller representation~\eqref{00poR}. 

This enables the derivation of many features characteristic of random matrix models, including:
\ben 
\item the ramp,
\item the plateau, 
\item rapid macroscopic oscillations in the density of states,
\item the spectral curve, and
\item single-eigenvalue instantons, 
\een
all within a single-system framework without ensemble or explicit spectral averages.

\item While it is natural to anticipate a generalization of the Gutzwiller formula to quantum many-body systems in the large-$N$ limit  and there has been notable progress~\cite{EngUrb15,RicUrb22}, a formulation applicable to the genuine many-body regime (e.g., capable of resolving the $e^{- O(N^2)}$ energy spacing in holographic systems) does not yet exist.

We therefore postulate a minimal Gutzwiller-like structure for quantum many-body systems in the large $N$ limit (with holographic systems as a subclass). The general derivation of random matrix-like features (a)--(e) listed in the previous item can then be imported directly to these many-body systems, again in a single-system setting.

Beyond deriving the ramp and the plateau, this Gutzwiller-like structure predicts a universal correction to the density of states of the form
\begin{equation}\label{1edos}
-\frac{1}{\pi} e^{\hat{g}(E) - c(E)} \cos \left( 2 \pi i \bar{N}(E) - 2 i \phi \right) \ ,
\end{equation}
where $\bar{N}(E) = \int^E_{E_0} dE' \, \bar{\rho}(E')$ is the total number of states below $E$ obtained from the coarse-grained density of states  $\bar{\rho}(E)$. The exponent functions $\hat{g}(E)$ and $c(E)$ arise from smooth filter projections of products of the resolvents.
For holographic systems, $\log \bar \rho (E)$ can be identified with the Bekenstein-Hawking entropy of a black hole of energy $E$, scaling as $O(N^2) \sim O({1 \ov G_N})$. Consequently,~\eqref{1edos} takes the form of a double exponential~\eqref{1dex}.  
As will be discussed more explicitly below, the plateau is also an example of~\eqref{1dex}.  

Furthermore, this structure predicts the possible existence of hyper-instantons in holographic systems---the analogues of single-eigenvalue instantons in matrix models---leading to corrections of the form~\eqref{1dex} in the partition functions.

\item The derivation of the ramp from the Gutzwiller structure provides direct support for the proposal~\eqref{worm} that the Saad-Shenker-Stanford double cone~\cite{SaaShe18} and the Cotler-Jensen off-shell wormholes~\cite{CotJen20,CotJen20b,CotJen21,CotJen22} in general dimensions can be interpreted within a single-system setting. Conversely, assuming a single-system holographic duality, we argue that these wormholes serve as indirect evidence for the existence of a Gutzwiller-like structure in the boundary theory.

\item We explore the gravitational descriptions of the double exponentials~\eqref{1dex}, which we refer to as ``hyper-structures.''
To contextualize~\eqref{1dex}, note that the semiclassical expansion of the bulk 
Euclidean path integral (where $\Phi$ collectively denotes the bulk fields and $\sI_E$ is the bulk Euclidean action), 
\be \label{EGP}
 \int D \Phi \, e^{ - {1 \ov G_N} \sI_E [\Phi]} ,
\ee
is expected to generate a conventional level-1 transseries in $G_N$
\begin{align} \label{EGP0}
A_{\rm GPI} &  = \sum_{i} e^{- {1 \ov G_N} W_0^{(i)}} W_1^{(i)} \left(1 + G_N W_2^{(i)} +  \cdots \right)  \ .
\end{align}

The existence of~\eqref{1dex} in the boundary CFT demands new non-perturbative structures in the bulk gravity. Specifically, it implies the emergence of level-2 transseries terms in $G_N$ of the form
\begin{equation}\label{dobG} 
e^{f e^{O(1/G_N)}} , \quad f \in \mathbb{C} \ ,
\end{equation}
which extend beyond the conventional semiclassical transseries expansion~\eqref{EGP0}.

Hints of such double exponentials~\eqref{dobG} have already surfaced in JT gravity via its random matrix formulation~\cite{SaaShe19} (see also~\cite{AltSon20,PosVan22,AltPos22}). Here we argue for their universal presence in general higher-dimensional gravities, following both from the postulated boundary Gutzwiller structure and from an independent construction within the bulk gravity itself.

We demonstrate that ``hyper-non-perturbative structures'' (or simply hyper-structures) capable of generating~\eqref{dobG} can be constructed by utilizing the third-quantized Hilbert space of all baby universes, $\sH_{\text{multiverse}}$. While standard applications of the AdS/CFT correspondence fix the number of asymptotic boundaries---often to just a single one---introducing $\sH_{\text{multiverse}}$ to accommodate an arbitrary number of boundaries is a natural consequence of wormholes, as was already recognized in the classic literature on Euclidean wormholes~\cite{Haw87,LavRub87,Col88,GidStr88a,GidStr88,BanKle89}. 

Recent discussions of baby-universe Hilbert spaces~(see, e.g.,~\cite{MarMax20,McNVaf20,Sta20,PosVan22,AbdAnt26,Har26}) are almost exclusively topological in nature (as applicable to two-dimensional bulks). Here, we follow the construction of~\cite{Liu25c}, which applies to general higher-dimensional gravity theories\footnote{As emphasized in~\cite{Liu25c}, $\sH_{\text{multiverse}}$ is expected to be non-separable in higher dimensions. For a non-separable Hilbert space, while boundary-generating operators can be simultaneously diagonalized, they cannot be diagonalized by an orthonormal basis; their common diagonal vectors are non-normalizable and do not live in the Hilbert space. In other words, alpha-states are not expected to exist beyond simple topological situations. \label{ft:sep}}.

By constructing ``coherent states''\footnote{In the context of topological models or JT gravity, such objects in a third-quantized Hilbert space have been discussed before in~\cite{MarMax20,PosVan22,AltPos22}.} in $\sH_{\text{multiverse}}$---defined by exponentiating single-boundary states to produce a superposition over all possible numbers of boundaries---we introduce two classes of hyper-structures: baby-universe condensates (and anti-condensates) and multiverse instantons. These serve as higher-dimensional counterparts to the FZZT and ZZ branes~\cite{FatZam00,Tes00,ZamZam01} of two-dimensional Liouville gravity.

The overlap of such objects can then be used to derive level-2 transseries effects, including the plateau of the spectral form factor and the term~\eqref{1edos}.
In this framework, Cotler-Jensen off-shell wormholes~\cite{CotJen20,CotJen20b,CotJen21,CotJen22}~(and the double trumpet in JT gravity~\cite{SaaShe19}) act as ``baby-universe propagators,'' mediating macroscopic ``contractions'' among boundaries to yield the exponent functions $\hat{g}(E)$ and $c(E)$ in~\eqref{1edos}.

\begin{figure}
\begin{center}
\includegraphics[width=14cm]{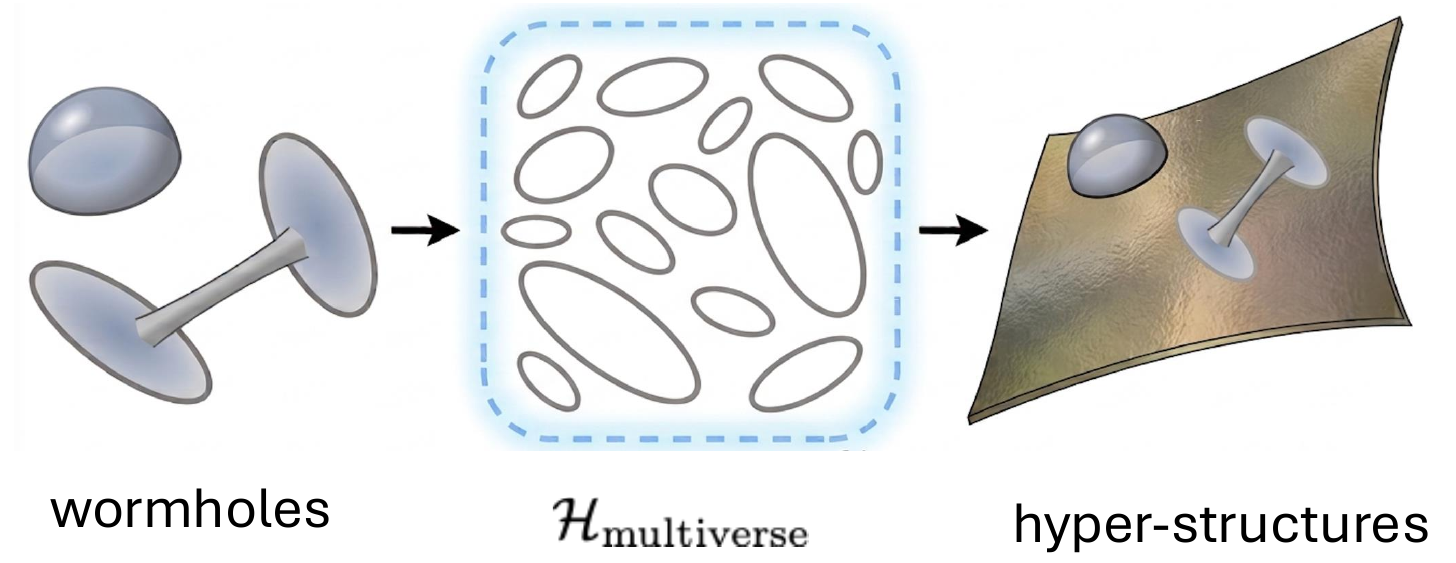}
\caption{\small From topology to hyper-structures: wormholes demand a multiverse; from the multiverse emerge hyper-structures.
The inclusion of wormhole topologies naturally necessitates an arbitrary number of disconnected boundaries, described by the multiverse Hilbert space $\sH_{\text{multiverse}}$. Within $\sH_{\text{multiverse}}$, one can construct ``coherent states'' corresponding to emergent hyper-structures, analogous to the construction of D-branes as boundary states in the closed string Hilbert space.
In the cartoon above, a hyper-structure is depicted as a ``brane'' on which spacetime boundaries can anchor. This is purely a visual metaphor; in higher-dimensional gravities, there is no physical target space in which the concept of a brane can be defined.
}
\label{fig:cartoon}
\end{center}
\end{figure}

The underlying logic is inherently hierarchical: the topological necessity of wormholes naturally leads to $\sH_{\text{multiverse}}$, which in turn hosts the higher-level configurations that extend beyond conventional, single-universe gravitational excitations. In other words, wormholes, despite being level-1 transseries objects themselves, point toward the existence of higher-level objects.\footnote{We emphasize that this encoding is a consequence of the topological and algebraic structure of the multiverse Hilbert space, and is entirely distinct from the standard mechanisms of resurgence theory where perturbative asymptotics decode non-perturbative effects.} 
See Fig.~\ref{fig:cartoon} for a cartoon illustration.

\item JT gravity provides a simple example where such double exponential effects can be calculated explicitly (without relying on a matrix model). Another example where these effects can be obtained explicitly is the large angular momentum regime of AdS$_3$ gravity. We caution, however, that this discussion is somewhat speculative and yields rather unconventional results. For instance, we find that the existence of multiverse instantons---which would generate corrections of the form~\eqref{dobG} in the partition function---depends sensitively on the central charge $c$.

\een

\subsection{Ramp and plateau with or without averaging}

Here we briefly review and elaborate on the ramp and plateau phenomena in quantum chaotic systems, which will play an important role in our discussion.

Consider a quantum system with density of states
\be\label{ddos}
\brho(E) = \Tr \delta(E-H) = \sum_n \delta(E - E_n)   \ .
\ee
The spectral two-point correlation function at energy separation $\ep \ll E$ is defined by
\be \label{av01}
R_2 (\ep) = \vev{\brho \le(E + \ha \ep \ri) \brho \le(E-\ha \ep\ri)},  
\ee
with its connected part given by
\be \label{av02}
R_2^c (\ep) =  \vev{\brho \le(E + \ha \ep \ri) \brho \le(E-\ha \ep\ri)} - \vev{\brho \le(E+ \ha \ep\ri)} \vev{\brho \le(E- \ha \ep\ri)} \ .
\ee
In standard treatments, the brackets $\vev{\cdots}$ denote either: (i) an average over an energy window centered around $E$, or (ii) an ensemble average over Hamiltonians, including disorder averages when applicable.

For quantum chaotic systems, $R^c_2(\ep)$  is expected to exhibit the universal behavior predicted by random matrix theory for the appropriate universality class~\cite{BohGia84,Haa10}. For example, in the unitary class, one finds\footnote{The full two-point spectral correlation function includes also a ``trivial'' diagonal contact term $\bar \rho (E) \de (\ep)$. Equation~\eqref{av02} captures ``off-diagonal'' correlations among different levels. } 
\be \label{ekn}
R_2^c  (\ep) 
= -{1 \ov 2  \pi^2 \ep^2} + {\cos (2 \pi \bar \rho (E)  \ep) \ov 2  \pi^2 \ep^2} \ ,
\ee
where $\bar \rho(E)$ denotes the smooth density of states obtained by coarse-graining over an energy window much larger than the mean level spacing. The Fourier transform of~\eqref{ekn} defines the spectral form factor (SFF) 
\be \label{speF}
\vev{Z_{it} Z_{-it}}_c \propto K (t) = \int d \ep \, e^{-i \ep t} R_2^c (\ep) , \quad Z_{it} \equiv \Tr e^{- i Ht}  ,
\ee
where the first term in~\eqref{ekn} generates the linear-$t$ ramp, while the oscillatory cosine term is responsible for the subsequent plateau.

For holographic systems, the ramp has been reproduced on the gravity side by double-cone wormhole geometries~\cite{SaaShe18}, and in pure AdS$_3$ gravity by the off-shell wormhole construction of~\cite{CotJen20}. In particular, in the AdS$_3$ context, the appearance of such random-matrix-like behavior has often been interpreted as evidence that the dual description of pure AdS$_3$ gravity involves an ensemble average over boundary theories.

In contrast, the smooth filter framework of~\eqref{Decom}--\eqref{gid} and~\eqref{worm} implies spectral correlations and wormholes need not originate from statistical averaging, rendering the traditional averages appearing in~~\eqref{av01}--\eqref{av02} and~\eqref{speF} unnecessary. Rather, the entirety of~\eqref{ekn}---both the ramp and the plateau terms---may be obtained from the smooth filter projection of the product of the densities of states:\footnote{$\rho$ below is a Tier-II description. There are subtleties involving the $i\ep$ prescription in the formula below, which will be discussed in detail in the main text.}
 \be \label{filS}
R_2^c (\ep) = \FF\le\{\rho \le(E + \ha \ep \ri) \rho \le(E- \ha \ep \ri)\ri\}_c = -{1 \ov 2  \pi^2 \ep^2} + {\cos (2 \pi \bar \rho (E)  \ep) \ov 2  \pi^2 \ep^2} , \quad \ep \to 0  \ .
\ee
We will derive~\eqref{filS} for chaotic quantum few-body systems by using the Gutzwiller trace formula~\eqref{00poR}, and discuss its  extension to quantum many-body systems. 

For holographic systems, the smooth density of states scales as
\be \label{basl}
\bar \rho (E) \sim e^{S_{\rm micro} (E)} = e^{O(N^2)},
\ee
where the microcanonical entropy satisfies $S_{\rm micro} (E) \sim O(N^2)$  for energy $E \sim O(N^2)$. 
Thus, the plateau term in~\eqref{filS} provides a concrete example of the complex double exponentials~\eqref{1dex} discussed earlier.

Equation~\eqref{filS} may appear surprising at first sight. It is well known that the spectral form factor of a single chaotic system is not self-averaging~\cite{Pra97} (see also~\cite{BraHaa15,AltDiv25} for recent discussions); if one is given the exact eigenvalue spectrum of a system, exposing the smooth ramp and plateau {\it requires} ensemble or energy averaging. How, then, is it possible to extract them in~\eqref{filS} without ensemble or explicit energy averaging? The key is that the Gutzwiller representation~\eqref{00poR} (together with other inputs discussed in detail in Sec.~\ref{sec:II}) operates on a fundamentally different type of ``data" compared to the raw, discrete quantum spectrum. While~\eqref{filS} does not involve an explicit energy average, it secretly involves one: the very act of working with classical periodic orbits in~\eqref{00poR} inherently implies that a level of coarse-graining over the exact quantum spectrum has already taken place. In the context of holography, the proposal~\eqref{gid} suggests that it is precisely this asymptotic boundary data that should be mapped to semiclassical gravity, and this explains how semiclassical gravity reproduces the ramp and plateau without requiring an explicit average over the quantum spectrum in a single system (as we shall discuss in Sec.~\ref{sec:gravity}).

Equation~\eqref{filS} also offers a different perspective on the common view that spectral correlations contain information not directly encoded in the density of states itself: namely, that while $\brho(E)$ merely counts eigenvalues, the correlator $R_2(\ep)$ probes correlations among nearby levels (extracted via ensemble or spectral averages). Equation~\eqref{filS} says that the information contained in the spectral correlator is already encoded in the erratic fine structure of the density of states. A heuristic picture for this is provided by wave interference among closely spaced modes. The  density of states  $\rho(E)$ contains highly erratic components with effective 	``frequencies" determined by typical level spacings. The product $\rho (E+\ep/2)\,\rho (E-\ep/2)$ generates both sums and differences of these frequencies. The sum frequencies oscillate wildly as functions of $E$ and are removed by the smooth projection, whereas the difference frequencies vary relatively smoothly and are controlled by pairs of eigenvalues separated by $\sim \ep$. The smooth projection therefore isolates the information about how eigenvalue pairs are distributed at a given separation---precisely yielding the two-point spectral correlator.

\subsection{Connections with earlier work} 

Here we briefly comment on how our approach relates to earlier work on the single-system interpretation of wormholes and non-averaging in gravity.

In the Marolf-Maxfield model~\cite{MarMax20}, factorization is achieved in an $\alpha$-state, requiring the holographic dual pair to reside in a specific $\alpha$-superselection sector---a structure absent in standard top-down examples of AdS/CFT. The discussion of~\cite{McNVaf20} maintains the $\alpha$-parameter framework while preserving factorization by postulating that the baby universe Hilbert space is strictly one-dimensional. In the current work, which follows~\cite{Liu25c} (see also footnote~\ref{ft:sep} and Sec.~\ref{sec:hmul}), the multiverse Hilbert space $\mathcal{H}_{\text{multiverse}}$ in higher-dimensional gravitational systems is constructed intrinsically from a single boundary system. Our proposal addresses the factorization issue regardless of whether $\mathcal{H}_{\text{multiverse}}$ is non-separable (i.e., regardless of whether $\alpha$-states exist or not). Furthermore, baby universes in this setting possess a nontrivial semiclassical Hilbert space.

In~\cite{SaaShe21}, in the context of 0-dimensional SYK, half-wormhole saddles were discovered that, when added to the standard wormhole saddle, can closely approximate a factorized partition function with fixed couplings (see~\cite{Muk21,Yan25} for further developments). In JT gravity, candidates for half-wormholes are no longer saddles, involving eigenbranes~\cite{BloMer19,BloKru21} or spacetime D-branes~\cite{GarGod21}. For more general systems, half-wormholes can be postulated as whatever gravity configurations restore factorization for bulk gravity dual to a single boundary system, though their precise physical nature remains elusive. 
In our framework, such postulated half-wormholes would correspond to a collective description of the gravity dual to the erratic components of the boundary partition functions. 

In~\cite{HanUrb24}, it was argued that JT gravity can be understood from a single-system perspective by utilizing the spectrum and exact discreteness of periodic orbits on a hyperbolic space, which is conceptually aligned with our approach. See also~\cite{GarZac20}. 

In~\cite{DiUPer23a}, it was shown that the wormhole of~\cite{CotJen20}, as well as the emergence of the ramp in chaotic CFT$_2$ (for arbitrary central charges), admits a Gutzwiller-like representation in a single-system sense---a mechanism closely related to the one advocated here. Their work defined an erratic part of the partition function and introduced a projection procedure (the Hecke projection) to extract the ramp and the wormhole. A crucial difference between their discussion and the one presented here (and in~\cite{Liu25c}) is that the construction in~\cite{DiUPer23a} does not assume a large-$c$ limit, whereas the large-$c$ limit is essential for our decomposition~\eqref{Decom}. It will be valuable to explore these connections more deeply in future work.

In~\cite{PelSon24}, it was argued that the gravitational path integral serves as an effective ``mesoscopic" description of quantum gravity from a single-system perspective, with wormholes capturing moments of erratic microscopic behavior. Our discussion is consistent with this  picture.

In~\cite{KudWit26}, a Mellin-Laplace transform in $N$ was introduced to isolate the asymptotic part of a boundary observable in the large-$N$ limit. This can be regarded as an explicit implementation of a smooth filter. In particular, they analyzed the correlations in $N$ required for such a smooth extraction to emerge. Their procedure can also be shown to be consistent with the projection discussed in the Gutzwiller context of Sec.~\ref{sec:II}.\footnote{J. Kudler-Flam and E. Witten, private communication.}


\bigskip

\noindent{\bf Plan of the paper} 

\medskip

The organization of the paper is as follows. In Sec.~\ref{sec:II}, we review the Gutzwiller trace formula and then introduce smooth filter projection, and then use it to derive the ramp, the rapid macroscopic oscillations and the plateau without averages. In Sec.~\ref{sec:further}, we discuss how to derive some additional features characteristic of random matrix models in a single-system setting such as the spectral curve, and the single-eigenvalue instantons. In Sec.~\ref{sec:gravity}, we first postulate a minimal Gutzwiller-like structure for general chaotic quantum many-body systems, and then discuss gravity constructions of hyper-structures. In Sec.~\ref{sec:ads3}, we explore possible existence of multiverse instantons for quantum gravity in AdS$_3$. In Sec.~\ref{sec:disc} we conclude with some future directions.

\bigskip

\noindent 
{\bf Notations and conventions} 

\medskip

\noindent $\bullet\,$ All quantum systems considered in this paper are assumed to have a discrete spectrum. For a holographic CFT in $d$ spacetime dimensions, this means that the CFT is defined on a compact space such as $S^{d-1}$. 
$H$ always denotes the Hamiltonian of the quantum system under consideration. 

\noindent $\bullet\,$ $z$ refers to a complex coordinate, while $E$ always refers to real energy values. A function $R(z)$ is defined on the full complex plane, whereas $R_+ (E)$ should be understood as being defined from $R(z)$ by taking $z = E+i 0$ with $E \in \RR$.

\noindent $\bullet\,$ The large $N$ (or $N \to \infty$) limit is not a strict mathematical limit in the usual sense, but rather refers to the general behavior of observables when the parameter $N$ is taken to be large. Mathematically, it refers to the structure of~\eqref{Decom} in an asymptotic sense.

\section{Ramp and plateau without averaging} \label{sec:II}

In this section, we illustrate how the ramp and the plateau of a chaotic system can be recovered from a smooth projection 
of the product of density of states, using  the Gutzwiller trace formula.

\subsection{Gutzwiller trace formula}  \label{sec:Gutz}

Consider a few-body quantum chaotic system with Hamiltonian $H(\bx,\bp)$, where $\bx$ and $\bp$ collectively denote the phase-space coordinates. The Gutzwiller trace formula~\cite{Gut71,BalBlo72,Gut90} relates the quantum energy spectrum of the system, in the semiclassical $\hbar \to 0$ limit, to the system's classical periodic orbits.

Consider the resolvent (Tier-I) of the system
\bega \label{rude0}
\bR (z) \equiv \Tr {1 \ov z - H}  = \int dE\, {\brho (E) \ov z - E}  \ .
\end{gather}
The Gutzwiller formula is derived by expressing the values along the real axis $\bR_+(E) \equiv \bR(E_+)$~(where $E_+ =  E+i0$ with $E \in \RR$) in terms of the Green's function,
\be
\bR_+(E) = \int d\bx\, \vev{\bx\le|{1 \ov E_+-H}\ri|\bx} = \int d\bx\, G(\bx,\bx;E_+),
\ee
and in the semiclassical $\hbar \to 0$ limit, evaluating its path-integral representation via the saddle-point approximation. This involves summing over classical paths starting and ending at $\bx$; a further saddle-point evaluation of the $\bx$-integral ensures that the initial and final momenta match, yielding periodic orbits.

This procedure yields a Tier-II representation of the resolvent along the real axis, which we denote with ordinary letters:
\be \label{poR}
R_+(E)  = R_-^* (E) = \bar R (E) - {i \ov \hbar} \sum_a T_a F_a e^{{i \ov \hbar} S_a(E)} + \cdots \ .
\ee
The smooth Weyl approximation $\bar R (E)$ arises from ``zero-length'' paths to the Green's function, where the particle only probes the ``local'' phase-space volume:
\bega \label{weyl0}
\bar R (E) =  \int {d \bp \, d \bx \ov (2 \pi \hbar)^\sN} \, {1 \ov E + i 0 - H (\bx, \bp)} , \\
\quad \Im \, \bar R (E) =- \pi \bar \rho (E), \quad \bar \rho (E) =  \int {d \bp \, d \bx \ov (2 \pi \hbar)^\sN} \, \de (E - H (\bx, \bp)), 
 \label{weyl}
\end{gather} 
where $\sN$ is the number of degrees of freedom. The sum in the second term of~\eqref{poR} runs over {\it primitive periodic orbits} of energy $E$, where $T_a$ and $S_a$ are the period and action of orbit $a$, respectively, with $\p_E S_a(E)=T_a(E)$. The  prefactors $T_a \,F_a $ in the sum arise from the saddle-point integrations around an orbit: integration over coordinates transverse to the orbit yields the stability amplitude $F_a$, while integration along the longitudinal direction~(the zero mode associated with shifts of the base point) evaluates to the primitive period $T_a$. In~\eqref{poR}, $\cdots$  denotes contributions from multiple traversals of primitive periodic orbits. For the observables we consider below, they will not play an important role. 

Equation~\eqref{poR} takes the form of~\eqref{Decom}, with the identification 
\be 
R^{\rm (sm)}_+ (E) = \bar R (E), \quad R^{\rm (err)}_+ (E) =- {i \ov \hbar} \sum_a T_a F_a e^{{i \ov \hbar} S_a(E)}+ \cdots  \ .
\ee
As already mentioned in the Introduction, periodic orbits represent microscopic, or at most mesoscopic, dynamical data of the system, and that the corresponding phase factors $e^{\pm iS_a(E)/\hbar}$ fluctuate wildly in the semiclassical limit. 
Thus, the sum over periodic orbits encodes fine-grained information about the dynamics, albeit using a continuum language. 
By contrast, the Weyl part~\eqref{weyl0}--\eqref{weyl} contains only smooth, coarse-grained information about the phase-space volume.


From~\eqref{poR}, the corresponding Tier-II density of states $\rho(E)$ then takes the form
\bln \label{dosD}
\rho (E) &= {1 \ov \pi} \, {\rm Im}\, R_- (E) = {1 \ov 2 \pi i} \bigl(R_- (E) - R_+ (E)\bigr) \\
\label{dosD1}
&= \bar \rho (E) + {1 \ov \pi \hbar} \sum_a T_a |F_a| \cos \le({S_a (E) \ov \hbar} - {\pi \ov 2} \mu_a \ri) \\
& = \rho^{(\rm sm)}(E)+\rho^{(\rm err)}(E)
 \ .
 \label{dosD2}
\end{align}
Here we have written $F_a = |F_a| e^{- i {\pi \ov 2} \mu_a}$, where $\mu_a$ is the Maslov index, which plays no further role in our discussion.
In~\eqref{dosD2}, we have again identified the two terms in~\eqref{dosD1} as the smooth and erratic parts, respectively. 

The smooth Weyl term  $\bar R (E)$ (and consequently $\bar \rho (E)$) receive further higher-order ``smooth'' corrections as a power series in $\hbar$. They can therefore be identified as the leading-order contributions to the macroscopic Tier-III asymptotic expansion in $\hbar$.
We define the action of filtering projection on these objects as 
\bega \label{cfil0}
\Fil{\bar R (E) } = \bar R (E)  , \quad \Fil{R_+ (E) } = \bar R (E)  , \quad \Fil{\bar \rho (E) } = \bar \rho (E),   \quad \Fil{\rho^{\rm (err)} (E)}=0 \ .
\end{gather}


In the semiclassical limit, the mean level spacing becomes small on fixed classical energy scales, so that the spectrum may be treated as continuous. The semiclassical resolvents $R_\pm (E)$~\eqref{poR} are complex for $E > E_0$, where $E_0$
is the threshold for classically allowed energies. In the classically forbidden region $E< E_0$, we have 
\be\label{forB}
R_+ (E)= R_- (E)= \text{real}, 
\ee 
which can be obtained from~\eqref{poR} via analytic continuation. 

By integrating~\eqref{poR} over $E$, we obtain the Gutzwiller representation for the integrated resolvent
\be\label{inRu}
\bX (z) \equiv \Tr \log (z - H) 
\ee
in the following form,
\bega  \label{defFE0}
X_+ (E) = X_-^* (E) = \bar Q (E) + Y (E) + a ,  \\
\bar Q  (E) = \int^E_{E_0} dE' \, \bar R (E') , \quad  Y (E)  = - \sum_a F_a e^{{i\ov \hbar} S_a (E)} ,
\label{defFE}
\end{gather} 
where in obtaining $Y(E)$ we have assumed $F_a (E)$ are weakly dependent on $E$. Note that $X_\pm (E)$ are defined up to an $E$-independent constant. Here we have fixed this constant by choosing the integration to start from $E_0$.  $a$ is the integration constant 
resulted from the integration of the second term in~\eqref{poR} under this prescription.

We conclude this brief review with some further remarks: 
\ben

\item For a general quantum system, the definition of the resolvent requires a regulator. 
We will discuss this in more detail in Sec~\ref{sec:spec}. In particular, there we show that it is possible to adopt a regularization procedure such that the imaginary part of the resolvent is regulator-independent, and thus only depends on the intrinsic properties of the system. Indeed, the Weyl density of states $\bar \rho$ is manifestly regulator-independence, and so is its integration $\Im \bar Q (E)$. 
A more nontrivial statement is that in~\eqref{defFE0}, $\Im a$ is also regulator-independent. 
For notational simplicity, we will always suppress the explicit regularization. \label{re:1}

\item 
The stability amplitudes $F_a$ satisfy the Hannay--Ozorio de Almeida sum rule~\cite{HanOzo84}:
\be
\sum_a |F_a|^2 f(T_a) = \int_{T_0}^{\infty} {dT \over T}\, f(T)  \ .
\label{sumR}
\ee
Here $f(T)$ is an arbitrary, sufficiently smooth function and $T_0$ is a minimal period beyond which the orbits may be treated as ergodic. The factor $1/T$ in the sum rule has a simple geometric interpretation: a period-$T$ orbit may be based at any point along the trajectory, producing a redundancy proportional to $T$.

\item In the semiclassical limit $\hbar \to 0$, the sum over $a$ in~\eqref{poR} is expected to be convergent for  $E + i 0$ with any infinitesimal $\hbar$-independent $i 0$.  
More explicitly, it can be argued that the sum over $a$ is convergent for ${\rm Im}\,E > \ga$, where $\ga \sim \hbar \left(h_{\rm top} - {1 \over 2}\lambda_u\right)$. Here $h_{\rm top}$ is the topological entropy, governing the exponential growth in the number of periodic orbits, while $\lambda_u$ denotes a typical total unstable expansion rate, characterized by the sum of the positive Lyapunov exponents and controlling the decay of the periodic-orbit stability amplitude. In the $\hbar \to 0$ limit, $\ga$ is thus smaller than any $\hbar$-independent $i 0$. 


\item  There is a celebrated setting where the semiclassical Gutzwiller trace formula becomes mathematically exact: the quantum mechanics of a free particle moving on a compact hyperbolic surface  $\Sigma = \mathbb{H}/\Gamma$, where $\mathbb{H}$ is the two-dimensional hyperbolic space and $\Gamma$ is a discrete, co-compact subgroup  of $\text{PSL}(2,\mathbb{R})$. In this context, the trace formula is known as the Selberg trace formula, and the primitive periodic orbits (closed geodesics) define the Selberg zeta function. Therefore, in this case, despite the continuum appearance of the trace formula, it in fact captures a genuinely discrete spectrum.

\item A main goal of the paper is to explore possible implications of the Gutzwiller representation for chaotic many-body systems. With that in mind,  even in the discussion of few-body systems, we will adopt a language that will naturally generalize to  many-body systems.

In the $\hbar \to 0$ limit (with the number of degrees of freedom $\sN$ fixed), 
the microcanonical entropy $\sS(E)$, defined by 
the average density of states~\eqref{weyl} as $\bar \rho (E) \equiv e^{\sS(E)}$, scales with $\hbar$ as $e^{\sS(E)} \sim {1 \ov \hbar^\sN}$. 
That is,  $\hbar \sim e^{-O(\sS)}$. Consequently, the Weyl term $\bar R (E)$ and its smooth perturbative corrections in powers of $\hbar$ give a sum of exponentials in $\sS$, i.e. a level-1 transseries in $\sS$. In contrast, the oscillatory phase 
\be\label{scW}
e^{iS_a(E)/\hbar} \sim e^{i e^{O(\sS)}}
\ee
for a periodic orbit  scales as a~(complex) double exponential in $\sS$.

The Gutzwiller formula is traditionally derived for few-body systems in the semiclassical regime $\hbar \to 0$ with $\sN$ fixed. For quantum many-body systems, including those relevant to holography, one is instead interested in the limit
\be \label{manB}
\sN \to \infty \quad \text{with $\hbar$ fixed} \ .
\ee
In particular, we focus on the extensive many-body regime, where both the energy and microcanonical entropy scale linearly with the system size, $E \sim \mathcal{O}(\sN)$ and $\sS(E) \sim \mathcal{O}(\sN)$, resulting in an average density of states $\bar\rho(E) = e^{\sS(E)} \sim e^{\mathcal{O}(\sN)}$.

Assuming a trace formula like~\eqref{poR} holds in the many-body regime, we again expect that the oscillatory phase associated with a periodic orbit to scale as~\eqref{scW}. 
 In a many-body microstate, a collective periodic orbit requires the system to return to its initial configuration in all degrees of freedom. For instance, in a toy model of non-interacting particles with incommensurate single-particle periods $T_i$, the period $T_a$ of a collective many-body orbit becomes exponentially  large as one increases $\sN$, and it is natural to expect exponential growth with $\sN$ in generic interacting systems as well. We therefore have $T_a\sim e^{c_a \sN}$. Since $\p_E S_a = T_a$, this implies that $S_a$ grows like $\sN e^{c_a \sN}$. Therefore, the oscillatory terms involve ``double exponentials'' of the form $e^{i\,\sN e^{c_a \sN}} \sim e^{i e^{O(\sS)}}$. This fine-grained, extreme oscillatory behavior is not a mathematical artifact; it is precisely the kind of fine-grained structure needed to probe physics at the many-body level-spacing scale, which is exponentially small, scaling as $e^{-O(\sN)}$.

Therefore, in both few-body and many-body systems, $\bar \rho$ constitutes a level-1 transseries with respect to the microcanonical entropy $\sS$, whereas expressions such as $e^{i \bar \rho}$ and $e^{- \bar \rho}$ correspond to level-2 transseries.\footnote{While we formally use the microcanonical entropy $\sS \equiv \ln \bar{\rho}$ as a unifying organizing parameter for both few-body and many-body systems, we note a crucial difference in their scaling behaviors. In many-body systems, both energy $E$ and entropy $\sS$ are extensive, leading to the linear relation $E \sim \sS$. In contrast, for few-body systems in the semiclassical limit, $E \sim {1 \ov \hbar} \sim e^{O(\sS)}$.}

We emphasize that these transseries terms strictly refer to smooth, macroscopic quantities. For instance, the double exponential $e^{iS_a(E)/\hbar}$ does not qualify as a transseries term, as it inherently belongs to an erratic, microscopic description.

\een

\subsection{Ramp from smooth filter projection} \label{sec:ramp1}

We now examine the product of $\rho^{(\rm err)}$, as defined in~\eqref{dosD1}--\eqref{dosD2}, demonstrating that a projection to its smooth component precisely yields the ramp.
At the technical level, the analysis is a quick adaptation of the classic result of~\cite{Ber85}.


For this purpose, it is convenient to write the erratic part as an explicit sum over conjugate phases:
\bln  \label{yup0}
\rho^{(\rm err)} \le(E + {\ep \ov 2} \ri) & ={1 \ov 2 \pi \hbar} \sum_{a} \sum_{n=-}^+  A_{na}  
 \exp  \le[i n {S_{a} (E + {\ep_n \ov 2}) \ov \hbar} \ri] \\
 & = {1 \ov 2 \pi \hbar} \sum_{a} \sum_{n=-}^+  A_{n a}  
 \exp  \le[i n {S_{a} (E) \ov \hbar}  + {i n \ov 2 \hbar} T_a \ep_n  \ri], 
 \label{yup}
  \end{align} 
where we have taken $\ep \ll E$ and introduced 
\be
A_{+a} = T_a F_a, \quad A_{-a} = T_a F_a^* , \quad \ep_\pm = \ep \pm i 0 \ .
\ee
In~\eqref{yup0}--\eqref{yup}, the signs of the imaginary parts of $\ep_\pm$ come from their origins in~\eqref{poR}. 
It then follows that 
  \bln
 & \rho^{(\rm err)} \le(E + {\ep \ov 2} \ri)  \rho^{(\rm err)} \le(E - {\ep \ov 2} \ri) \cr
 & = {1 \ov (2 \pi \hbar)^2} \sum_{a_1, a_2, n_1, n_2} A_{n_1 a_1 } A_{n_2 a_2 } 
e^{{i\ov \hbar} (n_1 S_{a_1} (E)  + n_2 S_{a_2} (E) )  + {i  \ov 2 \hbar} (n_1T_{a_1} \ep_{n_1} - n_2 T_{a_2} \ep_{n_2}) } \ .
\label{rPro}
  \end{align}
   
We now extract the smooth part of~\eqref{rPro} by keeping only terms where the rapidly oscillating phases  ${i\ov \hbar} (n_1 S_{a_1} (E)  + n_2 S_{a_2} (E) )$ are canceled. Assuming the periodic orbits are non-degenerate---meaning $S_{a_1} (E) = S_{a_2} (E)$  if and only if $a_1 = a_2$, as is generically expected for chaotic systems lacking time-reversal or other specific symmetries---this cancellation occurs strictly when $n_1 = - n_2$ and $a_1 = a_2$. We then find  
     \bln
& \FF \le\{\rho^{(\rm err)} \le(E + {\ep \ov 2} \ri)  \rho^{(\rm err)} \le(E - {\ep \ov 2} \ri) \ri\} \cr
\label{der0} 
 & =  {1 \ov (2 \pi \hbar)^2}  \sum_{a} |A_{+a}|^2 \le(e^{{i  \ov \hbar} \ep_+ T_a} + e^{-{i  \ov \hbar} \ep_- T_a} \ri) \\
  & = {1 \ov (2 \pi \hbar)^2} \int_0^\infty {dT \ov T} T^2  \le(e^{{i  \ov \hbar} \ep_+ T} + e^{-{i  \ov \hbar} \ep_- T} \ri) 
= - {1 \ov 2 \pi^2 \ep^2}  ,
\label{der1}
  \end{align}
precisely recovering the universal ramp term in~\eqref{ekn}. In the first equality of~\eqref{der1}, we used the sum rule~\eqref{sumR} (noting $|A_{+a}|^2 = T_a^2 |F_a|^2$) and extended the lower integration limit $T_0$ 
to $0$. This extension is justified because for sufficiently small $\ep$, the integral is heavily dominated by long periodic orbits, with corrections suppressed by powers of $\ep T_0$. 

While the Fourier transform of~\eqref{der1} leads directly to the linear-$t$ ramp, it is instructive to examine the calculation  
more explicitly in the time domain, which essentially repeats Berry's original calculation~\cite{Ber85}.
With a complex inverse temperature $\b+i t$ (where, for definiteness, we take $t$ is positive), 
the erratic part of the thermal partition function is
\be \label{uen} 
 Z_{\b+ i t}^{(\rm err)}   = \int dE \, \rho^{\text{(err)}} (E) e^{-i  E t - \b E} = 
   {1 \ov 2 \pi \hbar}\sum_{a} \sum_{n=-}^+ \int dE \, A_{na}e^{i n {S_a (E) \ov \hbar} } e^{-i  E t - \b E}  \ .
\ee
It then follows that the product representing the spectral form factor is (with $t$ large)
\bln
 Z_{\b+i t}^{(\rm err)}  Z_{\b-i t}^{(\rm err)}  & =  {1 \ov (2 \pi \hbar)^2} \sum_{a_1, n_1, a_2, n_2}\int dE_1 dE_2 \,  A_{n_1 a_1} A^*_{n_2a_2} 
e^{i n_1 {S_{a_1} (E_1) \ov \hbar} - 
i n_2 {S_{a_2} (E_2 ) \ov \hbar} }  e^{- i t (E_1 - E_2) - \b (E_1+E_2)}  \cr
& = {1 \ov (2 \pi \hbar)^2}   \sum_{a_1, n_1, a_2, n_2}\int dE \, e^{-2 \b E} \cr
& \times \int d \om \,  A_{n_1 a_1} A^*_{n_2a_2} 
e^{i n_1 {S_{a_1} (E) \ov \hbar} - 
i n_2 {S_{a_2} (E ) \ov \hbar} }  e^{i {\om \ov 2\hbar } (n_1 T_{a_1} + n_2 T_{a_2}) - i t \om } 
\end{align} 
where in the second line we have introduced $E_1 = E + {\om \ov 2}$, $E_2 = E-{\om \ov 2}$,
and expanded $S_{a_1} (E_1)  = S_{a_1} (E) + {\om T_{a_1} \ov 2}$. 
For sufficiently large $t$, the rapid oscillations of $e^{-i \om t}$ heavily suppress the $\om$-integral everywhere except in the small-$\om$ region,  justifying this linear expansion.
Extracting the smooth component via the projection $\Fil{\cdot}$ again isolates the diagonal $a_1 = a_2$ and $n_1 = n_2$ terms. This leads to 
\bln
\Fil{Z_{\b + i t}^{(\rm err)}  Z_{\b-i t}^{(\rm err)}} & = {1 \ov (2 \pi \hbar)^2} \int dE \, e^{-2 \b E} \sum_{a, n} |A_{na}|^2  
\int d \om \, e^{{i \ov \hbar} \om n T_{a}  - i t  \om }  \\
\label{ber1}
& =  {1 \ov 2 \pi \hbar^2}  \int dE \, e^{-2 \b E} \sum_{a} |A_{+a}|^2  
\de\le({T_{a} \ov \hbar}  -  t \ri) \\
\label{ber2}
& = {1 \ov 2 \pi \hbar^2}  \int dE \,  e^{-2 \b E} \int_{T_0}^\infty {dT \ov T} T^2 \de\le({T \ov \hbar}  -  t \ri)  = 
 {t \ov 2 \pi} \int d E \,  e^{-2 \b E}
\end{align} 
where in the last line we have again used the sum rule~\eqref{sumR}. 
Notice that because we chose $t>0$ and periods $T_a$ 
are strictly positive, only the $n=+$ term survives the delta function constraint in~\eqref{ber1}. We also note that the delta function 
$\de\le({T_{a} \ov \hbar}  -  t \ri)$ can only be saturated at physical time scales of order $t \sim O(1/\hbar)$.

There is a simple geometric reason for the linear $t$ dependence~\eqref{ber2}. The diagonal projection of the spectral form factor pairs two copies of the same orbit with period $T= \hbar t$. Each copy brings a factor of $T$ from the integration over its zero mode, giving $T^2$. We can view the $1/T$ factor in the periodic orbit measure as canceling out the common zero mode between the two copies. The remaining factor of $T$ giving the linear ramp corresponds to the integration over their relative zero mode.\footnote{This matches well with the gravity description of the linear ramp~\cite{SaaShe19,CheIvo23}, see further discussion in Sec.~\ref{sec:holo}.}

Note that for $t \gg \b$ we can obtain an approximation to $Z_{\b + i t}^{\rm (err)}$ by doing a saddle point approximation in~\eqref{uen}. Taking the product of the resulting object does not yield the linear ramp, as it misses other contributions from the full product of the erratic part. 

We can also consider the product of Euclidean thermal partition functions, with $\b_1, \b_2 > 0$, 
\bln
 Z_{\b_1}^{(\rm err)}  Z_{\b_2}^{(\rm err)}  & =  {1 \ov (2 \pi \hbar)^2} \sum_{a_1, n_1, a_2, n_2}\int dE_1 dE_2 \,  A_{n_1 a_1} A^*_{n_2a_2} 
e^{i n_1 {S_{a_1} (E_1) \ov \hbar} - 
i n_2 {S_{a_2} (E_2 ) \ov \hbar} }  e^{- \b_1 E_1 - \b_2  E_2 } \ .
\end{align} 
While we still expect a smooth diagonal projection to exist, the purely Euclidean nature of the product means the $\om$-integral no longer possesses a saddle point. Consequently, the filtered product $\Fil{  Z_{\b_1}^{(\rm err)}  Z_{\b_2}^{(\rm err)}}$ does not yield a universal expression.

To conclude this discussion, we make a few further remarks:
\ben

\item In the periodic orbit literature, the mechanism of pairing $a_1 = a_2$ and canceling the phases is known as the ``diagonal approximation,'' which is traditionally justified via explicit energy averaging~\cite{Ber85}. 
Here, although the mathematical procedure is similar, we would like to emphasize that smooth projection and averaging are conceptually distinct.

\item One can also include the contributions from multiple traversals of primitive periodic orbits. The stability amplitude $|F_{na}|$ for $n>1$ traversals of an orbit $a$ typically decays as $|F_{na}|/|F_{a}| \sim e^{-\lam (n-1) T_a}$ (where $\lam>0$ is governed by the Lyapunov exponents of the system). Consequently, these higher-winding contributions yield only exponentially small, dynamically negligible corrections to the ramp.

\item The abstract definition~\eqref{0CC0} of the smooth filter is a procedure that isolates the transseries part of a quantity, as determined by the decomposition~\eqref{Decom}. While the Gutzwiller representation provides this decomposition for the density of states, we currently lack a first-principles method for decomposing products or arbitrary functions of the density. In~\eqref{der0}, we only retained strictly diagonal terms where the rapidly oscillating phases perfectly cancel; this should be understood as the minimal contribution that any proper projection must include. Indeed, we expect that the projection $\Fil{\cdot}$ should also capture off-diagonal pairs $(a_1, a_2)$ whose action difference $S_{a_1}(E)-S_{a_2}(E)$ is non-zero but of order $O(\hbar)$. Such correlated orbit pairs (known as Sieber-Richter pairs~\cite{SieRic01})---which arise from close self-encounters in phase space---do indeed exist, but they do not affect our result, as they only generate higher-order corrections to~\eqref{der1}~(see, e.g.,~\cite{HeuMul07,MulHeu09}). In the subsequent discussion, we will again focus solely on the diagonal part, which suffices for our purposes.

\item If the system possesses time-reversal symmetry, every generic orbit $a$ is paired with a distinct time-reversed partner having the exact same action and stability. This exact degeneracy yields an overall factor of $2$ in~\eqref{der1}, which perfectly matches the universal ramp coefficient for orthogonal random-matrix ensemble  as opposed to unitary ones.


\item  There is an important special case: the quantum mechanics of a free particle moving on $\Sigma = \mathbb{H}/\Gamma$, where $\Ga$ is an arithmetic subgroup of $\text{PSL}(2,\mathbb{R})$, such as the modular group $\text{PSL}(2,\mathbb{Z})$. In this context---known as arithmetic quantum chaos (see~\cite{BogGeo97} for a review)---the standard non-degeneracy assumption that $S_{a_1} (E) = S_{a_2} (E)$ if and only if $a_1 = a_2$ is drastically violated. Instead, arithmetic surfaces feature massive exact degeneracies, where exponentially many distinct primitive closed geodesics share precisely the same action (which is proportional to the geodesic length).

Under the smooth filtering projection $\Fil{\cdot}$, the standard diagonal contribution is now accompanied by contributions from an enormous number of off-diagonal pairs ($a_1 \neq a_2$, but $S_{a_1} = S_{a_2}$), whose rapidly oscillating phases also perfectly cancel. As a result, the spectral form factor is significantly modified by these arithmetic degeneracies and does not exhibit the universal  linear ramp, a result  consistent with the known anomalous spectral statistics of arithmetic chaotic systems.

\item Even after cancelling the rapidly oscillating phases via the projection in~\eqref{der0}, a ``microscopic'' sum over periodic orbits remains. It is the powerful sum rule~\eqref{sumR} that ultimately transforms this discrete sum into a smooth, macroscopic quantity.

\een

The statistics of the Riemann zeros also exhibit a ramp~\cite{Mon73,Odl87}.
In Appendix~\ref{app:Rie}, we show that the smooth filter projection can
also be used to derive this ramp.

\subsection{Dressed resolvents and rapid macroscopical oscillations in the density of states}\label{sec:MDOS}

As preparation for deriving the plateau using the smooth filter projection, we first establish a result of a different nature: a ``macroscopically determined oscillatory component'' of the density of states. The techniques introduced to isolate this term will lay the necessary groundwork for the subsequent plateau derivation.

Applying the smooth filter to the standard Gutzwiller trace formula~\eqref{dosD1} for the density of states (or~\eqref{poR} for the resolvent) yields the Weyl term $\bar \rho$, i.e., we find 
\be \label{egn} 
\brho^{\rm (sm)} (E) = \Fil{\rho (E)} =  \bar \rho (E)  \ .
\ee
As discussed following~\eqref{dosD2}, $\bar \rho (E)$ should be viewed as the leading term in a power series expansion in $\hbar$. 
Below, we introduce an alternative Tier-II representation for the resolvent, from which we can improve~\eqref{egn} by including a highly oscillatory expression of the form 
\be\label{eub}
\boldsymbol{\rho}^{\rm (sm)} (E)= \bar \rho (E) - {e^{-c} \ov \pi} \cos (2\pi \bar N (E) - 2 \phi) , \quad \bar N (E) = \int^E_{E_0} d E' \, \bar \rho (E') ,
\ee
where $c$ is some (non-universal) constant, and $\bar N (E)$ is the (mean) total number of states below $E$ (with $\bar N' (E) = \bar \rho (E)$). The argument $2\pi \bar{N}(E) - 2\phi$ of the cosine has a simple interpretation:  the oscillation completes one full cycle each time $E$ passes through a coarse-grained eigenvalue position. This is intuitive: the actual level staircase jumps by 1 at each eigenvalue, while the smooth approximation increases continuously, so their difference oscillates once per level.

As discussed earlier, $\bar \rho (E)$ scales exponentially with the microcanonical entropy $\sS$. Consequently, the argument of the cosine term $\bar N(E)$ also scales exponentially, resulting in exponentially rapid oscillations. Since the cosine decomposes into phase factors of the form $e^{\pm i 2\pi \bar{N}(E)} \sim e^{\pm i e^{O(\sS)}}$, this contribution manifests fundamentally as a complex double-exponential. It therefore acts as a level-2 transseries term\footnote{Recall the discussion around~\eqref{cftE}.}, contrasting with standard power series in $\hbar$ (which scale as single exponentials in the entropy) that constitute the level-1 transseries. Nevertheless, because this new oscillatory term depends exclusively on the smooth mean density $\bar \rho (E)$, it should be viewed as a macroscopic Tier-III quantity. Such oscillatory behavior has a well-known counterpart in random matrix theory---see, e.g., \cite{MalMoo04,For10} and~\cite{SaaShe19} for a recent discussion---and our approach draws inspiration from them.

To derive~\eqref{eub}, we introduce ``dressed'' resolvents and a dressed density of states defined by\footnote{The following definition is mathematically equivalent to $R (E_\pm) = - \p_{E_2} Z_2 (E_1, E_{2\pm}) |_{E_1=E_2=E}$, which is also frequently used in the literature. However, as we will discuss later, the definition provided here is physically more transparent. It was used in the context of random matrix models in~\cite{SaaShe19}.}
\begin{gather} \label{ienP0}
\tilde R_\pm (E) = \lim_{\ep \to 0} R(E + \ep_\pm) Z_2 (E, E+ \ep_\pm)  , \quad \ep_\pm = \ep \pm i 0, \quad E \in \mathbb{R}, \\
\tilde \rho (E) \equiv  - {1 \ov \pi} \, {\rm Im}\, \tilde R_+ (E) = {1 \ov 2 \pi i} \bigl(\tilde R_- (E) - \tilde R_+ (E)\bigr) ,
 \label{ienD0}
\end{gather} 
where we place the resolvent in a ``dipole background'' $Z_2 (E, E+ \ep_\pm)$ consisting of the ratio of spectral determinants 
\be \label{genF0}
Z_2 (E_1, E_2) \equiv D (E_1) \tilde D (E_2)\ .
\ee
Here $D (E)$ and  $\tilde D (E)$ are Gutzwiller representations of the spectral determinant $\bD (z)$ and its inverse $\tilde \bD (z)$ evaluated on the real axis, respectively,\footnote{Equation~\eqref{defBD} prescribes a specific normalization for $\bD (z)$ with the assumption that a specific additive constant has been fixed for $\bX (z)$. See Sec.~\ref{sec:spec} for more elaboration.}
\be\label{defBD}
\bD (z) \equiv \det (z-H) = e^{\bX (z)}, \quad \tilde \bD (z) \equiv {1 \ov \det (z-H)} = e^{-\bX (z)}  \ .
\ee

Naively, in the $\ep \to 0$ limit, the second factor on the right-hand side of~\eqref{ienP0} approaches unity, rendering the operation seemingly trivial. This is indeed the case in the exact Tier-I treatment. Here, however, the quantities on the right-hand side are the corresponding Gutzwiller representations. While the exact quantities satisfy by definition
\be 
\bD (z) \tilde \bD (z) =1, 
\ee
as we shall see below, their Gutzwiller representations do not invert exactly, 
\be \label{inED}
D (z) \tilde D(z) \neq 1 ,
\ee
and only do so under the smooth filtering projection 
\be \label{Fineq}
\Fil{D(z) \tilde D(z)} = 1 \ .
\ee

Consequently,~\eqref{ienP0} should be understood as a new Tier-II representation for the resolvent. 
In particular, as we shall see, under the smooth filter projection $\Fil{\cdot}$, this dressing yields a finite residual effect---the oscillatory term of~\eqref{eub}---that survives the $\ep \to 0$ limit.

We now proceed to give the Gutzwiller representations $D(E)$ and $\tilde D(E)$. As discussed earlier, in the semiclassical limit, the spectrum of a quantum chaotic system is treated as continuous, resulting in a branch cut in the resolvent $R(z)$ and its integrated version $X(z)$ along the real axis. One might naively expect a similar branch cut for the determinant $D(z)$ and its inverse $\tilde D(z)$, given the relation~\eqref{defBD} between their exact counterparts and $\bX (z)$. 
However, as we shall see momentarily, $D(E)$ and $\tilde D(E)$  behave very differently. 

At finite $\hbar$, we have 
\be\label{seNd}
\bD (z) =  \det (z-H)= \prod_i (z-E_i), \quad \tilde \bD (z) = {1 \ov \det (z-H)} 
= \prod_i {1 \ov z-E_i} \ .
\ee
The determinant $\bD (z)$ is an entire function with zeros along the real $E$-axis. In the semiclassical limit, where the spectrum becomes dense and the mean level spacing goes to zero, the values of $\bD (z)$ evaluated at $E + i\epsilon$ and $E - i\epsilon$ remain identical throughout the limit process. We thus expect that the branch cut discussed earlier for the resolvent $R(z)$ is absent in $D(z)$; i.e., $D(E)$ is well-defined on the real axis and is strictly real. This explains why there is no $\pm$ label for the first argument of~\eqref{ienP0}. 

In contrast, for $\tilde \bD (z)$, just as in the case of $\bR(z)$, the densely packed singularities in the continuum limit cause the function's amplitude to fluctuate wildly and diverge. We are therefore forced to introduce a finite $\pm i\eta$ regulator that explicitly breaks the symmetry between the upper and lower half-planes to extract any meaningful physics. The underlying mechanism is governed by the order of limits: when $\eta \ll \de$ (where $\de$ is the mean level spacing), the discrete states are well-resolved and no macroscopic discontinuity exists across the real axis. The continuum limit, conversely, requires fixing $\eta$ while taking $\de \to 0$. In this regime, the $\pm i \eta$ regulator effectively smears over many neighboring eigenvalues, yielding inequivalent limits for the upper and lower half-planes.

 Given~\eqref{defBD}, it is convenient to introduce the exponential of~\eqref{defFE0} 
\be
\De_+ (E) \equiv e^{X_+ (E)} = C (E)   \exp \le(- i \pi \bar N (E) +  Y (E)  \ri), \quad \De_- (E) = \De_+^* (E) ,
\ee
where we have introduced $C(E) = e^a e^{\Re \bar Q (E)}$ and used  that $\Im \bar Q (E) = - \pi \bar N (E)$. Note that the phase of 
$C(E)$ 
\be 
C (E) = |C(E)| e^{i \phi}, \quad \phi = \Im a, 
\ee
is regularization-independent (recall remark~\ref{re:1} of Sec.~\ref{sec:Gutz}). 
Because they are constructed directly from $R_\pm (E)$, the functions $\De_\pm (E)$ are initially defined only within their respective upper and lower half-planes.

The inverse determinant can be identified with these half-plane functions via 
\be \label{inD}
\tilde D_\pm (E) =  \De^{-1}_\pm (E)  \ .
\ee
However, neither $\De_+ (E)$ nor $\De_- (E)$ can individually represent the full determinant $\det (E-H)$, which is strictly real-valued along the real axis. Instead, we formally reconstruct $\det(E-H)$ on the real axis by summing its limiting boundary values from both half-planes:
\be \label{DD}
D(E)= \De_+ (E) + \De_- (E)   \ .
\ee
This additive prescription is natural, as it combines the convergent semiclassical data from the upper and lower half-planes to yield a manifestly real expression---a structure familiar from the Stokes phenomenon\footnote{Equation~\eqref{DD} can be checked explicitly in the WKB limit in simple solvable models.}. It can be considered as a slight variant\footnote{The proposal of~\cite{BerKea90,BerKea92,Kea92} placed an upper cutoff on the periods of pseudo-orbits, which we do not impose here. The presence or absence of this cutoff does not affect our subsequent discussion.} of the Berry--Keating framework~\cite{BerKea90,BerKea92,Kea92}, which is based on the Riemann--Siegel type formula for the Riemann $\ze$-function. Furthermore, in the context of random matrix theory, this additive prescription matches explicit calculations established in~\cite{MalMoo04}.

With these notations and identifications, the ratio of determinants in~\eqref{ienP0} takes the form
\be\label{Dip}
Z_2 (E, E+ \ep_+)  = \sZ^+_+  + \sZ^-_+, \quad   Z_2 (E, E+\ep_-)  = \sZ^+_-  + \sZ^-_-,
\ee
where we have defined the individual components as
\be 
\sZ^{s_1}_{s_2} (E_1, E_2) \equiv \De_{s_1} (E_1) \De_{s_2}^{-1} (E_2), \quad s_i \in \{+, -\} \ .
\ee
From~\eqref{Dip}, we see that the dipole background $Z_2 (E, E+ \ep_\pm)$ decomposes into a sum over two distinct structural classes:

\bi
\item Cis-dipoles ($\sZ^+_+$ or $\sZ^-_-$): The functions $\De$ and $\De^{-1}$ are evaluated on the same side of the branch cut, meaning they live on the same branch.

\begin{figure}
\begin{center}
\includegraphics[width=14cm]{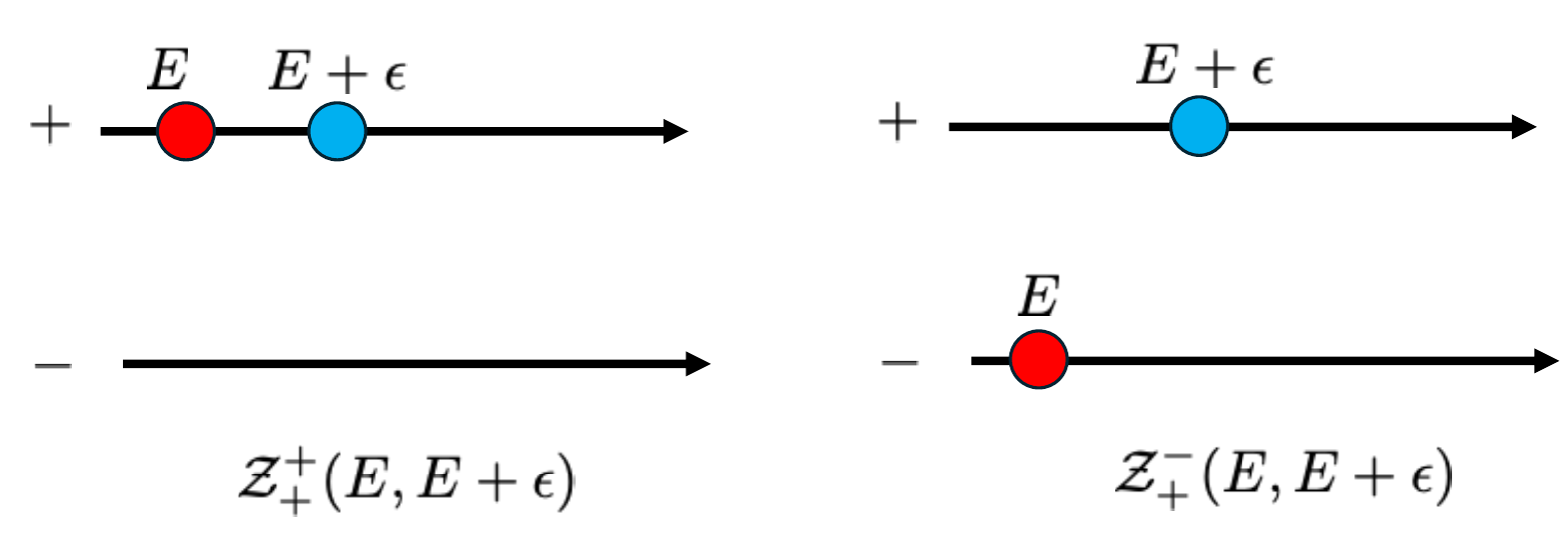}
\caption{\small Cis-dipole and trans-dipole. (a): $ \sZ^{+}_+ (E, E+\ep)$. (b): $ \sZ^{-}_+  (E, E+\ep)$. The two horizontal lines represent the upper $(+)$ and lower $(-)$ banks of the cut along the $E$-axis. Alternatively, by analytically continuing the values from the lower bank under the cut onto the adjacent Riemann sheet, these lines can also be viewed as representing the $E$-axis on two neighboring sheets. The filled red and blue circles denote $\De (E)$ and  $\De^{-1} (E)$, respectively.
}
\label{fig:sheet}
\end{center}
\end{figure}

\item Trans-dipoles ($\sZ^+_-$ or $\sZ^-_+$): The functions $\De$ and $\De^{-1}$ are evaluated on opposite sides of the branch cut, meaning they reside on different branches.
\ei
See Fig~\ref{fig:sheet} for an illustration.

Adopting the shorthand notations $\bar N_i \equiv \bar N(E_i)$ and $Y_i \equiv Y(E_i)$, the explicit expressions for the dipole components are given by
\bln \label{zz1}
 \sZ^{+}_+ (E_1, E_2)  &= e^{i \pi (\bar N_2 - \bar N_1 )}  e^{Y_1 - Y_2} =   (\sZ^{-}_-  (E_1, E_2))^* , \\
 \sZ^{-}_+  (E_1, E_2)  &= e^{-2 i \phi} e^{i \pi (\bar N_2 + \bar N_1)} e^{Y_1^* - Y_2} =  ( \sZ^{+}_-  (E_1, E_2) )^* \ .
 \label{zz2}
\end{align} 
In writing these expressions, we have suppressed the regularization-dependent amplitude ratio ${|C (E_1)| \ov |C(E_2)|}$, which safely approaches unity and drops out of the analysis in the $E_1 \to E_2$ limit of interest.

We now evaluate how the diagonal filter acts on these configurations. Following the discussion of Sec.~\ref{sec:ramp1}, we isolate the diagonal piece for a cross-conjugated product such as $Y_1 Y_2^*$:
\be\label{sp1}
\Fil{Y_1 Y_2^*}  = \sum_a  |F_a|^2 e^{{i \ov \hbar} (S_a (E_1) - S_a (E_2))} \equiv g_{+-} (E_1, E_2 ) \ .
\ee
Taking the complex conjugate implies that $g_{+-}$ should satisfy the condition 
\be\label{csn1} 
 g_{+-}^* (E_1, E_2 ) = g_{+-} (E_2, E_1) \ .
 \ee
It must be kept in mind that $Y(E)$ and $Y^* (E)$ inherit their analytic structures directly from $R_+ (E)$ and $R_- (E)$, respectively. Consequently, $Y(E)$ is implicitly evaluated at $E+i0$, whereas $Y^* (E)$ requires $E-i0$. 

In the limit $E_1 \to E_2$, we can evaluate $g_{+-}(E_1,E_2)$ by invoking the Hannay--Ozorio de Almeida sum rule~\eqref{sumR}:
\be \label{sp2}
g_{+-}(E_1,E_2) = \int_{T_0}^\infty {dT \ov T}  e^{{i \ov \hbar} T (E_1 -E_2)}
= - \log (- i (E_1 -E_2 +i 0)) + c + O(E_1-E_2) \ ,
\ee
where $c \equiv -\ga - \log (T_0/\hbar)$ is a non-universal real constant depending on the minimal ergodic period $T_0$, and $\ga$ is the Euler--Mascheroni constant.

Because there is no corresponding diagonal pairing available for non-conjugated combinations, the diagonal filter yields $\Fil{Y_1 Y_2} = 0$. Nevertheless, with future generalizations to quantum many-body systems in mind, we parameterize these correlations more generally as
\bega\label{spff}
\Fil{Y_1 Y_2}  = g_{++} (E_1, E_2 ) = g_{++} (E_2, E_1), \\ 
\Fil{Y_1^* Y_2^*}  = g_{--} (E_1, E_2 )  = (g_{++} (E_1, E_2))^* ,
\label{spff1}
\end{gather}
and require $g_{++}$ to be regular in the limit $E_1 \to E_2$. This implies that 
\bega \label{jhs0}
 g_{++} (E+\ep/2, E-\ep/2) = g (E) +O(\ep^2) \quad \text{as } \ep \to 0,
 \end{gather} 
for some regular function $g(E)$. We will also slightly generalize~\eqref{sp2} by writing it as 
 \bega
g_{+-} (E+\ep/2, E-\ep/2) =  - \log (- i (\ep +i 0)) + c (E)+ i O(\ep) \quad \text{as } \ep \to 0 ,
\label{jhs}
\end{gather} 
allowing the constant $c$ there to depend on energy. It follows from~\eqref{csn1} that $c(E)$ is real and the $O(\ep)$ correction must be purely imaginary.

The filtered expressions for the exponentials in~\eqref{zz1}--\eqref{zz2} are obtained by expanding the terms involving $Y$, applying the diagonal projection term by term, and re-exponentiating the result. Since~\eqref{sp1} and~\eqref{spff}--\eqref{spff1} involve only  two-point ``correlations,'' we thus find for any expression $A$ consisting of a linear sum of these quantities\footnote{There can in principle be contribution from non-Gaussian correlations, which we assume is suppressed.},
\be\label{Faex}
\Fil{e^A} = e^{\Fil{A} + \ha \Fil{A^2}_c}  \ .
\ee 
For example (for $E_1 = E + \ep/2$ and $E_2 = E-\ep/2$),
 \bln\label{sp22}
\Fil{e^{Y_1 - Y_2^*} }  = e^{\ha \Fil{Y_1^2} + \ha \Fil{(Y_2^*)^2} -\Fil{Y_1 Y_2^*}} 
\stackrel{\ep \to 0}{=} - i e^{\hat g (E) -c (E)} \ep , \quad \hat g (E) \equiv \Re g (E) \ .
\end{align}

Applying this procedure to~\eqref{zz1}--\eqref{zz2}, we find 
\bega \label{eonr}
\Fil{\sZ^+_+ (E_1, E_2)} =e^{i \pi (\bar N_2 - \bar N_1 ) + \ha g^{++}_{11} + \ha g^{++}_{22} - g_{12}^{++}},   \\
\Fil{\sZ^-_+ (E_1, E_2)} = e^{i \pi (\bar N_2 + \bar N_1) - 2 i \phi} e^{\ha g_{11}^{--} + \ha g_{22}^{++}- g^{+-}_{21}}  ,
\label{eonr1}
\end{gather} 
where we have abbreviated $g_{+-} (E_1 , E_2)$ as $g_{12}^{+-}$, and similarly with the others. 
As $\ep \to 0$, equation~\eqref{eonr} gives $1$, while~\eqref{eonr1} vanishes  linearly in $\ep$. Thus $\lim_{\ep \to 0} \Fil{Z_2 (E, E+ \ep_+)} =1$, and~\eqref{Fineq} is satisfied.  

We can now readily find the filter projection of the dressed resolvent~\eqref{ienP0}:
\be 
\Fil{\tilde R_+ (E)} = \lim_{\ep \to 0} \le[\Fil{R_+ (E+\ep) \sZ^+_+ (E, E+\ep)} + \Fil{R_+ (E+\ep) \sZ^-_+ (E, E+\ep)}  \ri],
\ee
where (evaluating at $E_1 =E, E_2 = E+\ep$):
\bln
& \lim_{\ep \to 0} \Fil{R_+ (E+\ep) \sZ^+_+ (E, E+\ep)}  =\Fil{R_{+} (E)} ,  \\
& \lim_{\ep \to 0}  \Fil{R_+ (E+\ep) \sZ^-_+ (E, E+\ep)}  
= e^{2 \pi i \bar N (E)-2 i \phi} \Fil{\p_2 Y_{2} Y_{1}^*}_c \Fil{e^{Y_{1}^* - Y_{2}}} \cr
& \qquad = \lim_{\ep \to 0} e^{2 \pi i \bar N (E)-2 i \phi} \p_2 g_{21}^{+-}  e^{\hat g (E) - g_{21}^{+-}}
= i e^{\hat g (E)- c (E)} e^{2 \pi i \bar N (E)-2 i \phi}, 
\label{transC}
\end{align} 
and we have used the dentity $\Fil{B e^A} = \le(\Fil{B} + \Fil{BA}_c\ri) \Fil{e^A}$.

\begin{figure}
\begin{center}
\includegraphics[width=14cm]{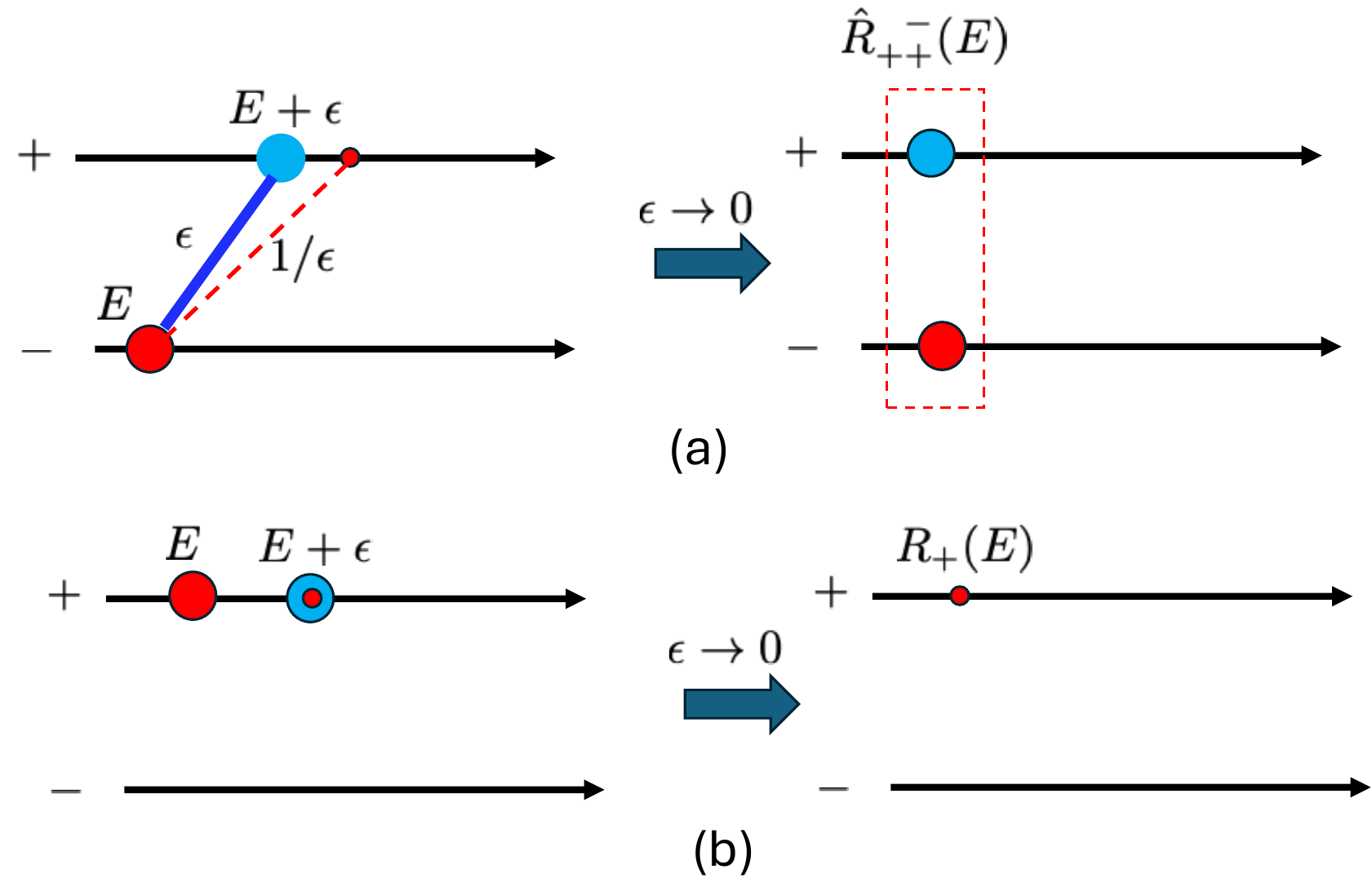}
\caption{\small The conventions are the same as those in Fig.~\ref{fig:sheet}, with $R_+ (E+\ep)$ represented by a red dot. 
(a) $R_+ (E + \ep)$ and the trans-dipole $\sZ^-_+ (E, E+\ep)$ form a nontrivial three-body bound configuration~\eqref{bound1} in the $\ep \to 0$ limit, represented on the right by the dashed box. $R_+ (E+\ep)$ is deliberately separated from $\De_+^{-1} (E+\ep)$ for visualization purposes. The contractions between $\De_- (E)$ and $\De_+^{-1} (E+\ep)$ are represented by a thick blue line (yielding a contribution proportional to $\ep$), while the contraction between $R_+ (E+\ep)$ and $\De_- (E)$ is represented by a red dashed line~(yielding a contribution proportional to $1/\ep$). Note that after contraction $R_+ (E+\ep)$ has been ``absorbed'' by the dipole background, which explains the absence of the red dot in the right plot. 
(b) The cis-dipole trivially cancels in the $\ep \to 0$ limit.
}
\label{fig:bound}
\end{center}
\end{figure}

The contribution from the cis-dipole thus becomes trivial in the $\ep \to 0$ limit, simply reducing to the original $R_+ (E)$. In contrast, the trans-dipole yields a non-trivial contribution: as $\ep \to 0$, the simple zero originating from $\sZ_+^- (E, E+\ep)$ is exactly canceled by a simple pole (in $\ep$) arising from the ``contraction'' between the resolvent $R_+$ and the trans-dipole background. This exact zero-pole cancellation results in a finite, non-vanishing term. Consequently, $R_+(E)$ and the dipole background effectively ``bind'' together, forming a distinct three-body bound configuration, which we shall refer to as the trans-bound resolvent. See Fig.~\ref{fig:bound} for an illustration.
We can thus decompose the dressed resolvent as
\be \label{bound}
\Fil{\tilde R_+ (E)} = \Fil{R_{+} (E) + \hat R^{\;\;\,-}_{++} (E)},
\ee
where $ \hat R^{\;\;\,-}_{++} (E)$ denotes the trans-bound resolvent, 
\be \label{bound1}
\hat R^{\;\;\,-}_{++} (E) \equiv i e^{\hat g (E)- c (E)} e^{2 \pi i \bar N (E)-2 i \phi} :e^{Y^* - Y}:  \ .
\ee
Here, the notation $:e^A:$ indicates that there are no ``self-contractions'' among the $A$ variables, meaning $\Fil{:e^A:} = e^{\Fil{A}}$. Similarly, we can introduce the trans-bound resolvent of the opposite ``orientation'',
\be \label{bound2}
\hat R^{\;\;\,+}_{--} (E) = \left(\hat R^{\;\;\,-}_{++} (E)\right)^* = - i e^{\hat g (E)- c (E)} e^{-2 \pi i \bar N (E)+2 i \phi} :e^{Y - Y^*}:  \ .
\ee

As a result, we have 
\bln
\boldsymbol{\rho}^{\rm (sm)} (E) & =\Fil{\tilde \rho (E)} =  -{1 \ov \pi} \Im \Fil{\tilde R_+ (E)} =
  -{1 \ov \pi} \Im \Fil{R_{+} (E) + \hat R^{\;\;\,-}_{++} (E)} \\
& = \bar \rho (E) - {f(E) \ov \pi} \cos (2\pi \bar N (E) - 2 \phi) ,
\label{uj1}
\end{align} 
with $f(E) = e^{\hat g (E) - c(E)}$. For the diagonal projection, we have $\hat g =0$ and $c(E) =c$, then $f = e^{-c}$, giving~\eqref{eub}. 

We conclude this discussion with some remarks:

\ben 

\item In the contribution from the cis-dipole, the phase factors $e^{\pm i \pi \bar N (E)}$ exactly cancel, reflecting a local operation on a single Riemann sheet. In contrast, for the trans-dipole, these macroscopic phases add up. This ``phase accumulation'' between the determinant and inverse determinant---resulted from separating the configurations across different sheets---is elevated via the contraction between the resolvent and the trans-dipole into the macroscopic oscillation seen in~\eqref{uj1}. 

\item We emphasize that the survival of the new oscillatory term in~\eqref{uj1} hinges fundamentally on two ingredients: the additive prescription~\eqref{DD}, which explicitly opens the trans-dipole channel, and the singular nature of the contraction between the resolvent and the trans-dipole background, which in turn arises from the logarithmic singularity in the cross-sheet filtering behavior~\eqref{sp2}.

\item Both the original Gutzwiller density of states $\rho (E)$~\eqref{dosD1} and the dressed density of states $\tilde \rho (E)$~\eqref{ienD0} provide semiclassical (Tier-II) approximations for the exact density of states $\brho (E)$. However, by systematically incorporating the phase coherence across Riemann sheets, the dressed representation $\tilde \rho (E)$ provides a fundamentally finer description of the spectrum. Specifically, while the naive filter projection of the standard Gutzwiller formula yields only the smooth Weyl background $\bar \rho(E)$, the dressed formulation incorporates the macroscopic oscillations, through the formation of a resolvent-determinant-anti-determinant bound configuration $ \hat R^{\;\;\,-}_{++} (E)$.

\een

\subsection{Plateau without averaging} \label{sec:plateau}

We now proceed to derive the plateau term in~\eqref{ekn} through the smooth filter projection discussed in Sec.~\ref{sec:ramp1}. The derivation uses the Gutzwiller formula~\eqref{poR}, with the crucial input~\eqref{DD}, and does not involve ensemble or explicit spectral averages (as clarified below~\eqref{basl} in the Introduction). Our approach draws inspiration from the classic results of~\cite{HeuMul07,KeaMul07,MulHeu09} (which inherently rely on energy-window averaging).

Consider the ``correlation'' between the dressed density of states~\eqref{ienP0}--\eqref{ienD0}: 
\be \label{proD}
\Fil{\tilde \rho (E_1) \tilde \rho (E_2)} =  - {1 \ov 2 \pi^2} {\rm Re} \, \Fil{\tilde R_+ (E_1) \tilde R_+ (E_2) - \tilde R_+ (E_1) \tilde R_- (E_2) } \ .
\ee
We are interested in the singular behavior of~\eqref{proD} as $E_1 \to E_2$.  More explicitly, 
\bln\label{ij1}
\Fil{\tilde R_+ (E_1) \tilde R_+ (E_2)} & = \Fil{R_+ (E_1 + \ep_1) R_+ (E_2 + \ep_2) Z_2 (E_1, E_1 + \ep_{1+}) Z_2 (E_2, E_2 + \ep_{2+})}
\\
 & = \Fil{R_{3+} R_{4+}  \le(\sZ^{++}_{++} +\sZ^{--}_{++} + \sZ^{+-}_{++} + \sZ^{-+}_{++}\ri)}  ,
 \label{ij2}
\end{align} 
where we have introduced the shorthand notations 
\bega 
R_{3\pm} = R_\pm (E_3), \quad E_3 =E_1 + \ep_1, \quad E_4 = E_2 + \ep_2 , \\
\sZ^{s_1 s_2}_{s_3 s_4} = \sZ^{s_1}_{s_3}  (E_1, E_3) \sZ^{s_2}_{s_4} (E_2, E_4)
, \quad
s_i \in \{+,-\} \ .
\end{gather} 
In~\eqref{ij1}--\eqref{ij2}, the $\ep_1, \ep_2 \to 0$ limit should be implicitly understood. Similarly, we have 
\be  \label{ij3}
\Fil{\tilde R_+ (E_1) \tilde R_- (E_2)} = \Fil{R_{3+} R_{4-}  \le(\sZ^{++}_{+-} +\sZ^{--}_{+-} + \sZ^{+-}_{+-} + \sZ^{-+}_{+-}\ri)} \ . 
\ee

Equations~\eqref{ij2} and~\eqref{ij3} can be evaluated using the filtering rules~\eqref{sp1}--\eqref{jhs}. We find 
\be\label{Fyen}
\Fil{\tilde \rho (E_1)\tilde  \rho (E_2)}  = \Fil{\tilde \rho(E_1)} \Fil{\tilde \rho(E_2)}
+ C_{\rm sing} (E_1, E_2) + C_{\rm non-sing} (E_1, E_2)  ,
\ee
where $C_{\rm sing} (E_1, E_2)$ and $C_{\rm non-sing} (E_1, E_2)$ represent the connected contributions, which are singular and non-singular in the limit $\ep = E_1 - E_2 \to 0$, respectively. 

The singular part $C_{\rm sing} (E_1, E_2)$ originates from the terms proportional to $\sZ^{+-}_{+-}$ and $\sZ^{-+}_{+-}$ in~\eqref{ij3}~(and their complex conjugates). More explicitly, 
\bega \label{00ram}
\lim_{\ep_1, \ep_2 \to 0} \Fil{R_{3+} R_{4-}  \sZ^{+-}_{+-} }= \Fil{R_{1+} R_{2-}} 
= \pi^2 \bar \rho(E_1) \bar \rho(E_2) - {1 \ov (\ep + i 0)^2}  + O(\ep^0) \ .
\end{gather} 
In this sector, $\sZ^{+-}_{+-}$ consists of two cis-dipoles ($++$ for $E_1$ and $--$ for $E_2$). In the $\ep_{1,2} \to 0$ limit, both dipoles reduce to $1$,  recovering the ramp result derived in Sec.~\ref{sec:ramp1}. 

In contrast, $\sZ^{-+}_{+-}$ consists of two trans-dipoles in opposite ``orientations'' ($-+$ for $E_1$ and $+-$ for $E_2$), and has the form\footnote{Note $\Fil{B C e^A} = \le(\Fil{BC}_c  + (\Fil{B} + \Fil{BA}_c)( \Fil{C} +  \Fil{CA}_c )   \ri) \Fil{e^A}$.}, 
 \bln
\lim_{\ep_1 , \ep_2 \to 0} \Fil{R_{3+} R_{4-}  \sZ^{-+}_{+-} } & = \lim_{\ep_1 , \ep_2 \to 0}  \Fil{R_{3+} R_{4-} e^{Y_{1}^* + Y_{2} - Y_{3} - Y^*_{4}}} \\
\label{Plsa0}
& =\lim_{\ep_1 , \ep_2 \to 0}  \Fil{R_{3+} Y_{1}^* }_c \Fil{R_{4-}Y_{2} }_c \Fil{e^{Y_{1}^* + Y_{2} - Y_{3} - Y^*_{4}}}  \ .
\end{align} 
Much like the calculation of~\eqref{transC}, the factors $\Fil{R_{3+} Y_{1}^* }_c$ and $\Fil{R_{4-}Y_{2} }_c$ in~\eqref{Plsa0} generate simple poles proportional to $1/\ep_1$ and $1/\ep_2$. These respective poles exactly cancel the first-order zeros in $\ep_1$ and $\ep_2$ arising from the contractions between $Y_1^*$ and $-Y_3$, and between $Y_2$ and $-Y_4^*$, ultimately rendering the $\ep_{1,2} \to 0$ limit finite and non-vanishing. In other words, we can write~\eqref{Plsa0} as 
 \bln \label{plads0}
\lim_{\ep_1 , \ep_2 \to 0} \Fil{R_{3+} R_{4-}  \sZ^{-+}_{+-} } & = \Fil{\hat R^{\;\;\,-}_{++} (E_1) \hat R^{\;\;\,+}_{--} (E_2)} \\
& = e^{\hat g (E_1) + \hat g(E_2)  - c(E_1) - c (E_2)}  e^{2\pi i (\bar N_1 - \bar N_2) } \Fil{:e^{Y_1^*- Y_1}: :  e^{Y_2- Y_2^*}:} \\
& =  e^{- c(E_2) - c (E_1)}  e^{2\pi i (\bar N_1 - \bar N_2) } e^{\hat g (E_1) + \hat g(E_2) }  
e^{g^{+-}_{21}  - g^{--}_{12} - g^{++}_{12}  + g^{+-}_{12}} \\ 
& = e^{2 \pi i \bar \rho (E) \ep + O(\ep^2)} {1 \ov (\ep + i 0) (\ep - i 0)}  +O (\ep^0) ,
\label{plads}
 \end{align} 
 where the trans-bound resolvents $\hat R^{\;\;\,-}_{++} (E)$ and $\hat R^{\;\;\,+}_{--} (E)$ were defined in~\eqref{bound1}--\eqref{bound2}, and we have used the abbreviated notations introduced below~\eqref{eonr}--\eqref{eonr1}. In the last line we have taken $E={E_1 + E_2 \ov 2}$ and taken $\ep = E_1 -E_2 \to 0$. Note that all the non-universal functions $c (E)$ and $\hat g (E)$ have dropped out, and it is important that the only $O(\ep)$ term in the exponent is the indicated one. 
 
\begin{figure}
\begin{center}
\includegraphics[width=14cm]{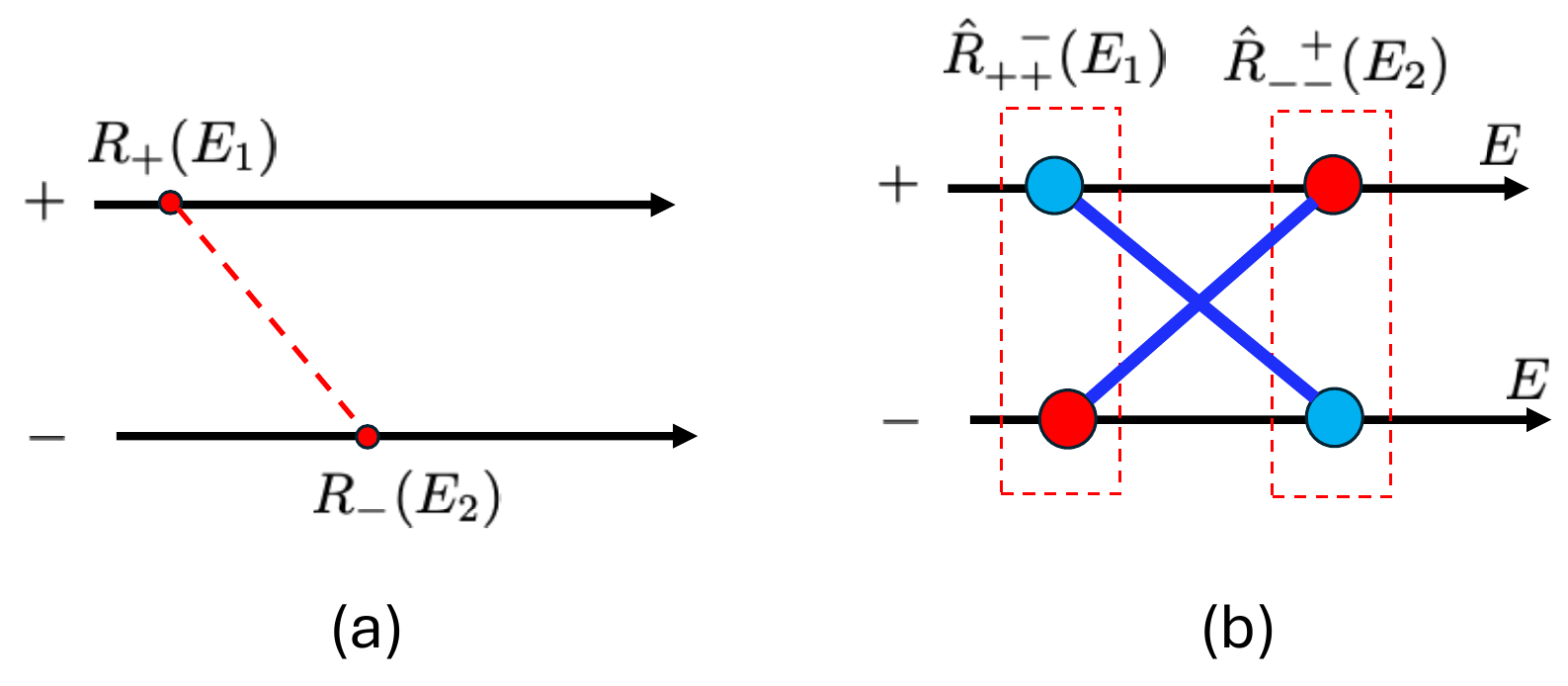}
\caption{\small While the ramp arises from correlations between two resolvents residing on different branches (as illustrated in (a)), the plateau arises from correlations between two trans-bound resolvents of opposite orientations (as illustrated in (b)).
Conventions follow those of Fig.~\ref{fig:sheet} and Fig.~\ref{fig:bound}. 
}
\label{fig:sheet1}
\end{center}
\end{figure}

 We thus see that the plateau term~\eqref{plads} can be interpreted as arising from ``correlations''~\eqref{plads0} between two trans-bound resolvents of opposite orientation, just as the ramp results from correlations~\eqref{00ram} between two standard Gutzwiller resolvents of opposite orientation.  See Fig.~\ref{fig:sheet1} for an illustration.  We emphasize that both the poles and their specific $i0$ structures in~\eqref{plads} exist strictly because the two trans-dipoles possess opposite orientations; no singular term in $E_1 - E_2$ emerges if they share the same orientation.   
 
The macroscopic phase factor $e^{2 \pi i \bar \rho (E_1-E_2)} = e^{2 \pi i \bar N (E_1) - 2\pi i \bar N (E_2)}$ in~\eqref{plads} arises from the nearly complete cancellation  between the macroscopic components of the two oppositely oriented trans-dipoles.  We can rewrite this factor as $e^{i (E_1 - E_2) \de t}$, where $\de t$ is naturally interpreted as a time shift between the two systems. This shift 
\be
\de t = 2 \pi \bar \rho (E) \equiv t_H,
\ee
corresponds precisely to the Heisenberg time $t_H$ at energy $E$.

Altogether, we obtain 
\bln\label{c11}
C_{\rm sing} (E_1, E_2) & = -{1 \ov 2 \pi^2} {\rm Re} \le[ {1 \ov (\ep+ i 0)^2} -{e^{2 \pi i \bar \rho(E)  \ep} \ov (\ep + i 0) (\ep-i0)}\ri] \\
& = -{1 - \cos (2 \pi \bar \rho (E) \ep) \ov 2 \pi^2 \ep^2} + \lim_{\eta \to 0} {\de (\ep) \ov 2 \pi \eta}  \ .
\label{c12}
\end{align} 
The singular delta-function term in~\eqref{c12}  originates from the $i0$ structure of the plateau term in~\eqref{c11}, via the identity 
\be 
{1 \ov (\ep + i0) (\ep - i 0)} = \lim_{ \eta \to 0} {1 \ov \ep^2 + \eta^2} = P {1 \ov \ep^2} + \lim_{ \eta \to 0} {\pi \ov \eta} \de (\ep)   \ .
\ee

 In quantum chaos, the ramp and the plateau have often been treated through entirely different conceptual lenses. Here, we find that the plateau emerges from the exact same type of two-point correlation as the ramp---simply by swapping the standard Gutzwiller resolvent for the trans-bound resolvent, thus providing a unified framework. In fact, writing the numerator of the first term of~\eqref{c12} as 
\be
\ha + \ha - \ha e^{i t_H \ep} -  \ha e^{- i t_H \ep}
\ee
we see that the plateau terms share exactly the same structure as the ramp terms, except that they carry a macroscopic time shift of $\pm t_H$. This can be made even more explicit by considering the Fourier transform of~\eqref{c12}. The ramp term gives (with $t > 0$)
 \be \label{ramp_FT}
 -{1 \ov 2 \pi^2}  \int_{-\infty}^\infty d\ep \, e^{i \ep t} {1 \ov \ep^2 + \eta^2}
 = -{1 \ov 2 \pi \eta} +  {t \ov 2 \pi} + O(\eta)  \ ,
\ee
where we have introduced an infinitesimal regulator $\eta \to 0$ to regulate the singularity at $\ep =0$. 

The plateau term gives 
\begin{gather} \label{plateau_FT}
 {1 \ov 2 \pi^2}  \int_{-\infty}^\infty d\ep \, e^{i \ep t} {\cos (t_H \ep) \ov \ep^2 + \eta^2}
 = {1 \ov 4 \pi^2} \int_{-\infty}^\infty d\ep \, {e^{i (t +t_H) \ep} + e^{i (t - t_H) \ep} \ov \ep^2 + \eta^2} 
\\
 = {1 \ov 2 \pi \eta} -  {t + t_H + |t -t_H| \ov 4 \pi} 
    = {1 \ov 2 \pi \eta} - {1 \ov 2 \pi} \max \left(t, t_H\right) \ .
\end{gather}
Summing the two contributions perfectly cancels the $1/\eta$ divergence and yields the spectral form factor $(t - \max(t, t_H)) / 2\pi$, demonstrating how the trans-bound resolvent forces the linear ramp to saturate precisely at the Heisenberg time.

We conclude this subsection with some further remarks:

\ben 

\item The plateau term~\eqref{plads} is a complex exponential in the density of states $\bar \rho = e^{\sS}$ and thus, like~\eqref{eub}, represents a smooth level-2 transseries term.

\item The presence of the divergent contact term in~\eqref{c12} appears to be an artifact of our continuum approximation, signaling that this continuous treatment is not equipped to resolve the purely discrete energy levels as $\ep \to 0$, despite successfully recovering both the macroscopic ramp and the plateau. The full two-point spectral correlation function includes a diagonal contact term $\bar \rho (E) \de (\ep)$, which is recovered if we regulate the singularity by setting the cutoff $\eta = 1/t_H = 1/(2 \pi \bar \rho (E))$.

\item The mathematical mechanism by which the plateau arises here is conceptually reminiscent of the Altshuler-Andreev instanton~\cite{AltAnd95a}---the non-local saddle point in the nonlinear sigma model approach to quantum chaos that couples the advanced and retarded sectors. We believe this  parallel is not accidental, even though the nonlinear sigma model relies on ensemble/disorder averaging.

\een

\section{Further parallels with random matrix theory via the Gutzwiller representation}\label{sec:further} 

In the preceding section, we demonstrated that key spectral features characteristic of random matrix theory---namely the ramp, the plateau, and the highly oscillatory density of states---can be reproduced for a single quantum chaotic system using the Gutzwiller representation, without the need for any ensemble or spectral averaging. 

In this section, we shall push these parallels even further. We first formalize the notion of the ``spectral curve'' for chaotic systems. Subsequently, we explore the possible existence of a chaotic counterpart to the single-eigenvalue ``instanton'' found in matrix models, a non-perturbative configuration that can induce doubly exponential ``hyper-instanton'' effects in the thermal partition function.

\subsection{Spectral curve} \label{sec:spec}

Consider the Gutzwiller representation
\be
R (z) = R^{(\rm sm)} (z) + R^{\rm (err)} (z)
\ee
of the exact resolvent~\eqref{rude0}, where $R(z)$ for general complex $z$ is obtained by analytic continuation of~\eqref{poR}. An important feature of the Gutzwiller formula, which has played a crucial role in the discussions of the preceding section, is that the global analytic structure of $R(z)$ is  dictated by its smooth part, $R^{(\rm sm)} (z)$.

This feature arises because both the smooth Weyl part and the highly oscillatory periodic orbit contributions share the same underlying semiclassical (WKB) origin. Within the Gutzwiller framework, both $R^{(\rm sm)}(z)$ and $R^{(\rm err)}(z)$ are constructed from the same classical phase-space dynamics. Consequently, both components are subject to the exact same classically allowed and forbidden energy regions. The branch points that dictate the global analytic structure of the resolvent thus correspond physically to the thresholds of the allowed energy spectrum.

Let $\Sig_{\bar \rho}$ denote the abstract Riemann surface defined by $R^{(\rm sm)} (z)$. This surface is fully specified by the discontinuity of $R^{(\rm sm)} (z)$ across the branch cut on the real axis of the physical sheet, which is given by the smooth Weyl density of states $\bar \rho(E)$. We can therefore view the erratic part $R^{\rm (err)} (z)$ and all functions derived from it (such as its integral or the spectral determinant), along with their smooth filter projections, as living on the base manifold $\Sig_{\bar \rho}$. This geometric hierarchy is structurally identical to topological recursion in random matrix models, where the analytic structures of all higher-order, multi-point correlation functions are completely governed by the spectral curve specified by the leading-order resolvent~(see e.g.,~\cite{EynOra07, Eyn16, AkeBai11}). In this spirit, we refer to $\Sig_{\bar \rho}$ as the ``spectral curve'' of the chaotic system.

There is, however, an important subtlety here. For a quantum system with an infinite-dimensional Hilbert space, the operator trace defining the exact resolvent~\eqref{rude0}  is in general ill-defined, and the integral over the density of states is UV divergent. 
Equivalently, when expressing the resolvent in terms of the thermal partition function via the Laplace transform,
\be
\bR (z) = - \int_0^\infty d \b \, e^{\b z}\,  \bZ (\b) , \quad \bZ (\b) = \Tr e^{-\b H} , 
\ee
this UV catastrophe manifests as a divergence in the high-temperature ($\b \to 0$) limit. For a simple harmonic oscillator, the partition function scales algebraically as $\bZ (\b) \sim {1 \ov \b}$, whereas for a $d$-dimensional local quantum field theory, one encounters a far more severe essential singularity of the form $\bZ (\b) \sim e^{O(\b^{-d})}$ (assuming the theory flows to a well-defined UV fixed point).\footnote{The corresponding density of states grows at high energies as  $e^{O(E^{(d-1)/d})}$.} 
Consequently, a systematic UV regularization is required to tame the trace and render the exact resolvent well-defined.

To establish a sensible and rigorous definition of the spectral curve for a chaotic system, we therefore require a regularization scheme that allows $\Sig_{\bar \rho}$ to be defined intrinsically, independent of the chosen UV regulator.

For few-body quantum mechanical systems, one often uses the $\zeta$-function regularization or subtracts a suitable reference system. 
However, for quantum many-body systems and field theories, these methods generally fail to regularize the exponential-type essential singularities present in the high-energy density of states (or equivalently, the high-temperature partition function).
Another possibility is to multiply the density of states $\rho(E)$ by a regulator $f_\Lambda(E)$ that suppresses the high-energy part of the spectrum above a scale $\Lambda$. This procedure, however, is also unsatisfactory for our purposes, since it generally changes the analytic structure of the resolvent in a significant way.

We introduce a diagonal analytic regulator of the form
\be \label{Gres0}
\bR_{f_\Lam} (z) = \int dE \,  \brho (E)  {f_\Lam (E, z) \ov z- E} ,
\ee
where the regulator function $f_\Lam(E,z)$ satisfies the following conditions:
\ben
\item For a given $z$, its dependence on $E$ decays sufficiently fast to guarantee the absolute convergence of the spectral integral.
\item It satisfies the diagonal condition:
\be \label{diagC}
f_\Lambda(s,s)=1 \ .
\ee
\item It recovers the unregularized limit as $\Lam \to \infty$:
\be
f_{\Lam \to \infty} (E,z) \to 1 \ .
\ee
\item It is single-valued and holomorphic with respect to $z$, and remains real-valued when both $E$ and $z$ are real.
\een

An example satisfying these requirements is the Gaussian regulator,
\be \label{ejo}
f_\Lam (E,z) = e^{-{ (E-z)^2 \ov  2 \Lam^2}}  ,
\ee
which ensures the convergence of~\eqref{Gres0} even in local quantum field theories with exponential spectral growth. We can equivalently express~\eqref{Gres0} in terms of a regularized partition function via the integral transform
\be
\bR_{f_\Lambda} (z)= - \int_0^\infty d \b \, e^{\b z} \bZ_{f_\Lam} (\b,z), \quad
 \bZ_{f_\Lam} (\b,z) = \int dE \, \brho (E) e^{-\b E} f_\Lambda(E,z) \ .
\ee
Physically, for the Gaussian choice~\eqref{ejo}, the regularized partition function $\bZ_{f_\Lam} (\b,z)$ takes the form of a microcanonical partition function localized around an energy window centered at $z$ with a characteristic width $\Lam$.

The regularization~\eqref{Gres0} descends to the Gutzwiller representation, with, for example, the smooth part having the form
\be \label{Gres}
R^{(\rm sm)}_{f_\Lam} (z) = \int_{E_0}^\infty dE \, \bar \rho (E)  {f_\Lam (E, z) \ov z- E} \ .
\ee
Let $R^{(\rm sm)}_{f_\Lam, \pm} (E)$ denote the limiting values of $R^{(\rm sm)}_{f_\Lam} (z)$ as the real energy axis is approached from above and below, i.e., $z = E \pm i 0$ for $E \in \RR$. Crucially, the diagonal constraint~\eqref{diagC} guarantees that the imaginary parts---and thus the underlying physical spectral profile---remain strictly regularization-independent:
\be
\Im R^{(\rm sm)}_{f_\Lam, \pm} (E) = \mp \pi \bar \rho (E) , \quad
R^{(\rm sm)}_{f_\Lambda,+}(E)-R^{(\rm sm)}_{f_\Lambda,-}(E) = - 2\pi i \bar \rho (E)  \ .
\ee

While the real part of $R^{(\rm sm)}_{f_\Lam, \pm} (E)$ and the asymptotic large-$z$ behavior do depend on the specific choice of $f_\Lam$ (for instance, a finite $\Lam$ breaks the standard $1/z$ fall-off at infinity), this dependence is rather benign. Consider two distinct diagonal regulators, $f_\Lam$ and $\tilde f_{\tilde \Lambda}$. Their difference,
\be
R^{(\rm sm)}_{f_\Lambda}(z)-R^{(\rm sm)}_{\tilde f_{\tilde \Lambda}}(z) ,
\ee
possesses zero discontinuity across the real axis by virtue of~\eqref{diagC}, rendering it a globally holomorphic (entire) function of $z$. 

If we denote the analytic continuation of $R^{(\rm sm)}_{f_\Lambda} (z)$ to an arbitrary Riemann sheet $\al$ as $R^{(\rm sm)}_{f_\Lambda, \al} (z)$, it follows that any two regularization schemes differ by a uniform shift across all sheets:
\be
R^{(\rm sm)}_{f_\Lambda, \al} (z)
=
R^{(\rm sm)}_{\tilde f_{\tilde \Lambda}, \al} (z)
+
E_{f_\Lambda,\tilde f_{\tilde \Lambda}}(z),
\ee
where the entire function $E_{f_\Lambda,\tilde f_{\tilde \Lambda}}(z)$ is independent of the sheet index $\al$. Consequently, the abstract Riemann surface $\Sig_{\bar \rho}$---on which $R^{(\rm sm)}_{f_\Lambda, \al} (z)$ is single-valued---is uniquely dictated by the physical discontinuity $\bar \rho(E)$. The spectral curve is thus an intrinsic, scheme-independent geometric object. In what follows, we shall drop the explicit $f_\Lam$ subscript for notational simplicity.

Once the resolvent has been properly defined, we can define its integrated version $\bX(z)$, the spectral determinants $\bD(z)$ and $\tilde{\bD}(z)$, and their corresponding Gutzwiller representations as before. We can also analytically continue~\eqref{sp2} and its generalizations~\eqref{spff}--\eqref{jhs} to an arbitrary pair of points
$p$ and $p'$ on the spectral curve $\Sig_{\bar \rho}$, writing 
\be \label{cmPX}
\Fil{Y(p) Y (p')} = g(p, p') 
\ee
where $g(p,p')$ is a single-valued bi-local function defined on $\Sig_{\bar \rho} \times \Sig_{\bar \rho}$.

To summarize, the existence of a diagonal analytic regulator such as~\eqref{Gres} demonstrates that while the absolute value of the resolvent itself is scheme-dependent---and therefore does not constitute an unambiguous physical observable---its discontinuity across the real axis and the geometry of the corresponding spectral curve are  invariant, physical constructs. Furthermore, the correlation~\eqref{sp2} under the smooth filter projection between the erratic parts of the resolvents  is manifestly regulator-independent\footnote{It follows directly from the sum rule~\eqref{sumR} which is in turn derived from the ergodicity properties of the system at the classical level.}.

\subsection{Dressed resolvents in the ``forbidden'' region}\label{sec:forB}

Our discussion in Sec.~\ref{sec:II} has been restricted to the behavior of $R_\pm (E)$ for $E$ lying within the spectral support region $E > E_0$ of $\bar \rho (E)$. In general quantum systems, the behavior of the resolvent $R(E)$ on the real axis outside this region is also of fundamental importance. While the primary support region typically captures the dense, chaotic bulk of the spectrum, the classically forbidden region encodes non-perturbative features and finite-size fluctuations, such as tunneling phenomena, bound states, and resonances. In random matrix theory and disordered systems, the region outside the main support governs large deviations, rare-region fluctuations (such as Lifshitz tails), and non-perturbative effects, including single-eigenvalue instanton contributions. We will now generalize our analysis of the dressed resolvent~\eqref{ienP0} to the ``forbidden'' region $E < E_0$. As we shall see, the discussion here is of a more exploratory nature compared with the previous sections.

In the classically forbidden region $E < E_0$, the resolvent and its integrated counterpart are strictly real and single-valued, denoted simply by $R(E)$ and $X(E)$. Because the Gutzwiller representation $D(z)$ of the spectral determinant possesses no branch cut (as established in Sec.~\ref{sec:MDOS}), we naturally have\footnote{Equation~\eqref{forS} can be checked explicitly in simple systems.} 
\be \label{forS}
D(E) = \De (E) = e^{X (E)}  , \quad X(E) = \bar Q (E) + Y(E) +a, 
\ee
where both $\bar Q(E)$ and $Y (E) +a$ are real. 
In contrast, the Gutzwiller representation of the inverse determinant, $\tilde D_\pm (z)$, is sensitive to the branch cut. We can obtain $\tilde D_\pm (E)$ for $E < E_0$ from its defining expression $\De_\pm^{-1} (E) = e^{- X_\pm (E)}$ for $E > E_0$ via analytic continuation. The most natural choice is to continue $\tilde D_\pm (E)$ through the half-planes on the physical sheet where they are originally defined---i.e., through the lower half-plane for $\tilde D_- (E)$ and the upper half-plane for $\tilde D_+ (E)$. This direct path yields 
 \be  \label{foriS}
\tilde D_+ (E) = \tilde D_- (E) = {1 \ov \De (E)} = e^{- X (E)} = {1 \ov D (E)}  , \quad E < E_0 \ . 
\ee
Equation~\eqref{foriS} is indeed what one would expect for a simple quantum mechanical system below the threshold for supporting any energy eigenvalues.

\begin{figure}
\begin{center}
\includegraphics[width=14cm]{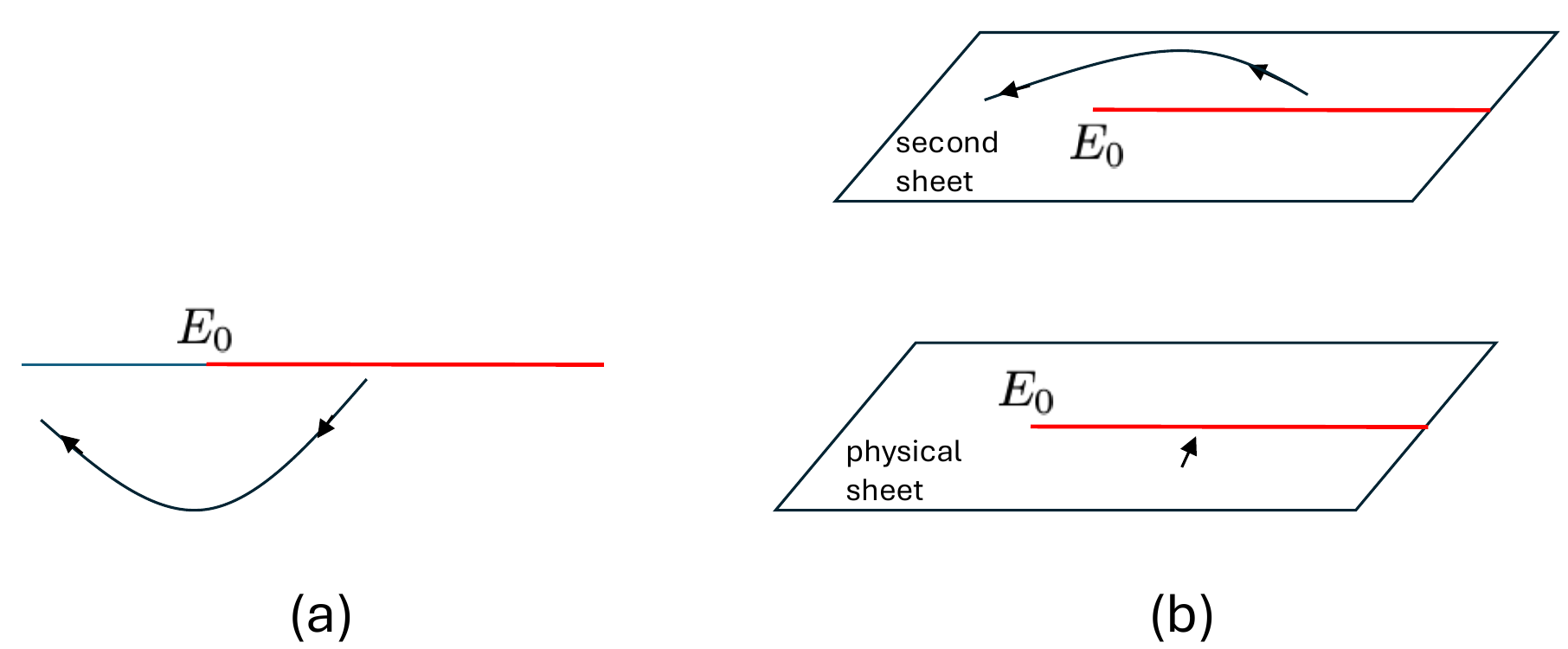}
\caption{\small (a) Analytic continuation of $\De_- (E)$ from $E > E_0$ to $E < E_0$ through the lower half-plane of the physical sheet (only the physical sheet shown). 
(b) Analytic continuation of $\De_- (E)$ from $E > E_0$ to $E < E_0$ by passing under the branch cut and moving through the upper half-plane of the second sheet. The branch cuts are represented by thick red lines. 
}
\label{fig:anaC}
\end{center}
\end{figure}

However, the analytic continuation of these semiclassical asymptotic representations is subject to the Stokes phenomenon. For example, to evaluate $\tilde D_-(E)$ in the forbidden region, we could alternatively continue $\De_-^{-1} (E)$ from $E > E_0$ by passing under the cut and moving through the upper half-plane of the second sheet to reach $E < E_0$, as indicated in Fig.~\ref{fig:anaC}.
The resulting expression along this path is given by $(\De^{(2)} (E))^{-1}$, where $\De^{(2)} (E) = \De (E_0 + (E- E_0) e^{2 \pi i}) = e^{X^{(2)} (E)}$. 
Because asymptotic expansions can abruptly switch their dominant exponential saddles across Stokes lines, the Gutzwiller representation for $\tilde D_- (E)$ at $E < E_0$ could, in principle, be a superposition of $\De^{-1} (E)$ and $(\De^{(2)} (E))^{-1}$:\footnote{For definiteness, we consider a two-sheet topology. In systems with more than two sheets, values from additional sheets could also, in principle, contribute to the superposition. \label{ft:2-sh}} 
 \be\label{tilDe}
 \tilde D_- (E) = e^{- X(E)} + b e^{- X^{(2)} (E)}, \quad E < E_0 \ ,
 \ee
where $b$ is a Stokes constant whose value depends on the specific system. Its precise value is not essential for our present purposes, as long as it is non-zero. While $X (E)$ is real, $X^{(2)} (E)$ is generally complex, implying that $\tilde D_- (E)$ can be complex when $b \neq 0$. The coefficient of the first term in~\eqref{tilDe} must be $1$ from a self-consistency requirement, as we shall see below. By definition, we must also have 
\be
\tilde D_+ (E) = \tilde D_-^* (E) \ .
 \ee
It is not immediately obvious whether there exist specific chaotic systems where the Stokes coefficient $b$ in~\eqref{tilDe} is 
non-zero.\footnote{In few-body systems, it may be possible to rigorously demonstrate that such Stokes mixing cannot occur for $E$ far below the ground state energy, because the analytic structure in that regime is relatively simple. However, for quantum many-body systems in the large-$\sN$ limit---which are our primary focus---the spectral and analytic structures are vastly more complex, making the absence of such non-perturbative phenomena far less obvious.} Nevertheless, it represents a natural possibility to explore.

A necessary consistency check for the validity of~\eqref{tilDe} is that under the smooth filter projection, we must still recover the identity 
\be \label{filter_check}
\Fil{D (E) \tilde D_\pm (E)} = 1, \quad E< E_0 \ .
\ee
More explicitly, consider the shifted product 
\be \label{pod1}
\Fil{D (E) \tilde D_- (E + \ep) }= \Fil{\sZ^1_{1} (E, E+\ep) }+ b \Fil{ \sZ^1_{2} (E, E+\ep) },
\ee
where we define 
\bega
\sZ^{1}_1 (E_1, E_2) \equiv \De (E_1) \De^{-1} (E_2) = e^{\bar Q_1 - \bar Q_2} e^{Y_1 - Y_2}, \\
\sZ^{1}_2 (E_1, E_2) \equiv \De (E_1) (\De^{(2)} (E_2))^{-1}
= e^{\bar Q_1 - \bar Q_2^{(2)}} e^{Y_1 - Y_2^{(2)}} \ .
\end{gather} 

To evaluate the second term of~\eqref{pod1}, we need to analytically continue~\eqref{sp1}--\eqref{jhs} to the full abstract curve $\Sig_{\bar \rho}$. In particular, $g_{+-} (E_1, E_2)$, as defined in~\eqref{sp1}, can be interpreted as the cross-correlation between 
$Y (E_1)$ on the physical sheet and $Y^{(2)} (E_2)$ on the second sheet (for $E_1, E_2 > E_0$).
We thus have\footnote{More generally, there should be a logarithmic singularity for any point $p$ on the physical sheet and its mirror point $\bar p$ on the second sheet. For simplicity, we will again restrict our attention to a two-sheet topology.}
\be \label{ons1}
\Fil{Y_1 Y_2^{(2)}} =  g_{+-} (E_0 + e^{i \pi} (E_0 -E_1) , E_0 + e^{i \pi} (E_0 -E_2))
= - \log (- i \ep) + \tilde c (E) + O(\ep),
\ee
where in the second equality we have taken $\ep = E_1 - E_2 \to 0$ and $E={E_1 + E_2 \ov 2}$.  Here $\tilde c (E)$ is in general not real any more.
Similarly, we have
\bega
\Fil{Y_1 Y_2} =g_{++} (E_0 + e^{i \pi} (E_0 -E_1) , E_0 + e^{i \pi} (E_0-E_2)) , 
 \\
\Fil{Y_1^{(2)} Y_2^{(2)}} = g_{--} (E_0 + e^{i \pi} (E_0 -E_1) , E_0 + e^{i \pi} (E_0 -E_2)) \ . 
\label{ons3}
\end{gather}
Both are non-singular as $E_1  \to E_2 = E$, we will denote their coincidental limits as $\tilde g (E)$ and $\tilde g^{(2)} (E)$, respectively.

It then follows that 
\bega 
\lim_{\ep \to 0}  \Fil{\sZ^1_{1} (E, E+\ep)} = 1 ,\\
\lim_{\ep \to 0}  \Fil{\sZ^1_{2} (E, E+\ep)} = i e^{\bar Q_1 - \bar Q_2^{(2)}}  e^{\ha \tilde g(E) + \ha \tilde g^{(2)} (E) - \tilde c(E)} \ep = 0
\end{gather} 
and~\eqref{filter_check} is satisfied for an arbitrary Stokes constant $b$, provided that the coefficient of the first term in~\eqref{tilDe} is $1$.

We now define the dressed resolvent for  for $E < E_0$ as 
\be
\tilde R_- (E) = R (E) D (E) \tilde D_- (E + \ep) 
= R (E)\le( \sZ^1_{1} (E, E+\ep) + b  \sZ^1_{2} (E, E+\ep) \ri),
\ee 
As before, the contribution from the cis-dipole is trivial
\be
\lim_{\ep \to 0} \Fil{R (E)\sZ^1_{1} (E, E+\ep)} = \Fil{R (E)} = \bar R (E) = {\rm real} ,
\ee
while the contribution from the trans-dipole 
\bln
& \lim_{\ep \to 0} \Fil{R (E)\sZ^1_{2} (E, E+\ep)} =e^{\bar Q_1 - \bar Q_2^{(2)}}   \Fil{R (E) e^{Y (E) - Y^{(2)} (E+\ep)} } \\
& =  i   e^{\bar Q_1 - \bar Q_2^{(2)}} 
e^{\ha \tilde g (E) +\ha \tilde g^{(2)} (E) - \tilde c(E)} \ .
\end{align}
We thus find
\be\label{exdos}
\Fil{\tilde \rho (E)} = {1 \ov \pi} \Im \, \Fil{\tilde R_- (E)} = {1 \ov \pi} \Re \le(\sE (E) e^{\lam (E)} \ri) , \quad E< E_0, 
\ee
where we have introduced 
\bega\label{Doenw}
\lam (E) \equiv \bar Q (E) - \bar Q^{(2)} (E) ,\\
\sE (E) = b  \, e^{\ha \tilde g (E) +\ha \tilde g^{(2)} (E) - \tilde c(E)}   \ .
\label{Doenw1}
\end{gather} 

Recall that 
\be 
\bar Q (E) = \int_{E_0}^E dE'  \, \bar R (E') 
\ee
and $\bar R^{(2)} (E)$ can be obtained from the standard relation\footnote{Recall that $R^{(2)} (z)$ is obtained from $R (E-i 0)$ by analytically continuing under the cut to the upper half plane.} 
\be\label{reSh}
\bar R^{(2)} (E) =\bar R(E) + 2 \pi i \bar \rho_{+} (E)
\ee
where $\bar \rho_+ (E)$ is obtained from $\bar \rho (E)$ for $E> E_0$ by continuing through the upper half plane. 
 We thus find that 
\be\label{oenw}
\lam (E) = - 2 \pi i \int_{E_0}^E dE' \,  \bar \rho_+ (E') \ .
\ee
Note that while $\bar Q (E)$ and $\bar Q^{(2)} (E)$ are separately regulator-dependent, their difference is not, and thus~\eqref{exdos} is regulator-independent.  

The expressions~\eqref{exdos}--\eqref{Doenw1} are again familiar in random matrix theory, see e.g.~\cite{SeiShi03,MalMoo04,SaaShe19}.

\subsection{Thermal partition function: threshold effects and hyper-instantons} \label{sec:doub}

We have discussed two types of ``non-perturbative'' corrections to the density of states: (i) rapid macroscopic oscillations in~\eqref{uj1}, which are supported for $E> E_0$, and (ii) a potentially non-vanishing support~\eqref{exdos} in the forbidden region $E< E_0$. Both corrections involve exponentials of the integral of the macroscopic density of states $\bar \rho \sim e^{\sS}$, making them doubly exponential in $\sS$. We now explore their implications for the thermal partition function. 
In the absence of an explicit expression for  $\bar \rho (E)$ and the various functions appearing in~\eqref{uj1} and~\eqref{Doenw1}, our discussion is unavoidably formal. We therefore focus on the simplest scenarios for illustrative purposes, rather than attempting an exhaustive classification of all possibilities.

\begin{figure}
\begin{center}
\includegraphics[width=5cm]{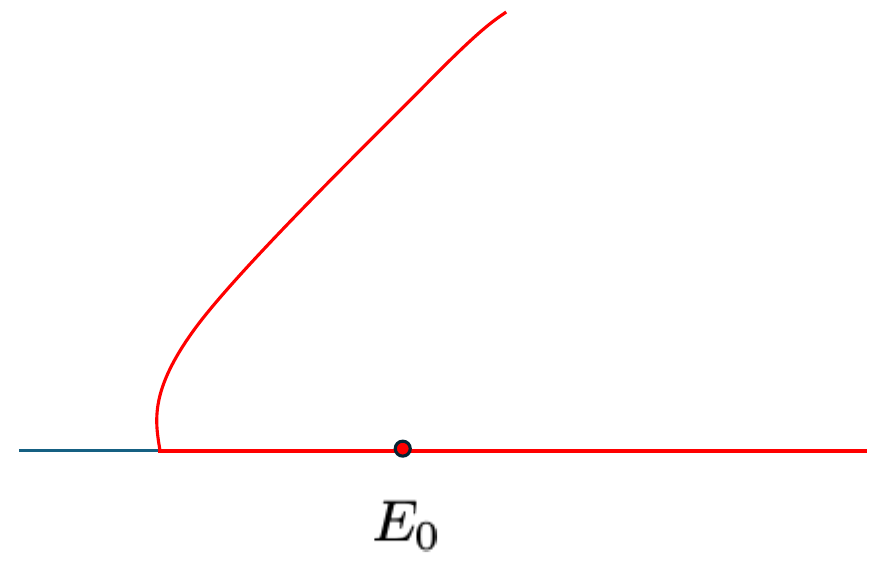}
\caption{\small Illustration of the integration contour $C$ (red line). The contour consists of the semi-infinite segment $(E_0, +\infty)$ along the real axis and extends into the ``forbidden'' region $E < E_0$, ultimately ending in the complex energy plane along a valley of steepest descent.
}
\label{fig:contour}
\end{center}
\end{figure}

Consider the thermal partition function resulting from the smooth part of the dressed resolvent:
\be \label{thPar}
\tilde Z (\b)= \int_{C} dE \, e^{-\b E} \, \Fil{\tilde \rho (E)} ,
\ee
where the integration contour $C$ includes the spectrally supported region $(E_0, \infty)$, but is also deformed to include at least part of the ``forbidden'' region $E < E_0$ so as to capture the contribution of $\Fil{\tilde \rho (E)}$ there, as illustrated in Fig.~\ref{fig:contour}.
Because the Boltzmann weight $e^{-\b E}$ diverges rapidly as $E \to -\infty$, the contour may not simply extend asymptotically along the negative real axis if the total integrand lacks sufficient suppression to ensure convergence. Consequently, the path may have to terminate somewhere else in the complex plane.  A rigorous topological specification of this path ultimately requires a complete understanding of the analytic structure of $\Fil{\tilde \rho (E)}$; therefore, the contour $C$ depicted here serves merely as an illustrative cartoon. 

We can write~\eqref{thPar} more explicitly as 
\be 
\tilde Z (\b) = \bar Z (\b) + Z_{\text{osc}} (\b) + Z_{\text{forb}} (\b) 
\ee
where $\bar Z (\b) = \int_{E_0}^\infty dE e^{-\b E} \bar \rho (E)$ is the contribution from $\bar \rho$, while $Z_{\text{osc}} (\b)$ and $Z_{\text{forb}} (\b)$ represent the contributions from the rapid macroscopic oscillations and the non-vanishing support in the forbidden region, respectively,
\bega\label{z0pa}
Z_{\text{osc}} (\b) = -{1 \ov \pi} \int_{E_0}^\infty dE \, e^{-\beta E} f(E) \cos(\Phi(E)),  \quad \Phi(E) = 2\pi \bar N(E) - 2 \phi, \\
Z_{\text{forb}} (\b) ={1 \ov \pi}  \int_{C_1} dE \, e^{-\b E} \sE (E) \, e^{\lam (E)}  \ .
\label{z1pa}
\end{gather} 
For simplicity, we have assumed in~\eqref{z1pa} that both $\lam(E)$ and $\sE(E)$ from~\eqref{exdos} are real,
and $C_1 \equiv C \backslash (E_0, \infty)$. The portion of the integrand extending into the complex plane is obtained via analytic continuation. 

\subsubsection{Threshold contributions from macroscopic oscillations}

Consider the integral~\eqref{z0pa}. Given $\Phi' (E) = 2 \pi \bar \rho (E) > 0$ for $E > E_0$, there are no stationary points in the interior of the integration domain. 
The integral is thus dominated by the contribution from the lower boundary $E_0$, i.e. from the region near the energy threshold. By definition,  
$\bar N (E_0) =0$. Depending on the behavior of $\bar \rho (E)$ and $f (E) = e^{\hat g (E)- c(E)}$, we can have the following scenarios:

\ben 

\item $\bar \rho (E_0) \neq 0$ and $f (E_0)$ is non-singular. The leading behavior of~\eqref{z0pa} can be then readily obtained via integration by parts, 
\be
Z_{\text{osc}} (\b) = {e^{-\b E_0} f (E_0) \ov 2 \pi  \bar  \rho (E_0)} \sin 2 \phi + \cdots    \ .
\ee

\item $\bar  \rho (E_0) = 0$ and $f (E_0)$ is non-singular. In this case, $E_0$ becomes a stationary point, and the leading behavior is given by 
\be
Z_{\text{osc}} (\b) = {e^{-\b E_0} f (E_0) \ov 2 \sqrt{\bar  \rho' (E_0)}} \cos \le({\pi \ov 4} - 2 \phi\ri)  + \cdots    \ .
\ee

\item If $f (E_0)$ is singular, but still integrable, the above expressions do not apply, but it is straightforward to derive the necessary modifications. For example, suppose $f(E) \propto (E-E_0)^{\nu -1}$ for $\nu \in (0,1)$ and $\bar  \rho (E_0) \neq 0$, we find $Z_{\text{osc}} (\b) \propto (\bar  \rho (E_0))^{-\nu}$. 

If $f(E)$ is sufficiently singular at $E_0$ that the integral~\eqref{z0pa} becomes un-integrable, we expect that equation~\eqref{uj1} 
must be modified in the immediate vicinity of $E_0$. Such details are expected to be system-dependent.

\een

In summary, the presence of rapid macroscopic oscillations in the density of states~\eqref{uj1} leads to partition function contributions that are entirely dominated by the region near the energy threshold $E_0$.

\subsubsection{Hyper-instantons} \label{sec:hyper-I}

We now examine~\eqref{z1pa}. Assume that the factor $e^{\lam (E)}$ completely dominates the integrand.\footnote{This is certainly the case in the many-body context, where it is doubly exponential in $\sS$ while the other factors are at most singly exponential. In a few-body system, this assumption may not be warranted.} 
Under this assumption, the integral is governed by the dominant saddle point of $\lam (E)$,
\be \label{hype1}
Z_{\text{forb}} (\b) = {1 \ov \pi} e^{-\b E_s} \sE (E_s) \sqrt{2 \pi \ov - \lam'' (E_s)} e^{\lam (E_s)} \sim e^{\# O(e^\sS)},
\ee
where the saddle point $E_s$ satisfies
\be \label{z1pa1}
\lam' (E_s) = 0  \quad \implies \quad \bar \rho_+ (E_s) =0 \ .
\ee

Equation~\eqref{z1pa1} implies that the saddle points of~\eqref{z1pa} correspond precisely to the zeroes of the analytically continued density of states in the forbidden region. Furthermore, by~\eqref{reSh}, the resolvent takes identical values on the physical and second sheets at such a zero:
\be 
\bar R (E_s) =  \bar R^{(2)} (E_s) \ .
\ee
This identity indicates that the two sheets intersect at $E_s$, giving rise to a node singularity on the Riemann surface $\Sig_{\bar \rho}$. Consequently, the existence of a potential saddle point is deeply tied to the presence of node singularities on the underlying Riemann surface $\Sig_{\bar \rho}$. 

This mechanism directly parallels the single-eigenvalue instantons in random matrix theory, as well as the target-space locations of ZZ-branes in minimal string theory~\cite{SeiShi03}. In a many-body system where the entropy scales extensively with the number of degrees of freedom~($\sS \propto \sN$), such a saddle yields a doubly exponential contribution that may be naturally termed a ``hyper-instanton.''

We note an important self-consistency condition regarding the sign of $\Re \lam(E_s)$. As $Z_{\text{forb}}(\b)$ represents a non-perturbative correction from the forbidden region, it must be exponentially suppressed,  requiring $\Re \lam(E_s) < 0$. If a formal saddle point exists with $\Re \lam(E_s) > 0$, by self-consistency we expect it should not lie on $C_1$ (or its legitimate deformations).

\section{Holographic Systems and Gravity}  \label{sec:gravity} 

It is widely expected that the deep infrared spectral statistics of quantum chaotic systems---whether few-body or many-body---are universally captured by random matrix theory (RMT). In the preceding sections, we demonstrated how random-matrix-like spectral correlations can be extracted for a single chaotic system from the few-body Gutzwiller trace formula via the smooth filter projection procedure.

It is natural to anticipate a many-body generalization of the Gutzwiller formula. Some notable progress has been made in~\cite{EngUrb15,RicUrb22}, which developed a trace formula for many-body systems such as Bose-Hubbard models. In their formulation, the large-$\sN$ limit~\eqref{manB} serves as a semiclassical limit, with the role of classical trajectories being played by periodic solutions to the non-linear, classical mean-field equations.  The analysis therein provides an existence proof that, similar to the closely related few-body semiclassical $\hbar \to 0$ limit, the large-$\sN$ limit~\eqref{manB} is generically singular and can be captured by a Gutzwiller-like representation.   
 However, the mean-field semiclassical approaches developed in these works do not capture the genuine quantum many-body regime; by constructing a collective macroscopic field, they effectively synchronize the individual degrees of freedom. As a consequence, the relevant collective timescales remain $O(1)$ rather than $e^{O(\sN)}$, and the corresponding action scales as $O(\sN)$ rather than exhibiting the exponential $\sN e^{O(\sN)}$ behavior discussed in Sec.~\ref{sec:Gutz}.

Thus, a rigorous derivation of a Gutzwiller trace formula in the fully many-body regime remains an open challenge. While one might formally attempt to generalize the standard formula by expressing the resolvent as a sum over periodic orbits whose complexity scales exponentially with the number of degrees of freedom $\sN$, essential ingredients from the few-body analysis are currently missing. Most importantly, in the many-body regime~\eqref{manB} where $\hbar$ is fixed, the classical notion of periodic orbits in the Gutzwiller sense does not apply. We require a ``classical'' configuration space in the limit~\eqref{manB} that is sufficiently ``large''  to capture timescales of order $e^{O(\sN)}$.  

Even if the explicit framework of summing over discrete periodic orbits does not survive in the many-body regime, the physical expectation that a Gutzwiller-like representation exists to capture the non-smooth microscopic features of the large-$\sN$ limit remains sound. Motivated by this intuition, the anticipated RMT universality of chaotic many-body systems, and the success of the few-body framework, we postulate that the core architecture of the Gutzwiller representation---alongside its associated smooth filter projection---can be generalized to arbitrary chaotic many-body systems, including holographic CFTs such as $\fn=4$ SYM theory.

In this section, we first collect the main elements of this assumed minimal Gutzwiller-like structure, effectively summarizing the analytic framework of the previous two sections.

We then turn to the gravitational side. While it remains unclear what precisely constitutes the gravitational analogue of summing over periodic orbits in the Gutzwiller trace formula, there exists an elegant algebraic framework to distinguish Tier-II and Tier-III objects~\cite{Liu25c}. Specifically, Tier-II objects are treated as vectors within a transseries-valued Hilbert space, while Tier-III quantities function as the complex scalars over this space. 

Physically, this Hilbert space has a compelling interpretation as a third-quantized space of closed universes~\cite{Liu25c}, where each asymptotic boundary represents a macroscopic, closed universe. This multiverse Hilbert space can be defined intrinsically from gravity using wormhole amplitudes\footnote{Assuming the positivity of such amplitudes, which is a consequence of the duality~\eqref{worm} and the positivity~\eqref{CC0} of the smooth filter projection.}. It is thus a direct logical consequence of summing over topologies in the gravitational path integral (corresponding to the first arrow in Fig.~\ref{fig:cartoon}). Furthermore, the multiverse Hilbert space provides a natural home for constructing the gravity duals of spectral determinants and dressed resolvents---hyper-structures that inherently involve an infinite number of boundaries (the second arrow in Fig.~\ref{fig:cartoon}). We demonstrate how these hyper-structures can be leveraged to extract universal level-two features, such as the rapid macroscopic oscillations~\eqref{1edos} and the spectral plateau.

We note that while the mechanism for a hyper-instanton is universal (as is its bulk dual), its actual realization is system-dependent. Therefore, its existence in a specific system cannot be determined strictly from the macroscopic gravitational framework at our current level of understanding. In this section, we use JT gravity as a simple, concrete illustration, treating it purely as a gravitational theory without relying on its random matrix dual interpretation. 

Explicit calculations can also be performed in certain regimes of AdS$_3$ gravity, which we leave to Sec.~\ref{sec:ads3}.

Since the bulk analysis is technically and conceptually independent of the boundary Gutzwiller-like construction, its capacity to yield the same structure offers compelling, indirect support for the proposed Gutzwiller-like framework for holographic systems.

\subsection{Minimal Gutzwiller-like structure for quantum many-body systems} \label{sec:summary}

Consider a chaotic quantum many-body system whose  number of degrees of freedom is
characterized by a parameter $\sN$. We focus on the limit~\eqref{manB} with the 
energy scaling as $E \sim O(\sN)$. For a holographic system, this corresponds to the regime in which the bulk gravity dual is described by a black hole. We will thus refer to this regime as the ``black hole sector'' to distinguish it from the sector where $E \sim O(\sN^0)$ in the large $\sN$ limit.

We postulate the following 
Gutzwiller-like structure for the black hole sector in the large $\sN$ limit:

\ben

\item The resolvent~\eqref{rude0} and its integrated form~\eqref{inRu} have Tier-II descriptions of the form    
\bega
R_\pm (E)  = R (E \pm i0) =\bar R_\pm (E) + U_\pm (E) , \quad  \Im  \bar R_\pm (E) =\mp \pi \bar \rho (E),  \\
\label{jxx}
X_\pm (E) \equiv  \int_{E_0}^E dE' \, R_\pm (E')  
= \bar Q_\pm (E)  + Y_\pm (E) + a_\pm,  \quad  \Im  \bar Q_\pm (E) = \mp \pi \bar N (E) , \\
\bar N (E) = \int_{E_0}^E dE' \, \bar \rho (E'), \quad Y'_\pm (E) = U_\pm (E), \\
 A_-  = A_+^*, \; A = \{\bar R (E), U (E), \bar Q (E), Y (E), a \}  \ . 
\end{gather} 
$\bar R_\pm (E)$ denotes the smooth part, with the macroscopic density of states $\bar \rho (E)$ supported for $E > E_0$, 
while $U_\pm (E)$ captures erratic, ``microscopic'' contributions. $a_\pm$ is an integration constant from the integration of $U_\pm (E)$.
$\bar \rho (E)$ (and its analytic continuation) defines a spectral curve $\Sig_{\bar \rho}$ on which the resolvent $R(z)$ is single-valued. $\bar \rho (E)$, the spectral curve $\Sig_{\bar \rho}$, and $\phi = \Im a_+$ are regulator-independent. In the forbidden region, 
\be 
R_+ (E) = R_- (E) = R (E), \quad X_+ (E) = X_- (E) = X(E), \quad E < E_0 \ .
\ee

Accordingly, the density of states has the following Tier-II description 
\be 
\rho (E) = \bar \rho (E) + \rho^{\rm (err)} (E), \quad \rho^{\rm (err)} (E) = - {1 \ov \pi} {\rm Im} \, U_+ (E)  \ .
\ee
We also introduce the ``local'' determinant as
\bega\label{uehn}
\De_\pm (E) \equiv e^{X_\pm (E)} = |C (E)| e^{\mp i \pi \bar N (E) \pm i \phi} e^{Y_\pm (E)} , \quad E > E_0,  \\
\De (E) \equiv  e^{X (E)} , \quad E < E_0 \ . 
\end{gather}

On general grounds, we expect that
\be\label{rhsc}
\bar \rho (E) , \bar N (E) \sim e^{O(\sN)} ,
\ee
and thus $\De_\pm (E)$ scales with $\sN$ as double exponential. 

\item There exists a smooth filter projection $\FF$, which for $E > E_0$ acts as\footnote{Note that with $Y$ defined as satisfying $\Fil{Y} =0$, we cannot absorb the constant $a$ in~\eqref{jxx} into the definition of $Y$.}
\bega \label{dfil1}
\Fil{\bar \rho} = \bar \rho, \quad \Fil{U}= \Fil{Y} = 0, \quad 
\Fil{Y_s (E_1) Y_{s'}(E_2)} = g_{ss'} (E_1, E_2) , \; s, s'=\pm, \\
g_{ss'} (E_1, E_2) = g_{s's} (E_2, E_1), \quad g_{s s'}^* (E_1, E_2)  = g_{\bar s \bar s'} (E_1, E_2),
\; \bar s = - s \ .
\label{csn11}
\end{gather} 
As $E_1 \to E_2$, $g_{++}$ is regular, while $g_{+-}$ has a logarithmic singularity  
\bega \label{00jhs}
 g_{++} (E+\ep/2, E-\ep/2) = g (E) +O(\ep^2) , \quad \ep \to 0 ,\\
g_{+-} (E+\ep/2, E-\ep/2) =  - \log (- i (\ep +i 0)) + c (E)+ i O(\ep) , \quad \ep \to 0  \ ,
\label{jhs1}
\end{gather}
where $ c(E)$ is real. The expressions~\eqref{dfil1}--\eqref{jhs1} can be extended to the full spectral curve $\Sig_{\bar \rho}$. For example,  
for $E< E_0$, we have the following correlations between $Y$'s on the physical and second sheets
\bega \label{dsco1}
\Fil{Y (E+\ep/2) Y^{(2)} (E-\ep/2)} 
= - \log (- i \ep) + \tilde c (E) , \\
\Fil{Y (E) Y (E)} =\tilde g (E) , \quad
\Fil{Y^{(2)} (E) Y^{(2)} (E)} = \tilde g^{(2)} (E) , 
\label{dsco2}
\end{gather} 
where $\tilde c(E)$, $\tilde g (E)$, and $\tilde g^{(2)} (E)$ are obtained from $c(E)$ and $g(E)$ by analytic continuation as in~\eqref{ons1}--\eqref{ons3}. 

From matching with its gravity counterpart in Sec.~\ref{sec:holo}, we will see that for holographic systems $g_{ss'}$ scale with $\sN$ as 
\be \label{gssc}
g_{ss'} \sim O(\sN^{0})  \ .
\ee

\item  The spectral determinant and its inverse~\eqref{seNd} have a Tier-II representation in terms of the local determinants~\eqref{uehn} 
as
\bln \label{ddet}
D (E) & = \bca  \De_+ (E) + \De_- (E) & E > E_0 \cr
  \De (E) & E < E_0
  \eca \\
  \label{tddet}
\tilde D_\pm (E) & =   \bca  \De_\pm^{-1} (E)  & E > E_0 \cr
  \De^{-1} (E)  + b (\De^{(2)} (E))^{-1} & E < E_0
  \eca \ .
  \end{align}
where $\De^{(2)} (E)$ denotes the expression of $\De (E)$ on the second sheet. 
We expect equation~\eqref{ddet} to be universal, whereas in~\eqref{tddet}, both the precise value of $b$ and whether it is nonzero may depend on the specific system.

We can introduce a ``spectral dipole'' by considering a pair of slightly displaced determinant and inverse determinant, such as
\be 
D(E) \tilde D_+ (E+\ep) = \De_+ (E) \De_+^{-1} (E+\ep) + \De_- (E) \De_+^{-1} (E+\ep), \; E > E_0 ,
\ee
which in turn can be decompose into the sum of a cis-dipole (the first term) and a trans-dipole~(the second term) depending on whether they lie on the same side or opposite sides of the branch cut. We can similarly introduce such a dipole for $E < E_0$ which has a similar decomposition when $b \neq 0$.

\item A ``dressed'' resolvent is defined by attaching a resolvent to a spectral dipole and then taking the ``size'' $\ep$ of the dipole to zero, e.g., 
\be\label{ejnl}
\tilde R_+ (E) \equiv \lim_{\ep \to 0} R_+ (E + \ep) D (E) \tilde D_+ ( E+ \ep)  ,  \quad E > E_0 \ .
\ee
The cis-dipole has a trivial limit, while due to the singular contraction between the resolvent and the trans-dipole (as a result 
of~\eqref{00jhs}--\eqref{jhs1}), they form a nontrivial trans-bound resolvent. That is, after taking care of self-contractions within $\tilde R_+ (E)$ itself and taking the $\ep \to 0$ limit, we can write it as 
\be
\tilde R_+ (E)  = R_{+} (E) + \hat R^{\;\;\,-}_{++} (E) 
\ee
where the trans-bound resolvent $\hat R^{\;\;\,-}_{++} (E)$ is defined as 
\be
\hat R^{\;\;\,-}_{++} (E) \equiv  i e^{\hat g (E)- c (E)} e^{2 \pi i \bar N (E) - 2 i \phi} :e^{Y_- - Y_+}: , \quad
\hat g (E) \equiv \Re g (E), 
\ee
with $:e^A:$ meaning no self-contraction among $A$'s. 
 
When $b \neq 0$, we can define a nontrivial dressed resolvent for $E < E_0$, e.g.,  
\be\label{r-12}
\tilde R_- (E) = \lim_{\ep \to 0} R (E) D (E) \tilde D_- (E + \ep) 
= R (E) + b \hat R^{\;\,1}_{12} (E)
\ee 
where the trans-bound resolvent $\hat R^{\;\,1}_{12} (E)$ involves  the local inverse determinant in the second sheet and can be written as 
\be \label{r-121}
\hat R^{\;\,1}_{12} (E) =i   e^{\bar Q_1 - \bar Q_2^{(2)}} 
e^{\ha \tilde g (E) +\ha \tilde g^{(2)} (E) - \tilde c(E)} : e^{Y (E) - Y^{(2)} (E)} : \ .
\ee

 \een 

The above structure has the following immediate consequences: 

\ben[(a)]

\item  The ramp follows from the $\ep \to 0$ limit of 
\bln \label{raa1}
{1 \ov 2 \pi^2} \Re \Fil{R_+ (E+\ep/2) R_- (E-\ep/2)}_c
& ={1 \ov 2\pi^2} \Re \Fil{ Y'_+ (E +\ep/2)   Y'_- (E -\ep/2)} \\
& = -{1 \ov 2 \pi^2} {\rm Re} \, {1 \ov (\ep+ i 0)^2}  \ .
\label{raa2}
\end{align}

\item From the dressed resolvent~\eqref{ejnl} we have the dressed density of states
\be\label{drdos}
\tilde \rho (E) = - {1 \ov \pi} \Im \tilde R_+ (E) ,
\ee
whose smooth projection includes level-2 transseries terms 
\bln
\Fil{\tilde \rho(E)} & = \Fil{ \rho (E) }- {1 \ov \pi} \Im \Fil{\hat R^{\;\;\,-}_{++} (E)}  \\
& =  \bar \rho (E)  -{1 \ov \pi}  e^{\hat g (E)- c (E)} \cos \le( 2 \pi  \bar N (E) -2  \phi \ri) 
 \ .
 \label{ekio}
\end{align}
From~\eqref{gssc}, the magnitude of the oscillatory term scales as $O(\sN^{0})$.
The rapid macroscopic oscillations leads to partition function contributions that are entirely dominated by the region near the energy threshold $E_0$.  

\item The plateau is  obtained by examining two-point correlations of the dressed density of states~\eqref{drdos} $\Fil{\tilde \rho (E+\ep/2) \tilde \rho (E-\ep/2)}$, whose singular terms (as $\ep \to 0$) includes~\eqref{raa1}--\eqref{raa2} and 
\bln
 {1 \ov 2 \pi^2} \Re \Fil{\hat R^{\;\;\,-}_{++}  (E+\ep/2) \hat R^{\;\;\,+}_{--}  (E-\ep/2)}_c = 
{1 \ov 2 \pi^2} \Re  \le( {e^{2 \pi i \bar \rho (E) \ep}  \ov (\ep + i 0) (\ep - i 0)} \ri) \ .
\label{plas}
\end{align}

\item Equations~\eqref{r-12}--\eqref{r-121} lead to a nonzero density of states in the ``forbidden'' region,
\bega \label{enk}
\Fil{\tilde \rho (E) }=  {1 \ov \pi} \Re \le(\sE (E) e^{\lam (E)} \ri) , \quad E < E_0 , \\
\lam (E) \equiv \bar Q (E) - \bar Q^{(2)} (E) = - 2 \pi i \int_{E_0}^E dE' \,  \bar \rho_+ (E') ,\\
\sE (E) =  b  \, e^{\ha \tilde g (E) +\ha \tilde g^{(2)} (E) - \tilde c(E)}  \ .
\end{gather} 
From~\eqref{rhsc} and~\eqref{gssc}, we have $\lam (E) \sim e^{O(\sN)}$ and $\sE \sim O(\sN^{0})$. 
Equation~\eqref{enk} can lead to double exponential corrections to the thermal partition function of the form 
\be \label{0eune}
Z (\b) = \cdots +{1 \ov \pi} \Re \le( e^{-\b E_s} \sE (E_s) \sqrt{2 \pi \ov - \lam'' (E_s)} e^{\lam (E_s)} \ri)  \sim e^{\# O(e^\sN)}
\ee
when there exists an $E_s$ such that 
\be \label{eune}
\lam' (E_s) = \bar R (E_s) - \bar R^{(2)} (E_s) = - 2 \pi i  \bar \rho_+ (E_s) =0  ,
\ee
and $\Re \lam (E_s) < 0$. 
We refer to such a contribution as arising from a hyper-instanton.

\een

Behaviors such as those seen in \eqref{ekio}, \eqref{plas}, and \eqref{0eune}--\eqref{eune} are well established in random matrix theory, characterizing non-perturbative effects that scale as single exponentials of the matrix rank. Here, we argue that analogous non-perturbative objects exist in specific holographic systems entirely free of ensemble averaging, where they instead generate level-2 double-exponential effects in $\sN$.

\subsection{The multiverse Hilbert space and higher-dimensional analogues of D-branes} \label{sec:hmul}

We now specialize to holographic systems, namely $d$-dimensional CFTs on a boundary manifold---which, for definiteness, we take to be $\RR \times S^{d-1}$---that possess an equivalent description in terms of quantum gravity in asymptotically AdS$_{d+1}$ spacetimes. The parameter $\sN$ is now used to characterize the number of field-theoretic degrees of freedom of the CFT, which is related to the bulk Newton constant by $G_N \propto 1/\sN$. For $d=2$, we identify $\sN = c$, the central charge of the CFT. For $\fn=4$ super Yang-Mills (SYM) theory with gauge group $SU(N)$, $\sN \propto N^2$.

The duality with the boundary CFT implies that there are three tiers of description for the bulk gravitational system (recall Fig.~\ref{fig:3tiers}). Tier-I corresponds to the exact quantum gravitational description, valid for any $G_N$. In analogy with the $\hbar \to 0$ limit of the Gutzwiller formula, and mirroring our earlier large-$\sN$ discussion, we expect that exact quantum gravitational quantities do not generally possess a smooth $G_N \to 0$ limit. As a result, Tier-II quantities decompose into a smooth part, which can be expressed as an asymptotic expansion in $G_N$, and an erratic part containing ``microscopic'' information that reflects the singular nature of this limit. 

While we do not currently have an explicit description of the erratic part, the ``Hilbert space of all closed universes'' introduced in~\cite{Liu25c} provides a convenient framework to capture the general structure of Tier-II quantities, as well as a natural mechanism to extract Tier-III quantities (which correspond to the inner products on this space) from them.

More explicitly, consider the Tier-II CFT partition function $Z_{\rm CFT}[\sX]$ on some compact manifold $M$ with possible operator insertions $X$ (collectively denoted as $\sX = (M, X)$). The products of these partition functions form a commutative algebra, upon which a GNS Hilbert space can be constructed, with the smooth filter projection $\FF$ acting as a state on the algebra. Each vector in this space corresponds to an element of the algebra~\cite{Liu25c}.\footnote{For brevity, we assume that degenerate elements have already been quotiented out.} For example, we can associate a vector $\ket{\sX}$ with $Z[\sX]$ and a vector $\ket{\hat \sX}$ with $Z^{(\rm err)}[\sX]$, which are related by\footnote{The discussion in~\cite{Liu25c} focused only on $Z^{(\rm err)}$ and their products, with $\ket{\sX}$ there corresponding to $\ket{\hat \sX}$ here.}
\be \label{keyde}
\ket{\sX} = Z^{\rm (sm)}[\sX] \ket{\Om} + \ket{\hat \sX} ,
\ee 
where $\ket{\Om}$ is the state vector associated with the identity. Similarly, the vector $\ket{\sX_1, \cdots , \sX_n}$ corresponds to the product of $n$ partition functions, $Z[\sX_1] \cdots Z [\sX_n]$. 

This GNS Hilbert space can be formulated independently on the gravity side, with the state on the algebra defined in terms of the gravitational path integral. In the gravity picture, the GNS inner product between two states is defined by the bulk path integral over geometries connecting boundaries specified by corresponding boundary data---specifically, the wormhole amplitudes. In particular, the connected wormhole amplitudes can be used to define the state $\ket{\hat \sX}$ in~\eqref{keyde} (while the part proportional to $\ket{\Om}$ is given by the disconnected ``disk'' amplitude). This allows us to recover what is interpreted on the boundary as the erratic part, without explicitly introducing an erratic component on the gravity side. 

The equivalence between the bulk and boundary formulations is established through the relations~\eqref{worm} between the boundary smooth projections and the bulk wormhole amplitudes (which, as discussed in~\cite{Liu25c}, predicts an infinite tower of positivity relations for wormhole amplitudes). 
For example, 
\bln \label{mast1}
\vev{\Om|\sX} & = \Fil{Z[\sX]} = Z^{(\rm sm)} [\sX] = Z_{\text{GPI}} [\sX] , \\
\label{mast11}
\vev{\sX_1|\sX_2}  & = \le(Z^{(\rm sm)} [\sX_1] \ri)^* Z^{(\rm sm)} [\sX_2] + \Fil{\le(Z^{\text{(err)}}[\sX_1] \ri)^* Z^{\text{(err)}} [\sX_2]}  \\
\label{mast12} 
& = Z^*_{\text{GPI}} [\sX_1] Z_{\text{GPI}} [\sX_2] +  Z^{\text{wormhole}} [\bar \sX_1, \sX_2] , \\
\vev{\Om|\sX_1, \sX_2} & =  Z^{(\rm sm)} [\sX_1]  Z^{(\rm sm)} [\sX_2] + \Fil{Z^{\text{(err)}}[\sX_1]  Z^{\text{(err)}} [\sX_2]}  \\
& = Z_{\text{GPI}} [\sX_1] Z_{\text{GPI}} [\sX_2] +  Z^{\text{wormhole}} [\sX_1, \sX_2] 
\label{mast2} 
\end{align} 
where by definition $\vev{\Om|\hat \sX} = 0$ and $\vev{\Om|\Om} =1$. The same logic applies to overlaps of vectors involving more than one boundary. The semiclassical gravitational path integral can therefore be regarded as the bulk dual of the smooth filter projection $\FF$. 

It should be emphasized that the GNS Hilbert space is transseries-valued in the sense that the field of complex numbers in the standard definition of a Hilbert space is replaced here by the field of transseries. For example, the coefficient of the first term on the right-hand side of~\eqref{keyde} should be viewed as a transseries in $G_N$ (or $1/\sN$). The same applies to the overlaps between vectors in the Hilbert space, which give rise to Tier-III quantities. 

The GNS Hilbert space contains states associated with an arbitrary number of boundaries. With each asymptotic boundary interpreted as a macroscopic, closed universe, it can therefore be understood as a third-quantized Hilbert space that accommodates an arbitrary number of closed universes, naturally capturing transitions, branching, and interactions among them. Referred to in~\cite{Liu25c} as the ``Hilbert space of all closed universes,'' we will simply call it the multiverse Hilbert space and denote it henceforth as $\sH_{\text{multiverse}}$. The subspace of $\sH_{\text{multiverse}}$ containing only a single boundary is naturally termed the single-universe Hilbert space.

We will now show that $\sH_{\text{multiverse}}$ provides a natural home for the gravitational counterparts of the spectral determinants and dressed resolvents that generate the doubly non-perturbative exponential structure.

Since every element of the commutative algebra of partition functions corresponds to a vector in $\sH_{\text{multiverse}}$, it is convenient for our purposes below to adopt a different labeling scheme from the one introduced in~\cite{Liu25c} and used above. Instead, we will label states directly by their associated macroscopic observables. For example, the vector associated with the thermal partition function $Z (\b)$ on $M_\b = S^1 \times S^{d-1}$ (with $S^{d-1}$ having unit radius and $S^1$ having size $\b$) is written as $\ket{Z (\b)}$.
Given the resolvent relation (with proper regularization understood)
\be 
X(z) = \int_0^\infty {d\b \ov \b} \, e^{z\b} Z(\b) ,
\ee
we define the corresponding vector $\ket{X(z)} \in \sH_{\text{multiverse}}$ by linear extension:
\be \label{yehh}
\ket{X(z)} =  \int_0^\infty {d\b \ov \b} \, e^{z\b} \ket{Z(\b)}  \ .
\ee
This notation extends naturally to products and functions of observables. For $z = E > E_0$, there are two branches on the physical sheet, which we will denote as $\ket{X_\pm(E)}$.
From~\eqref{keyde}, $\ket{X(z)} $ should have the decomposition, 
\be  \label{yehh1}
\ket{X (z)} = X^{(\rm sm)} (z) \ket{\Om} + \ket{X^{\rm (err)} (z)} , \quad \vev{\Om|X^{\rm (err)} (z)}
=0 \ .
\ee

We can now define the states for the local spectral determinants $\De(z)$ and $\De^{-1}(z)$, as well as the dressed resolvents. More explicitly, for $\De(z) = e^{X(z)} = e^{X^{(\rm sm)} (z)} e^{X_{\rm (err)} (z)}$, we define 
\be \label{seul0}
\ket{\De(z)} = e^{X^{(\rm sm)} (z)}  \le(\ket{\Om} + \ket{X_{\rm (err)}  (z)} + {1 \ov 2!} \ket{X^2_{\rm (err)} (z)} + \cdots + {1 \ov n!} \ket{X^n_{\rm (err)}  (z)} + \cdots  \ri),
\ee
where the first term is the vacuum with no closed universes, the second term corresponds to a single universe, the third term contains two universes, and so on. Similarly, we can define 
\be \label{seul}
\ket{\De^{-1}(z)} = e^{- X^{(\rm sm)} (z)}   \le(\ket{\Om} - \ket{X_{\rm (err)}  (z)} + {1 \ov 2!} \ket{X^2_{\rm (err)} (z)} + \cdots + {(-1)^n \ov n!} \ket{X^n_{\rm (err)}  (z)} + \cdots  \ri) \ .
\ee
For $z = E > E_0$, we have the corresponding states $\ket{\De_\pm(E)}$ and $\ket{\De^{-1}_\pm(E)}$. Because~\eqref{seul0} and~\eqref{seul} involve an arbitrary number of boundaries (closed universes), we will refer to them as a \emph{baby-universe condensate} and a \emph{baby-universe anti-condensate}, respectively. Note that since $X^{(\rm sm)} (z)$, which originates from the smooth part of $Z(\b)$, scales as $e^{O(\sN)}$, the normalization factors in~\eqref{seul0}--\eqref{seul} scale with $\sN$ as double exponentials.

Following a similar logic, it is straightforward to construct vectors corresponding to spectral dipoles, consisting of the product of a local determinant and an inverse determinant. A dressed resolvent can then be described by the vector 
\be \label{vdR}
\ket{\tilde R_+(E)} = \lim_{\ep \to 0} \ket{R_+(E + \ep) D(E) \tilde D_+( E+ \ep)}   \ .
\ee

The hyper-instanton saddle discussed in Sec.~\ref{sec:doub} motivates us to define the following object on the gravity side,
\bln \label{seul1}
\ket{\sI (E_s)} & \equiv \ket{e^{X (E_s) - X^{(2)} (E_s)}} , 
\end{align}
where $E_s$ denotes a zero of $\bar \rho_+ (E)$ in the forbidden region $E < E_0$. We interpret~\eqref{seul1} as the gravitational dual of a boundary hyper-instanton saddle and refer to it as a multiverse instanton.

Given the intrinsic bulk definition of $\sH_{\text{multiverse}}$, we stress that the states \eqref{yehh}--\eqref{seul1} can be understood as purely bulk constructs (although motivated by a dual boundary interpretation). They represent emergent higher-level gravitational configurations arising directly from the presence of wormholes. The underlying hierarchical structure is captured in Fig.~\ref{fig:cartoon}.

The spectral determinants in a random matrix model are familiar objects in the duality between the matrix model and minimal string theory (see, e.g.,~\cite{McGTes03,SeiShi03,MalMoo04}). They correspond to certain D-brane objects called FZZT or ZZ branes~\cite{FatZam00,Tes00,ZamZam01} in the dual string theory. In string theory, D-branes are geometric objects in the target spacetime where open strings can end. This characterization relies on a target-space formulation, which, however, is generally absent in higher-dimensional quantum gravity theories. An alternative approach that avoids this reliance is to express D-branes as ``boundary states'' in the closed-string Hilbert space, which can be interpreted as the third-quantized Hilbert space for one-dimensional universes~(strings). The definitions~\eqref{yehh}--\eqref{seul} and~\eqref{seul1} can therefore be viewed as higher-dimensional generalizations of D-brane boundary states, provided we treat string theory as a two-dimensional theory of quantum gravity. See Table~\ref{tab:comp} for a direct comparison between standard string theory terminology and our current definitions.

\begin{table}[htbp]
    \centering
    \renewcommand{\arraystretch}{1.4}
    \begin{tabular}{|c|c|p{4.5cm}|}
        \hline
        \textbf{string theory (2D Gravity)} & \textbf{multiverse framework} & \textbf{physical interpretation} \\ \hline
        Closed-string Hilbert space & Single-universe Hilbert space & States of a single macroscopic connected geometry \\ \hline
        String field theory Hilbert space & $\sH_{\text{multiverse}}$ & State  space of an arbitrary number of universes \\ \hline
        Boundary state (e.g., FZZT brane) & $\ket{X(z)}$ & Insertion of a single macroscopic boundary \\ \hline 
        FZZT brane in string field theory & $\ket{\De(z)}$ or $\ket{\De^{-1}(z)}$ & Coherent superposition (condensate) of baby universes \\ \hline
         ZZ-branes & $\ket{e^{X (E_s)} - e^{X^{(2)} (E_s)}}$ & multiverse instantons \\ \hline
    \end{tabular}
     \caption{Comparison between standard string theory (treated as 2D gravity)  concepts and their higher-dimensional gravitational counterparts within the third-quantized multiverse framework.}
      \label{tab:comp}
\end{table}





\subsection{Elementary overlaps in the multiverse Hilbert space}

We now examine in more detail the gravity description of the elementary overlaps~\eqref{mast1}--\eqref{mast2} in the multiverse Hilbert space. These  level-1 objects provide the basic building blocks for constructing higher-level amplitudes. 

\subsubsection{Macroscopic density of states and the spectral curve}

The inverse Laplace transform of equation~\eqref{mast1} identifies the macroscopic density of states $\bar \rho(E)$ with the exponential of the microcanonical black hole entropy, i.e., 
\be
\sS(E) \equiv \log \bar \rho(E) = S_0(E) + S_1(E) + S_2(E) + \cdots , 
\ee
where $S_n$ has the scaling form $S_n(E) = \sN^{1-n} s_n(\ep)$ with $\ep = (E - E_{\text{gs}})/\sN$. Here, $E_{\text{gs}}$ is the ground-state energy (the Casimir energy of the vacuum), and the branch point $E_0$ of the resolvent is identified as the threshold energy required to form a black hole solution. 

The leading term, $S_0 \propto \sN$, is given by the Bekenstein-Hawking entropy of the black hole with energy $E$,
\be 
S_0(E) = {A \ov 4 G_N} ,
\ee
where $A$ is the horizon area of the black hole. $S_1(E)$ is obtained from one-loop contributions by expanding around the black hole geometry,\footnote{$S_1$ can contain a logarithmic contribution, i.e., $S_1 = s_1(\ep) + \al \log \sN$ for some constant $\al$.} and $S_{n \geq 2}(E)$ comes from higher-loop contributions. The macroscopic density of states for holographic systems therefore takes the form 
\be \label{gdso}
\bar \rho(E) = \sN^\al s_1(\ep) e^{\sN s_0(\ep)} \le(1 + O(1/\sN) \ri), \quad \ep = {E - E_{\text{gs}} \ov \sN}  \ .
\ee

For $d \geq 3$, from the AdS-Schwarzschild solution, we find\footnote{We set the AdS radius to $1$, and the radius of the boundary spatial manifold $S^{d-1}$ is also set to $1$.}
\be\label{gdso1} 
s_0(\ep) = r_s^{d-1}(\ep), \quad \ep =  {d-1  \ov 4 \pi } \le( r_s^{d-2}  + r_s^d  \ri) , \quad
\sN = {\om_{d-1} \ov 4 G_N} ,
\ee
where the second equation implicitly defines $r_s(\ep)$ as a function of $\ep$, and the third equation expresses the large parameter $\sN$ in terms of gravitational quantities (with $\om_{d-1}$ being the area of a unit $S^{d-1}$). The threshold energy is $\ep_0 = 0$. 
Note that $s_0(\ep)$ is single-valued for $\ep \in (0, \infty)$ and is monotonically increasing in $\ep$, with $s_0(\ep) \to \le({4 \pi \ep \ov d-1}\ri)^{d-1 \ov d}$ as $\ep \to \infty$ and $s_0(\ep) \to \le({4 \pi \ep \ov d-1}\ri)^{d-1 \ov d-2}$ as $\ep \to 0$. 
For $d=2$, from the non-rotating BTZ black hole, 
\be\label{gdso2}
s_0(\ep) = {2\pi \ov \sqrt{3}} \sqrt{\ep - \ep_{0}} ,  \quad \ep_{0} = {1 \ov 12}  , \quad \sN = c , \quad E_0 - E_{\text{gs}} = {c \ov 12} \ .
\ee
In all these cases, the one-loop prefactor $s_1(\ep)$ and the parameter $\al$ depend on the precise bulk matter content. 

For JT, we have~\cite{SaaShe19}  ($\ga$ is a parameter which sets the energy scale)
\bega\label{jtex}
Z_{\rm GPI} (\b)  = e^{\sN} {\ga^{3 \ov 2} \ov  (2 \pi)^\ha \b^{3 \ov 2}} e^{{2 \pi^2 \ga \ov \b}}, \quad 
\bar \rho(E) = \frac{e^{\sN}\ga }{2\pi^2 } \sinh(2\pi\sqrt{2 \ga E}) ,
\end{gather} 
with $E_0 =0$. The square root branch point implies that the spectral curve $\Sig_{\bar \rho}$ has two sheets.

As discussed in Secs.~\ref{sec:spec} and~\ref{sec:summary}, the macroscopic density of states $\bar{\rho}(E)$ in~\eqref{gdso} can be used to define a spectral curve $\Sig_{\bar{\rho}}$ for the black hole sector. For a general holographic gravitational system, the lack of knowledge regarding $s_1(\ep)$ precludes a precise characterization of $\Sig_{\bar{\rho}}$ (with JT gravity being a notable exception; see Sec.~\ref{sec:ads3} for another example). Nevertheless, a few general remarks are in order.

First, equations~\eqref{gdso1}--\eqref{gdso2} show that $s_0(\ep)$ possesses a $d$-th root branch point at the threshold energy $\ep_0$. Assuming $s_1(\ep)$ shares this branch point structure\footnote{Which is a natural expectation, as the one-loop contribution is evaluated on the same background black hole geometry.}, the spectral curve $\Sig_{\bar{\rho}}$ for a $\text{CFT}_d$ black hole sector is given by a $d$-sheeted Riemann surface.

Second, the functional form of $\bar{\rho}(E)$ in~\eqref{gdso} differs fundamentally from standard random matrix theory (RMT) ensembles. In traditional matrix models---including JT gravity~\eqref{jtex}---the density of states factorizes as $\bar{\rho}(E) = M f(E)$, where the large parameter $M$ (e.g., the matrix rank or $e^{\sN}$) serves merely as an overall multiplicative prefactor, leaving the energy dependence governed entirely by the $\mathcal{O}(1)$ function $f(E)$. In contrast, the holographic density of states in~\eqref{gdso} features the large parameter $\sN$ inside the exponent: $\bar{\rho}(E) \sim \exp[\sN s_0(\ep)]$. Consequently, as $\ep$ is analytically continued onto different sheets of the complex spectral curve $\Sig_{\bar{\rho}}$, the parameter $\sN$ directly amplifies both the phase and magnitude of $s_0(\ep)$. This leads to severe Stokes phenomena, wild phase oscillations, and regions of extreme exponential suppression or enhancement. As we will see in Sec.~\ref{sec:ads3}, this exponentiated energy dependence leads to rather unusual features regarding the possible existence and action of multiverse instantons.

\subsubsection{Baby universe propagator} \label{sec:holo}

\begin{figure}
\begin{center}
\includegraphics[width=10cm]{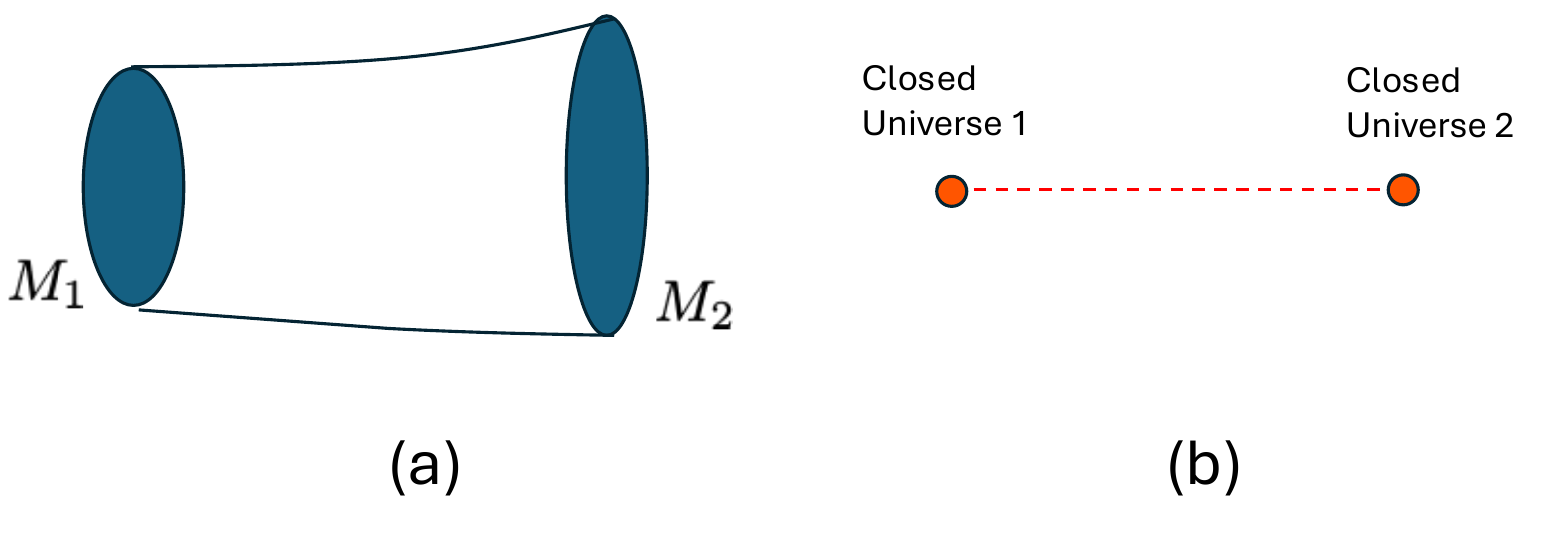}
\caption{\small The dual interpretations of a two-boundary wormhole. (a) The standard geometric perspective, where a Euclidean wormhole smoothly connects two asymptotic boundaries $M_1$ and $M_2$. (b) The third-quantized perspective in $\sH_{\text{multiverse}}$, where this same connected geometry is interpreted as a Feynman-like ``baby-universe propagator,'' mediating transitions (``contractions'') between closed universes.}
\label{fig:propa}
\end{center}
\end{figure}

From the perspective of the multiverse, the connected part of the overlap~\eqref{mast12} (similarly~\eqref{mast2}) plays the role of a ``propagator'' between different closed universes. The corresponding wormhole amplitude can thus be interpreted in the multiverse Hilbert space as a ``baby-universe propagator.'' See Fig.~\ref{fig:propa}.

We now specialize to the boundary manifold $M_\b = S^1 \times S^{d-1}$, with the wormhole amplitude connecting $M_{\b_1}$ and $M_{\b_2}$ denoted as $Z^{(\text{wormhole})}[\b_1,\b_2]$. 
From~\eqref{yehh}--\eqref{yehh1}, we then have\footnote{Proper regularizations at the lower limits may be needed.} 
\bega\label{jhb}
g (z_1, z_2) \equiv \vev{\Om|X^{\rm (err)} (z_1) X^{\rm (err)} (z_2)}= \int_0^\infty  {d\b_1 \ov \b_1} {d\b_2 \ov \b_2} \, e^{\b_1 z_1 + \b_2 z_2} \, Z^{(\text{wormhole})}[\b_1,\b_2] , \\
\vev{X^{\rm (err)} (z_1) |X^{\rm (err)} (z_2)} = g (z_1^*, z_2) , 
\end{gather} 
which may be called the baby-universe propagator in energy space. 

\begin{figure}
\begin{center}
\includegraphics[width=12cm]{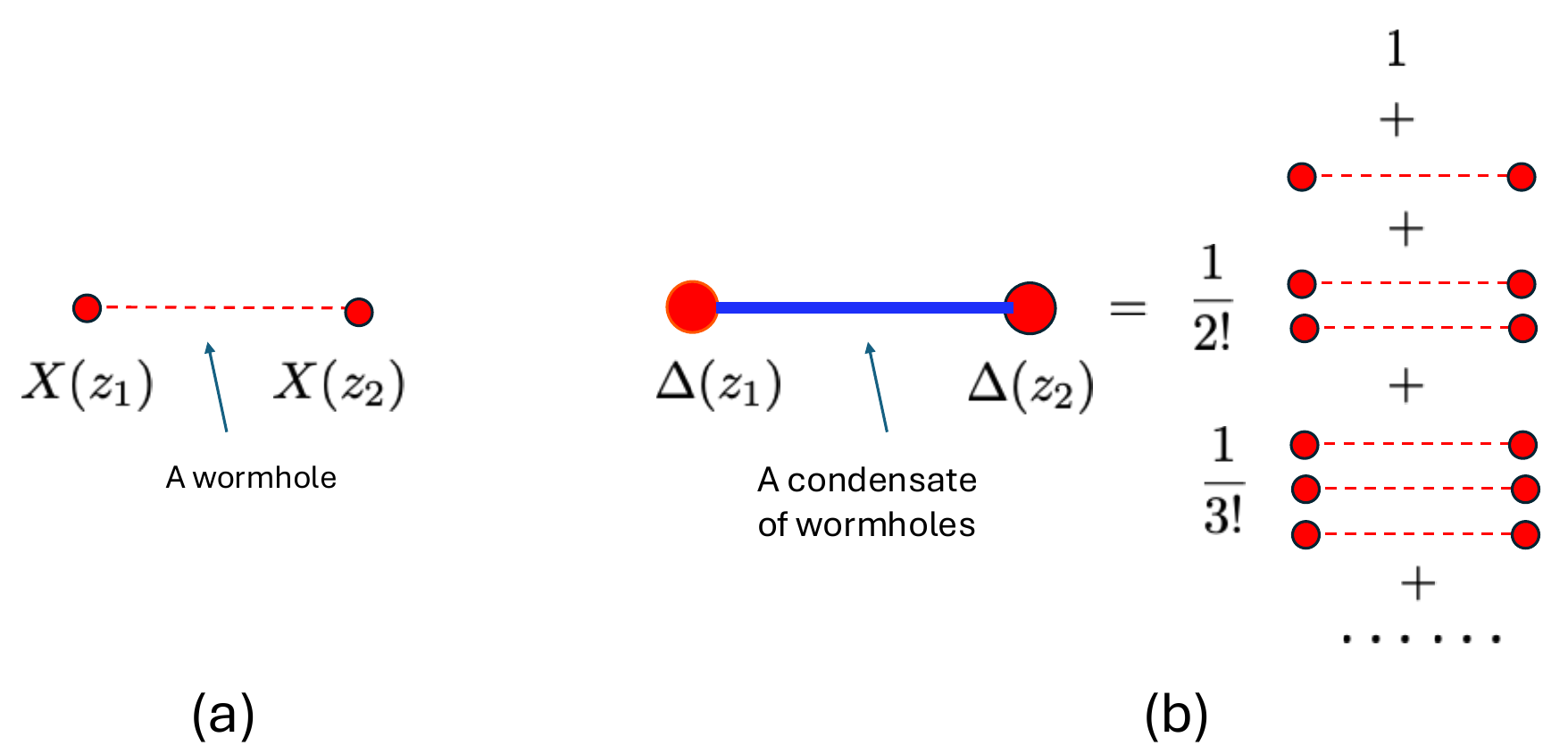}
\caption{\small  a) A single baby-universe propagator connecting two universes. (b) The overlap between baby-universe condensates $\De(z)$ is given by the exponentiation of this elementary propagator, yielding a sum over disconnected multi-wormhole configurations.
Note that the figure depicts only the cross-contractions between $\De (z_1)$ and $\De(z_2)$; the full overlap also includes self-contractions within $\De (z_1)$ and $\De(z_2)$ themselves, as given in~\eqref{selc1}--\eqref{selc3}.
}
\label{fig:expo}
\end{center}
\end{figure}

The overlaps between baby-universe condensates and anti-condensates~\eqref{seul0}--\eqref{seul} are then given by the exponentiation of~\eqref{jhb}, for example\footnote{As in~\eqref{Faex}, in the expressions below we assume that possible contributions from wormholes with more two boundaries are suppressed.} 
\bln\label{selc1}
\vev{\Om| \De (z)} & = e^{X^{(\rm sm)} (z)}  e^{\ha g (z, z)} \\
\vev{\De(z_1)|\De(z_2)} & = e^{X^{(\rm sm)*} (z_1) +X^{(\rm sm)} (z_2)} e^{\ha g^* (z_1, z_1) + \ha g (z_2, z_2) + g (z_1^*, z_2)} ,  \\
\vev{\Om|\De(z_1) \De^{-1} (z_2)} & = e^{X^{(\rm sm)} (z_1) - X^{(\rm sm)} (z_2)} e^{\ha g (z_1, z_1) + \ha g (z_2, z_2) - g (z_1, z_2)},
\label{selc3}
\end{align}  
i.e., by the condensates of baby-universe propagators. See Fig.~\ref{fig:expo} for an illustration.

The Witten--Yau theorem and its generalizations~\cite{WitYau99,CaiGal00} imply that there is no smooth \emph{on-shell} Euclidean wormhole with boundaries $M_{\b_1}$ and $M_{\b_2}$ for $d \geq 2$. Since an on-shell wormhole solution has an action $S_{\text{grav}} \sim 1 / G_N$, these theorems imply that 
\be
 Z^{(\text{wormhole})}[\b_1,\b_2] \not\sim e^{O(1/G_N)} \ .
\ee
The physical reason for the absence of a smooth Euclidean geometry that connects two copies of $S^1 \times S^{d-1}$ and solves the Einstein equations is that the two boundaries effectively exert a topological ``gravitational attraction'' on each other: the ``neck'' of the wormhole tends to pinch off, rendering the geometry singular. 

However, these theorems do not exclude the possibility that
\be\label{pejN2}
 Z^{(\text{wormhole})}[\b_1,\b_2]  \sim O(G_N^{O(1)}) \ .
\ee
Indeed, off-shell wormhole configurations with these topologies for $d \geq 2$ have been found by Cotler and Jensen (CJ)~\cite{CotJen20,CotJen20b,CotJen21,CotJen22}. The key is to manually fix the size of the wormhole throat~(or the total energy $2E$) to obtain a constrained saddle, and then integrate over the fixed parameter. The CJ wormhole amplitude $Z_{\text{CJ}}[\b_1,\b_2]$ scales with $G_N$ as $Z_{\text{CJ}}[\b_1,\b_2] \sim O(G_N^0)$. We thus find\footnote{As usual, the notation $O(G_N^{0})$ includes possible logarithmic dependence.}
\be  \label{pejN3}
g (z_1, z_2) \sim O(G_N^{0}) \ .
\ee
For pure AdS$_3$ gravity, $Z_{\text{CJ}}[\b_1,\b_2]$ can be written in closed form as a Poincar\'e series. In that case, we can find $g (z_1, z_2)$ explicitly in a specific kinematic regime (see Sec.~\ref{sec:ads3}). 

From~\eqref{worm} (or equivalently~\eqref{mast2}), we have the boundary identification 
\be\label{pejN1}
\Fil{Z (\b_1) Z (\b_2)}_c = Z^{(\text{wormhole})}[\b_1,\b_2]  \ ,
\ee
and equation~\eqref{jhb} can be identified with~\eqref{dfil1} (which justifies our use of the same symbol $g$ for both). 
The integral in~\eqref{jhb} should be well-defined for sufficiently negative $z_1 = E_1$ and $z_2 = E_2$. 
As a self-consistency check of this identification, we must find a branch cut when both $E_1$ and $E_2$ are greater than $E_0$ (where $E_0$ is determined by the support of $\bar \rho (E)$).\footnote{A suggestive support is that CJ wormholes only exist only when the energy associated with each boundary is greater than $E_0$~\cite{CotJen21,CotJen22}.} To evaluate $g(E_1, E_2)$ in the physical regime $E_1, E_2 > E_0$, we must choose a branch during analytic continuation. This naturally leads to the branch-dependent expression $g_{ss'} (E_1, E_2) = g (E_1 +i s 0 , E_2 + i s' 0)$.

Setting $\b_1 =\b + i t$ and $\b_2 =\b - i t$ with $t$ large, equation~\eqref{pejN1} evaluates to the spectral form factor, and the corresponding wormhole configuration can be identified with the double cone of~\cite{SaaShe18}. 
As discussed below~\eqref{ber2}, the ramp in the smooth projection of the Gutzwiller representation can be understood as the zero mode corresponding to the relative time shift. This precisely matches how the $t$ factor arises from the gravity side~\cite{SaaShe18,CheIvo23}---namely, as the relative time shift of a periodically identified eternal black hole geometry. Note that the gravity description does not correspond to individual periodic orbits (if such a boundary description exists), but rather to a collective effective description of a large number of microscopic periodic orbits.

We emphasize that in our earlier Gutzwiller-based discussion, the existence of the ramp is solely attributable to the short-distance structure in energy space~\eqref{00jhs}--\eqref{jhs1}, particularly the logarithmic singularity. The ramp behavior of the double cone, combined with the fact that it can be analytically continued from the CJ wormholes~\cite{CotJen21}, implies that $g_{ss'} (E_1, E_2)$ obtained from~\eqref{jhb} exhibits precisely the logarithmic singularity of~\eqref{jhs1}.

This can be seen explicitly in JT gravity, where the corresponding baby-universe propagator is the double trumpet of~\cite{SaaShe19}, which is also off-shell, with the amplitude given by
\be\label{dtr}
Z^{(\text{double trumpet})}[\b_1,\b_2]  = {\sqrt{\b_1 \b_2} \ov 2 \pi (\b_1 + \b_2)} \ .
\ee
Plugging it into~\eqref{jhb} leads to\footnote{Note that the integral is divergent for small $\b_1, \b_2$. To regularize it, we can replace $\b_1 + \b_2$ with $(\b_1 + \b_2)^s$ for $s<1$, and then expand around $s=1$.} (up to a choice of additive constant)
\be \label{dtr1}
g(E_1, E_2) = - \log (\sqrt{-E_1} + \sqrt{-E_2}) ,
\ee
giving for $E > 0$
\bega 
g_{++} (E_1, E_2) =- \log \le(-i (\sqrt{E_1} + \sqrt{E_2}) \ri) , \\
g_{+-} (E_1, E_2) =- \log \le(-i (\sqrt{E_1} - \sqrt{E_2}) \ri)  \ .
\end{gather} 
Taking $E_1 = E + \ep/2$ and $E_2 = E - \ep/2$, we have~\eqref{00jhs}--\eqref{jhs1} with 
\be\label{jtgx}
g(E) =- \log \le(-2 i \sqrt{E} \ri), \quad c (E) = \log (2 \sqrt{E})  \ .
\ee
For $E<0$, we find~\eqref{dsco1}--\eqref{dsco2} with 
\be \label{jtgx1}
\tilde g (E) = - \log 2 \sqrt{-E} , \quad \tilde g^{(2)} (E) = - \log 2 \sqrt{-E} - \pi i , \quad
\tilde c (E) =  \log 2 \sqrt{-E}  - {i \pi \ov 2}  \ .
\ee

\subsection{Hyper-non-perturbative effects from gravity} 

We now turn to how hyper-non-perturbative effects---namely, terms scaling as double exponentials of $G_N$---emerge within the gravitational theory.

The fundamental origin of this macroscopic double-exponential dependence lies in the definition of the condensate states $\ket{\De (z)}$ and $\ket{\De^{-1} (z)}$ in~\eqref{seul0}--\eqref{seul}. Because these states are defined via the exponentiation of the single-universe operator $X(z)$---which itself scales as $e^{\mathcal{O}(1/G_N)}$---their normalization factors naturally generate terms scaling as double exponentials in $G_N$. This exponentiation is the gravitational engine driving all hyper-non-perturbative phenomena.

To provide a purely gravitational derivation of the rapid macroscopic oscillations~\eqref{ekio} in the density of states and the spectral plateau~\eqref{plas}, we require the bulk description~\eqref{vdR} of the dressed resolvents. For convenience, we reproduce this here:
\begin{gather} \label{vdR1}
\ket{\tilde R_+(E)} = \lim_{\ep \to 0} \ket{R_+(E + \ep) D(E) \tilde D_+( E+ \ep)} , \\
R_+ (E) = X'_+ (E), \quad D (E) = \De_+ (E) + \De_- (E) , \quad \tilde D_+ (E) = \De_+^{-1} (E) \ .
\end{gather} 
From~\eqref{vdR1}, we can also introduce the dressed density of states $\ket{\tilde \rho (E)}$ and dressed thermal partition functions $\ket{\tilde Z (\b)}$ through appropriate linear superpositions. While~\eqref{vdR1} is formulated purely as a gravitational object residing in the multiverse Hilbert space $\sH_{\text{multiverse}}$, specifying its precise algebraic composition---such as the roles of $D(E)$ and $\tilde D_+ (E)$---does require some boundary input. We emphasize, however, that this boundary input is minimal and structurally universal; it does not depend on the microscopic details of the specific boundary system. 

\begin{figure}
\begin{center}
\includegraphics[width=6cm]{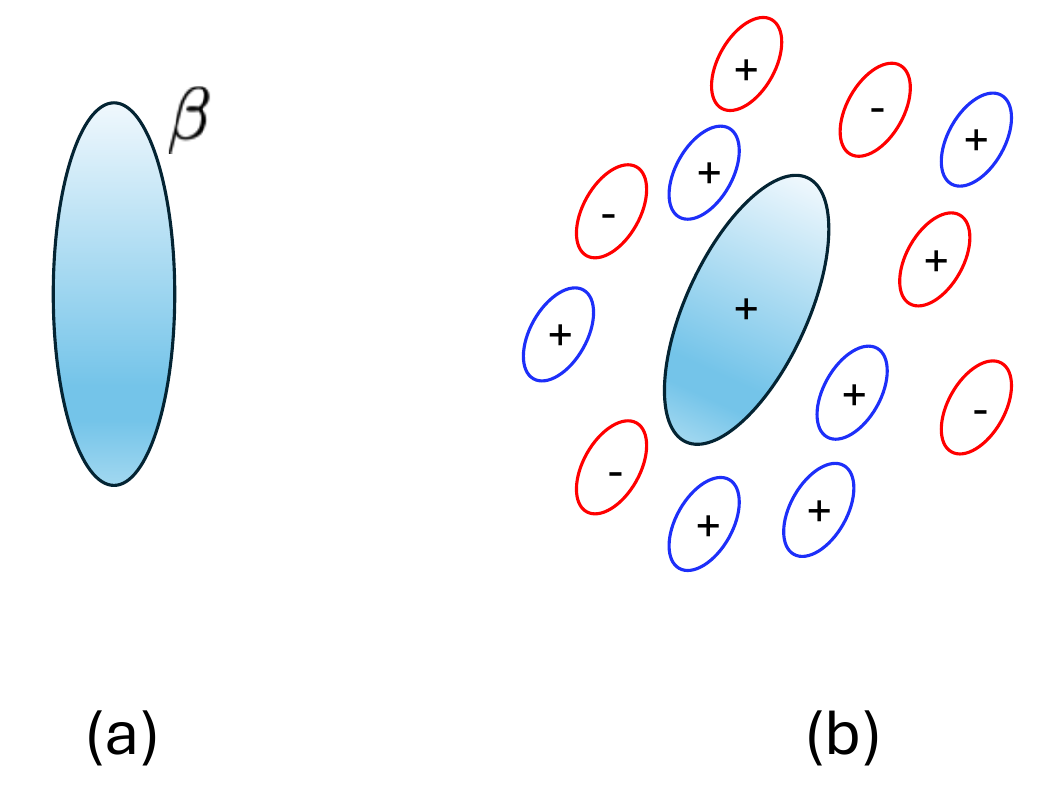}
\caption{\small {\bf Bare boundary:} The bare state $\ket{Z (\b)}$ describes the standard macroscopic boundary $S^1 \times S^{d-1}$, where the thermal circle $S^1$ has size $\b$. (b) {\bf Dressed boundary:} An intuitive visualization of the dressed resolvent $\ket{\tilde R_+ (E)}$. The standard boundary $S^1 \times S^{d-1}$ (with the energy conjugate to the size of $S^1$ held fixed) represented by a shaded circle is surrounded by a ``cloud'' of virtual boundaries. The colors indicate whether these virtual boundaries originate from the condensate $\De (E)$ (red) or its inverse $\De^{-1} (E)$ (blue), while the $\pm$ labels correspond to the respective sub-indices of $\De$ or $\De^{-1}$.}
\label{fig:dress}
\end{center}
\end{figure}

Geometrically, just as the bare state $\ket{Z (\b)}$ describes a single macroscopic universe (or boundary), we may interpret $\ket{\tilde R_+(E)}$ as describing a ``dressed'' boundary. It represents the standard resolvent boundary $R_+ (E)$ immersed in a coherent cloud of arbitrary ``virtual'' boundaries generated by the condensate operators $D(E)$ and $\tilde D(E)$. See Fig.~\ref{fig:dress} for a schematic illustration.

\subsubsection{Rapid macroscopic oscillations} \label{sec:macro} 

The rapid macroscopic oscillations~\eqref{ekio} in the density of states can be derived from gravity by considering the overlap of the dressed resolvent state~\eqref{vdR1} with the vacuum state $\ket{\Om}$:
\be\label{okeo}
\brho^{\text{(sm)}} (E) = -{1 \ov \pi} \Im \vev{\Om \biggr|\tilde R_+ (E)}  = -{1 \ov \pi} \Im  \lim_{\ep \to 0} \vev{\Om  \biggr| R_+ (E +\ep) D (E) \tilde D_+ (E+\ep)} \ .
\ee
This overlap can be evaluated using the bulk gravitational path integral, with the baby-universe propagator~\eqref{jhb} serving as the elementary contraction, precisely as in~\eqref{selc1}--\eqref{selc3}. Structurally, this calculation is identical to the boundary analysis of Sec.~\ref{sec:MDOS}, provided we identify~\eqref{jhb} with the corresponding boundary correlations~\eqref{dfil1}. Exponential factors such as 
\be 
e^{\hat g (E)} = e^{\ha g_{++} (E, E) + \ha g_{--} (E, E)} ,
\ee
which appear in the final expression of~\eqref{ekio}, are now physically reinterpreted: they represent a macroscopic wormhole condensate formed from an infinite sum over products of baby-universe propagators.

\begin{figure}[htbp]
\begin{center}
\includegraphics[width=10cm]{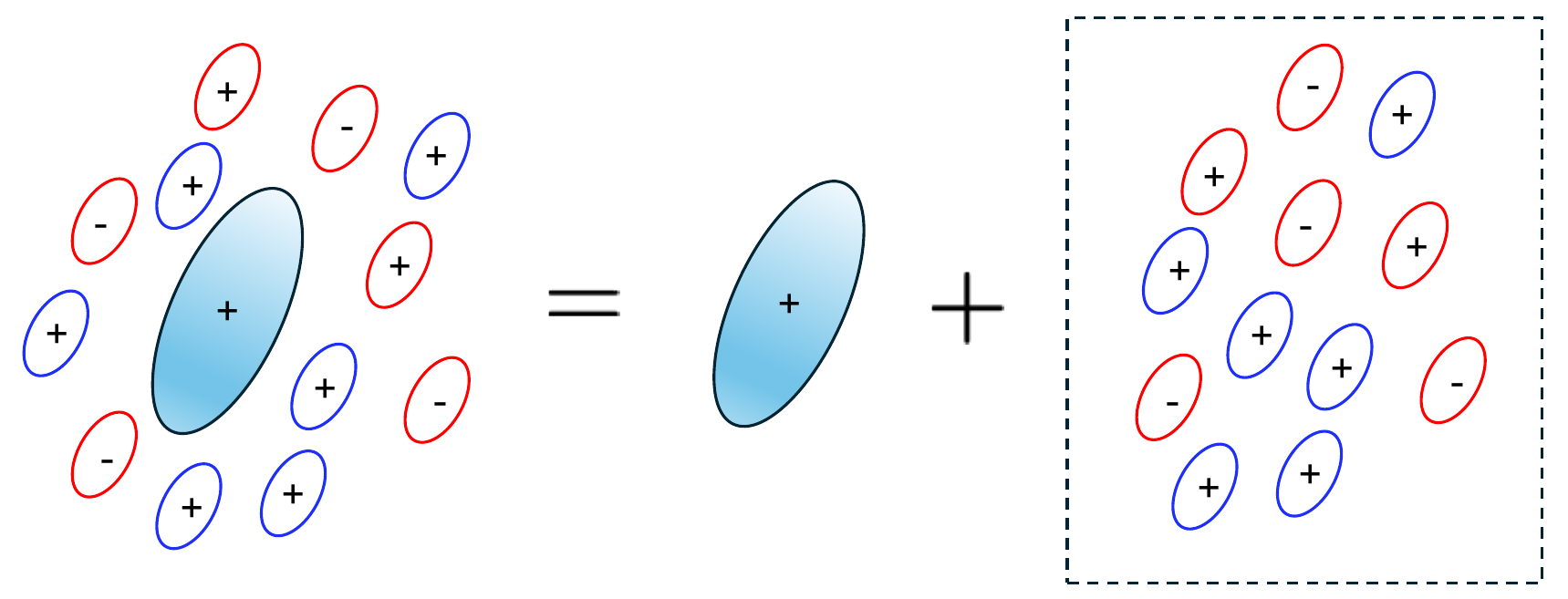}
\caption{\small  The dressed boundary in Fig.~\ref{fig:dress}(b) can be decomposed into the sum of a standard bare boundary and a three-body trans-bound resolvent, the latter being represented by a gas of boundaries inside the dashed box where the bare boundary has been ``resolved'' by the surrounding cloud of virtual boundaries. There are no contractions among the boundaries inside the dashed box.
}
\label{fig:dress1}
\end{center}
\end{figure}

Crucially, given that~\eqref{jhb} must exhibit the logarithmic singularity of~\eqref{jhs1} (as discussed in Sec.~\ref{sec:holo}), the delicate cancellation between the $1/\ep$ pole and the $\ep$ zero factor detailed in Sec.~\ref{sec:MDOS} must proceed in exactly the same manner in the computation of~\eqref{okeo}, as does the formation of the resulting three-body trans-bound resolvent,
\begin{gather}\label{bund0}
\ket{\tilde R_+ (E)} = \ket{R_+ (E)} + \ket{\hat R^{\;\;\,-}_{++} (E) } , \\
 \ket{\hat R^{\;\;\,-}_{++} (E) } =  i e^{\hat g (E)- c (E)} e^{2 \pi i \bar N (E) - 2 i \phi} \ket{:e^{X^{(\rm err)}_- (E) - X^{(\rm err)}_+ (E)}:} \ .
\label{bund}
\end{gather}
Note that in the bound configuration~\eqref{bund}, the bare boundary originally present in the definition of $\ket{\tilde R_+ (E)}$ has disappeared, leaving only the cloud of virtual boundaries. See Fig.~\ref{fig:dress1} for an illustration. Furthermore, because of the rapidly varying macroscopic phase factor $e^{2 \pi i \bar N (E)}$ in~\eqref{bund}, we can heuristically view this second term in~\eqref{bund0} as a much ``heavier'' object than the bare boundary, akin to the rapidly oscillating semiclassical phase of a massive, non-perturbative state.

For JT gravity, using~\eqref{jtgx}, the prefactor in~\eqref{ekio} can be explicitly evaluated, yielding 
\be
\Fil{\tilde \rho(E)}  =  \bar \rho (E)  -{1 \ov 4 \pi E}   \cos \le( 2 \pi  \bar N (E) -2  \phi \ri) ,
\ee
which agrees with the matrix model calculations of~\cite{SaaShe19}, where this effect was interpreted in terms of summing over an infinite of boundaries on FZZT branes on the JT gravity side.

\subsubsection{The plateau} \label{sec:Gplateau} 

The plateau can be derived purely from gravity by considering the correlation of dressed resolvents~\eqref{vdR},
\be\label{rre1}
\vev{\tilde R_+ (E_1)|\tilde R_+ (E_2)} \ ,
\ee
which, once again, can be evaluated using~\eqref{jhb} as the elementary contraction. The calculation proceeds in exactly the same manner as the boundary analysis in Sec.~\ref{sec:plateau}. Substituting the decomposition~\eqref{bund0}, the correlator~\eqref{rre1} naturally separates into two terms: the overlap between the bare boundaries, which yields the ramp, and the overlap between the trans-bound resolvents, which yields the plateau. 

\begin{figure}[htbp]
\begin{center}
\includegraphics[width=10cm]{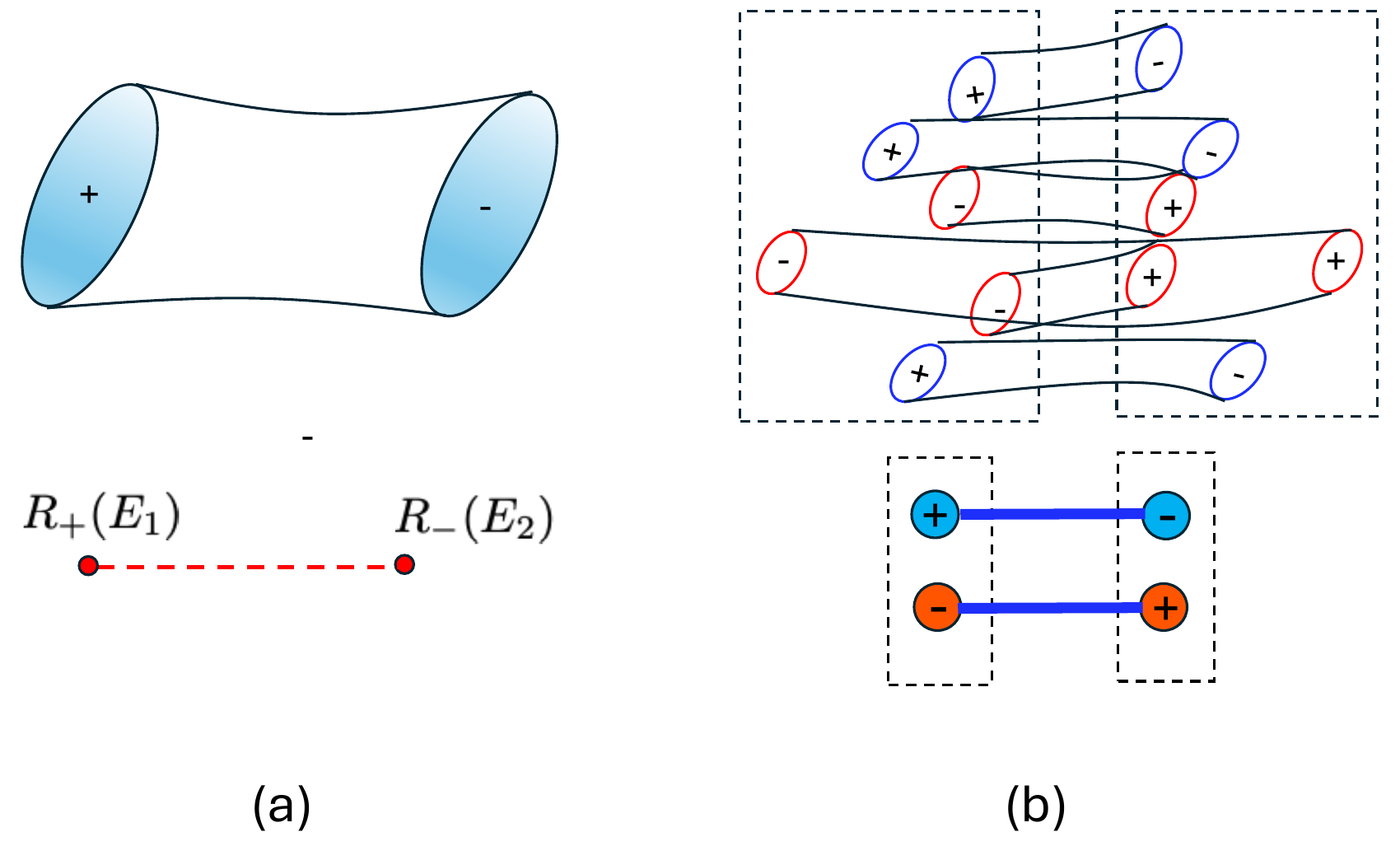}
\caption{\small Gravity counterpart of the ramp and plateau calculations previously illustrated from the boundary perspective in Fig.~\ref{fig:sheet1}. In both (a) and (b), the upper plot is a cartoon geometrically depicting the boundaries and wormholes, while the lower plot is a Feynman-like diagram. 
(a) The ramp arises from a single wormhole contribution, where the $1/\ep^2$ singularity results from acting with two derivatives on~\eqref{jhb}. 
(b) The plateau arises from the overlap of two trans-bound resolvents, each represented by a gas of boundaries inside a dashed box. In the corresponding lower plot, the condensate (anti-condensate) of baby universes is represented by a solid red circle (solid blue circle), with thick horizontal lines representing a condensate of wormholes that each contribute a factor of $1/\ep$. The semiclassical phase difference between these two ``heavy'' configurations leads to the emergence of the Heisenberg time scale $t_H$.
}
\label{fig:Gplat}
\end{center}
\end{figure}

Crucially, while both terms exhibit a $1/\ep^2$ singularity in the limit $E_1 - E_2 = \ep \to 0$, their physical origins are completely different. The former (ramp) arises from the derivative of a single wormhole contribution~\eqref{jhb}, whereas the latter (plateau) emerges from a macroscopic condensation of wormholes, as illustrated in Fig.~\ref{fig:Gplat}.

In the limit $E_1 - E_2 = \ep \to 0$, the macroscopic phase factors~\eqref{bund} associated with the trans-bound resolvents in the bra and ket of~\eqref{rre1} partially cancel. This near-cancellation produces the phase factor $e^{- 2 \pi i \bar \rho (E) \ep}$ (where $E = (E_1 +E_2)/2$), directly yielding the Heisenberg time scale $t_H (E) = 2 \pi \bar \rho (E)$. Thus, in the gravitational description, the emergence of the Heisenberg time scale stems fundamentally from the near-cancellation of the semiclassical phases associated with two heavy, non-perturbative objects. The plateau in the time domain is obtained from the Fourier transform of the sum of~\eqref{rre1} and its complex conjugate. As discussed around~\eqref{ramp_FT}--\eqref{plateau_FT}, this can be understood as the interference among three distinct contributions: a constant background (from the ramp), a time delay of $t_H$, and a time advance of $t_H$.

The plateau has been discussed previously in random matrix models and interpreted in JT gravity using FZZT branes~\cite{SaaShe19} (see also~\cite{BloMer19,MarMax20}). Our objective here is to develop a unifying framework that provides a derivation of the plateau intrinsically from gravity, valid in general dimensions. We also note that the plateau has been  captured in JT gravity within specific scaling regimes~\cite{SaaSta22,BloKru22,WebHan22}, where it can be evaluated through a convergent infinite series.

\subsubsection{Multiverse instantons}  \label{sec:hyper} 

We now consider the multiverse instanton~\eqref{seul1}. While this object is well-defined within $\sH_{\text{multiverse}}$, its physical significance relies on additional boundary data. First, we require the Stokes coefficient $b$ to be nonzero. Second, the integration contour for the partition function (or other observables involving an integration over $E$) must be deformed such that $E_s$ contributes as a saddle. We will proceed under the assumption that both conditions are satisfied. 

The overlaps of $\ket{\sI (E_s)}$ with other states can again be evaluated using~\eqref{jhb}. Because the singularity structure of~\eqref{jhb} is consistent with~\eqref{dsco1}, it follows that 
\be 
\vev{\Om|\sI (E_s)} = 0 \ ,
\ee
meaning the ``self-energy'' of the instanton possesses a zero mode. We can regularize this by defining $\ket{\sI_\ep (E_s)} = \ket{e^{X (E_s+\ep) - X^{(2)} (E_s )}}$, such that $\vev{\Om|\sI_\ep (E_s)} \propto \ep$. 
Probing the instanton with a bare boundary $\ket{R(E)}$ (where $E < E_0$) via the overlap $\vev{R (E)| \sI_\ep (E_s)}$ yields a pole at $E=E_s$. A finite result is obtained by coordinating the $E \to E_s$ and $\ep \to 0$ limits, yielding 
\be\label{qzs}
\vev{R (E_s)|\sI (E_s)} = i   e^{\lam (E_s)} 
e^{\ha \tilde g (E_s) +\ha \tilde g^{(2)} (E_s) - \tilde c(E_s)} ,
\ee
where all quantities are defined as in Sec.~\ref{sec:summary}, provided we identify~\eqref{jhb} with~\eqref{dfil1} and its analytic continuation. Equation~\eqref{qzs} can be interpreted as the multiverse instanton's correction to the resolvent at $E_s$. For self-consistency, the instanton action 
\be \label{mueI}
I_{MI} = - \Re \lam (E_s)
\ee
must be non-negative. The remaining exponential factor in~\eqref{qzs} naturally accounts for the self-energy of the instanton. 

For JT gravity, using~\eqref{jtex}, the density of states becomes 
\be 
\rho_+ (E) = i \frac{e^{\sN}\ga }{2\pi^2 } \sin (2\pi\sqrt{- 2 \ga E}) , \quad E < 0 \ .
\ee
This density has a zero at $E_s = - {1 \ov 8 \ga}$ (the root closest to $E=0$), which has been interpreted as a ZZ brane in JT gravity~\cite{SaaShe19}.
The corresponding instanton action evaluates to 
\be 
 I_{MI} = -  \lam (E_s) =   \frac{e^{\sN} }{\pi } \int_{0}^{{1 \ov 2}} dy   y \,  \sin (2\pi y) 
=  {e^{\sN} \ov 4 \pi^2}  \ .
\ee
From~\eqref{jtgx1}, we thus find the overlap to be
\be
\vev{R (E_s)|\sI (E_s)} = i   e^{- {e^{\sN} \ov 4 \pi^2} } {1 \ov (- 4E_s)} = 2 i \ga e^{- {e^{\sN} \ov 4 \pi^2} }  \ .
\ee 

In Sec.~\ref{sec:ads3}, we will explore a more nontrivial example of multiverse instantons in AdS$_3$.

\section{Multiverse instantons in AdS$_3$}\label{sec:ads3}

For general holographic systems (or any interacting quantum many-body systems), the prefactor $s_1 (\ep)$ in the macroscopic density of  states~\eqref{gdso} is not known, which prevents an explicit application of the general abstract discussion of the multiverse instanton.  A fortunate exception is for $d=2$, where we can use the modular invariance of the CFT partition function 
to find a closed form of $\bar \rho (E)$---including the precise prefactor---in a model independent way in certain regimes.

\subsection{General setup: primary partition function and density of states} 

In a CFT$_2$, the partition function on a torus with modular parameter $\tau$ is defined as 
\bega \label{edos}
\bZ (\tau, \bar \tau) = \Tr \le(q^{L_0 -{c \ov 24}} \bar q^{\bar L_0 -{c \ov 24}} \ri) 
= \int d h d \bar h \, \brho (h, \bar h)  q^{h - \frac{c}{24}} \bar{q}^{\bar{h} - \frac{c}{24}},
\end{gather} 
where $h, \bar h$ denote the eigenvalues of $L_0, \bar L_0$, and $q= e^{2 \pi i \tau},  \bar q= e^{-2 \pi i \bar \tau}$. While the partition function is original defined for $\bar \tau = \tau^*$, it is often convenient to treat $\tau, \bar \tau$ as independent quantities. 
 
 The full partition function $\bZ$ and density of states $\brho$ include the contributions from both the primaries and their descendants. 
 The descendants are fully determined by the Virasoro symmetries and are not ``chaotic.'' 
 Their presence ``contaminates'' the true chaotic dynamics of a system; we thus should ``quotient'' them out. 
 This can be achieved 
 by defining the primary partition function and primary density of states
 \be\label{epdos}
 \bZ_{\text{prim}}(\tau, \bar{\tau}) = \eta(\tau) \bar \eta(\bar{\tau}) \bZ(\tau, \bar{\tau}) = 
 \int d h d \bar h \, \brho_{\text{prim}} (h, \bar h)  q^{h - \frac{c-1}{24}} \bar{q}^{\bar{h} - \frac{c-1}{24}},
 \ee
 where $\eta (\tau) = q^{1 \ov 24} \prod_{n=1} (1-q^n)$ is the Dedekind $\eta$-function and $\bar \eta (\bar \tau) = \bar q^{1 \ov 24} \prod_{n=1} (1-\bar q^n)$. By comparing~\eqref{edos} and~\eqref{epdos}, it can be readily seen that the full density of states $\brho$ and its primary counterpart $\brho_{\text{prim}}$ are related by a convolution 
\be\label{hsna}
\brho (h, \bar h) = \sum_{n, \bar n=0}^{\infty} p(n) p (\bar n) \brho_{\text{prim}}(h - n, \bar h-\bar n) ,
\ee 
where $p(n)$ is the partition function of integer $n$ defined by
\be
{1 \ov  \prod_{n=1}^\infty (1 - q^n)} = \sum_{n=0}^{\infty} p(n) q^n  \ .
\ee 

A $\text{CFT}_2$ may possess other global symmetries whose effects should be quotiented out as well.  Because this is highly model-dependent, we restrict our attention to theories without additional global symmetries. Physically, this means that there are no extra conserved currents and any primary state other than the vacuum has conformal weights strictly greater than zero,
\begin{equation} \label{ejga}
h > 0 \quad \text{and} \quad \bar{h} > 0 \ .
\end{equation}

We are  interested in the leading behavior of $ \brho_{\text{prim}} (h, \bar h)$ in the black hole sector in the large $c$ limit, i.e., 
\begin{equation} \label{larc}
c \to \infty, \quad h \geq \frac{c-1}{24}, \quad \bar{h} \geq \frac{c-1}{24}  \ .
\end{equation}
For this purpose, it is convenient to use the parameterization
\begin{gather}
c = 1 + 6 Q^2 , \quad Q = b+ b^{-1} = \sqrt{\frac{c-1}{6}}, \\
E_L = h - \frac{c-1}{24} =  P^2  , \quad E_R = \bar h - \frac{c-1}{24} =  \bar P^2  \ .
\end{gather}
Note that the Liouville momentum $P$ is real for states above the black hole threshold ($h \geq \frac{c-1}{24}$) and becomes purely imaginary for light states sitting below the threshold.

The modular $S$-invariance of the partition function $\bZ (-1/\tau, -1/\bar \tau) = \bZ (\tau, \bar \tau)$ implies that the primary density of states must satisfy the integral constraint~\cite{Max19} 
\begin{equation} \label{xba}
\brho_{\text{prim}} (P', \bar P') =\int d P d \bar P \, S_{P'P} S_{\bar P' \bar P} \, \brho_{\text{prim}} (P, \bar P),
\end{equation}
where $S_{P'P}$ is the continuous Virasoro $S$-modular matrix~\cite{Tes01, PonTes99}, implicitly defined via the character transformation\footnote{Here, $\chi_P (\tau) = \frac{q^{P^2}}{\eta (\tau)}$ denotes the Virasoro character for a non-degenerate primary with Liouville momentum $P$.}
\begin{equation}\label{xba1}
\chi_P (-1/\tau) = \int_{0}^\infty d P'\,\chi_{P'} (\tau)  S_{P'P}  \ .
\end{equation}
In~\eqref{xba}, the integration contours are understood to extend into the complex plane along the imaginary axis to capture the contributions from light states. Equation~\eqref{xba1} maps a Gaussian to another Gaussian which uniquely determines $S_{P', ip_0} $ for a generic light operator with imaginary momentum $P = i p_0$ to be
\begin{equation}\label{msm1}
S_{P' , i p_0} = 2 \sqrt{2} \cosh (4 \pi p_0 P')  \ .
\end{equation}
For the identity operator $\mathbf{1}$, the vacuum character $\chi_{\mathbf{1}}(\tau) = \chi_{iQ/2}(\tau) - \chi_{i\tilde{Q}/2}(\tau)$ (with $\tilde Q = b^{-1} - b$)  yields the corresponding $S$-modular matrix element,
\begin{equation}\label{msm0}
S_{\mathbf{1}} (P') = S_{P' , i Q/2}  - S_{P' , i \tilde Q/2}  = 4 \sqrt{2} \sinh (2 \pi b P') \sinh (2 \pi b^{-1} P' )  \ .
\end{equation}

The primary density of states due to the $S$-modular transform of the identity operator then has the form 
\be \label{exsa}
\brho_{\text{prim}} (P, \bar P)  = S_{\bid} (P) S_{\bid} (\bar P) + \cdots
\ee
where $\cdots$ denotes those obtained from other possible light operators. For example, an operator with $P=i p_0, \bar P = i \bar p_0$ will contribute to a term $S_{P, i p_0} S_{\bar P, i \bar p_0}$. 

Now consider the $c \to \infty$ limit. Suppose $E_L = h - {c -1\ov 24}  \sim O(c)$, i.e. $P \sim O(c^\ha)$. We then find that 
\be
S_\bid (P) = 2 \sqrt{2} e^{- {13 \ov 24} \tilde \b_L} \sinh \le({\tilde \b_L \ov 2}\ri) 
e^{2\pi \sqrt{\frac{cE_L}{6}}}  + O(e^{- O(c)}), \quad \tilde \b_L \equiv 4 \pi \sqrt{6 E_L \ov c} \sim O(c^0)  \ .
\label{emq}
\ee
Introducing the leading microcanonical entropy $S_L$ and the corresponding inverse temperature $\b_L$,
\be\label{emq1} 
S_L = 2\pi \sqrt{\frac{cE_L}{6}} \sim O(c) , \quad \b_L \equiv {\p S_L \ov \p E_L}  = \pi \sqrt{c \ov 6 E_L} , 
\ee
we see that $\tilde \b_L$ is the dual inverse temperature. 
\be
\tilde \b_L = {4 \pi^2 \ov \b_L}  \ .
\ee
For a light operator of dimension $h_0 \sim O(c^0)$, we have $p_0 =  {Q \ov 2} - { h_0 \ov Q}$ and 
\be\label{azq}
S_{P, ip_0} = S_\bid (P)  {e^{- \tilde \b_L h_0 }  \ov 1 - e^{- \tilde \b_L}} + \cdots \ .
\ee

\subsection{A model-independent regime} 

 Since we cannot control the matter content, we will search for a regime within~\eqref{larc} where only the first term in~\eqref{exsa} dominates.

We can suppressed the contributions~\eqref{azq} by considering the limit $\tilde \b_L \propto \sqrt{E_L /c} \to \infty$.
Consequently, in a theory satisfying~\eqref{ejga}, in the regime 
\be \label{eya}
c \to \infty, \;\;  \tilde \b_L \propto \sqrt{E_L \ov c} \to \infty, \; \; \bar P = \text{arbitrary real} ,
\ee
the leading asymptotic behavior of $\brho_{\text{prim}}$ is given by (thus for the superscript ``(sm)'')
\bega\label{kk1}
\brho_{\text{prim}}^{(\text{sm})} (P, \bar P)  = S_{\bid} (P) S_{\bid} (\bar P) + \cdots,  \\
\label{kk2}
S_{\bid} (P) = \sqrt{2}  e^{2 \pi \sqrt{\frac{cE_L}{6}} - {\tilde \b_L \ov 24}} + \cdots , \quad
S_{\bid} (\bar P) = 4 \sqrt{2} \sinh (2 \pi b \bar P) \sinh (2 \pi b^{-1} \bar P) ,
\end{gather} 
where $\cdots$ denote terms that are either exponentially suppressed by $c$ or $\tilde \b_L$. 

In~\eqref{kk1}, $\bar P$  can take any real value. Depending on how it scales with $c$, we can separate~\eqref{kk1} into 
different sub-regimes (below $\tilde \beta_R \equiv 4 \pi \sqrt{6 E_R \ov c} $): 

\ben[(i)]

\item $E_R \sim O(c), \; \bar P \sim O(c^\ha), \; \tilde \b_R \sim O(c^0)$. In this regime, 
\be \label{kkk1}
S_{\bid}^{\rm (i)} (\bar P) = 2 \sqrt{2} e^{- {13 \ov 24} \tilde \b_R} \sinh \le({\tilde \b_R \ov 2}\ri) 
e^{2\pi \sqrt{\frac{cE_R}{6}}}  + e^{-O(c)}  \ .
\ee

\item $E_R \sim O(c^0), \; \bar P \sim O(c^0), \; \tilde \b_R \sim O(c^{-\ha})$. We then have 
\be \label{kkk2}
S_{\bid}^{\rm (ii)} (\bar P) =  \sqrt{2} \tilde \b_R  e^{2 \pi \sqrt{c E_R\ov 6} } + e^{-O(c^\ha)}  \ .
\ee

\item $E_R = \sE_R c^{-1}$ with $\sE_R \sim O(c^0)$ which corresponds to $\bar P \sim O(c^{-\ha})$ and $\tilde \b_R \sim O(c^{-1})$. 
We then find 
\be \label{kkk3}
S_{\bid}^{\rm (iii)} (\bar P) =  {8 \pi \sqrt{3 \sE_R} \ov c}  \sinh \le(2 \pi \sqrt{\sE_R \ov 6}\ri)   \ .
\ee
This regime, which reduces to the JT gravity, was studied previously in detail in~\cite{GhoMax20}.

\een

With the total energy $E$ and angular momentum $J$ defined as 
\bega 
E = E_L + E_R   , \quad J = E_L - E_R   ,
\quad  E_L = {E + J  \ov 2} , \quad E_R =  {E - J \ov 2} ,
\end{gather} 
the limit~\eqref{eya} can also be written as 
\be\label{eya1}
c \to \infty, \quad {E \ov c} \to \infty, \quad {J \ov c} \to \infty, \quad  E-J  = {\rm finite}  \times \bca O(c) \cr O(c^0) \cr O(c^{-1}) 
\eca ,
\ee
with the three possible sub-regimes for $E-J$ as described above in (i)--(iii). 

The primary density of states for $E, J$ can be obtained as 
\be 
\brho_{\text{prim}}  (E, J) =\ha \brho_{\text{prim}}  (E_R, E_L) =   {\brho_{\text{prim}}  (P, \bar P) \ov 8 P \bar P}  \ .
\ee
From~\eqref{kk1}--\eqref{kkk3} we then find the leading asymptotic behavior 
\be
\brho_{\text{prim}}^{(\text{sm})}  (E, J)  = \bca  M_1 (J)  {  e^{- {13 \ov 24} \tilde \b_R}  \sinh \le({\tilde \b_R \ov 2} \ri) e^{2 \pi \sqrt{c E_R \ov  6} }  \ov   \sqrt{E_R}} &  E_R \sim O(c)\\
M_2 (J) e^{2 \pi \sqrt{c E_R \ov  6}} & E_R \sim O(c^0) \\ 
  M_3 (J)  \sinh (2 \pi \sqrt{\sE_R \ov 6}) &  E_R= \sE_R c^{-1}
  \eca,
  \label{fdos} 
\ee
where we have introduced 
\be 
M_1 (J) = {e^{2 \pi \sqrt{c J \ov 6} - \pi \sqrt{J \ov 6c}} \ov 2 \sqrt{J}}, \quad M_2 (J) =M_3 (J) =  \pi \sqrt{6 \ov cJ} e^{2 \pi \sqrt{c J \ov 6} - \pi \sqrt{J \ov 6c}}  \ .
\ee
We can view~\eqref{fdos} as giving the primary density of states at a fixed spin $J \to \infty$ with excitation energy $E -J = 2 E_R$. 
In the third line of~\eqref{fdos}, the density of states reduces to that of JT gravity~\cite{GhoMax20}. 

Equation~\eqref{fdos} gives examples of the macroscopic density of states $\bar \rho (E)$ where the $O(1)$ prefactors are precisely fixed. 
We  can then use them explore possible existence of multiverse instantons, following the discussions of earlier sections. Before doing that, we briefly digress to discuss the connection with the BTZ black hole.

The BTZ black hole~\cite{BTZ} has the metric 
\bega 
ds^2 = - f dt^2 + {1 \ov f} dr^2 + r^2 (d\phi + n (r) dt)^2 , \\
f = {(r^2-r_+^2) (r^2 - r_-^2) \ov r^2} , \quad n(r) = {r_+ r_- \ov r^2}, 
\end{gather} 
where we have set the AdS radius to $1$. Various thermodynamic quantities associated with the black hole are given by 
\be
E = {r_+^2 + r_-^2 \ov 8 G_N} , \quad J = {r_+ r_- \ov 4 G_N} , \quad
\b = {2 \pi  r_+ \ov r_+^2 - r_-^2} , \quad 
S_{\rm BH} = {2 \pi r_+ \ov 4 G_N}, \quad \Om =  n (r_+) = {r_- \ov r_+} ,
\ee
where $E$ and $J$ are  the energy and angular momentum of the black hole,  $\b$ is the inverse temperature, $S_{\rm BH}$ is the black hole entropy, and $\Om$ is the angular velocity at the horizon which can be identified with the chemical potential for the angular momentum $J$. 

The left and right energies and the left and right inverse temperatures are then given by
\bega
E_L = {E+J \ov 2} = {1 \ov 16 G_N} \le({r_+ + r_-} \ri)^2, \quad E_R = {E-J \ov 2} =  {1 \ov 16 G_N} \le({r_+ - r_-} \ri)^2  , \\
\beta_L = \beta(1 - \Omega) ={2 \pi  \ov r_+ + r_-} , \quad
\beta_R = \beta(1 + \Omega) ={2 \pi  \ov r_+ - r_-} \ .
\end{gather} 
We can solve $r_+, r_-$ in terms of $E$ and $J$ and express the black hole entropy as 
\bega
S_{\rm BH}  = {2 \pi r_+ \ov 4 G_N} 
 =   4\pi \le(\sqrt{E_L \ov 16 G_N}
  +  \sqrt{E_R \ov 16 G_N} \ri) = S_L + S_R , \\
  S_L = 2 \pi \sqrt{c E_L \ov 6},  \quad S_R = 2 \pi \sqrt{c E_R \ov 6},
  \label{entr}
  \end{gather}
where we have used that ${1 \ov 16 G_N} = {c \ov 24}$. Similarly, we find 
\bega 
\beta_L = \pi  {\sqrt{c \ov 6 E_L}} , \quad
\beta_R = \pi  {\sqrt{c \ov 6 E_R}},
\label{LRT}
\end{gather}
Equations~\eqref{entr} and~\eqref{LRT} are identical to~\eqref{emq1}, i.e.,  the corresponding named quantities can be identified. 

It then immediately follows that the regime~\eqref{eya1} corresponds to a near extremal black hole with $r_+ , r_- \to \infty$ with $r_+ - r_-$ fixed. In particular, the three regimes of $E_R$ corresponds to 
\bln 
 \text{(i)}: & \quad r_+ - r_- \sim O(1), \quad \b = \ha \b_R =  O(1), \, \b_L \to 0, \\
 \text{(ii)}: &  \quad r_+ - r_- \sim O(c^{-\ha}) , \quad \b = \ha \b_R \sim O(c^\ha) ,  \, \b_L \to 0, \\
 \text{(iii)}: & \quad r_+ -  r_- \sim O(c^{-1}) , \quad \b = \ha \b_R \sim O(c)  ,  \, \b_L \to 0 \ .
\end{align}
Note that sub-regimes (ii) and (iii) are very close to extremality and are characterized by very low temperatures.

\subsection{Possible existence of multiverse instantons} 

With the density of states~\eqref{fdos} including the precise $\mathcal{O}(1)$ prefactor, we can now explore the possible existence of multiverse instantons and their actions. We should note that, given our current understanding of the boundary CFT$_2$, it is not clear whether the Stokes coefficient $b$ is nonzero, nor can we explicitly determine the integration contour for the partition function. We will therefore simply evaluate the would-be instanton action.

Note that the sub-regime (ii) does not have a zero, and the JT gravity which corresponds to sub-regime (iii) was already discussed in Sec.~\ref{sec:hyper},  We will thus focus here on sub-regime (i), with 
\bega 
\bar \rho_J  (E) = M_1 (J)  {  e^{- {13 \ov 24} \tilde \b_R}  \sinh \le({\tilde \b_R \ov 2} \ri) e^{2 \pi \sqrt{c E_R \ov  6} }  \ov   \sqrt{E_R}} , \\
\tilde \beta_R = 4 \pi \sqrt{6 E_R \ov c},
 \quad  E - J = 2E_R  \sim O(c) \ .
\end{gather} 
The density of states has a $\ZZ_2$ branch point at $E_R =0$. It is convenient to introduce a new variable
\be
E- J = 2 E_R = - c u^2  \ .
\ee
For $E < J$ from the upper half plane, we have $\sqrt{E_R} = i \sqrt{c \ov 2} u$ with $u> 0$. 
The analytically continued density of states can then be written as 
\bega\label{egw}
\bar \rho_J (-u^2) = M_1  \sqrt{2 \ov c} e^{- i b_1 u} \frac{\sin(b_2 u)}{u} e^{i c a u} ,\\
 a ={\pi \ov \sqrt{3}} , \quad b_1 = {13 \ov 6} \pi \sqrt{3}, \quad 
b_2 = 2 \pi \sqrt{3} \ .
\end{gather} 
With the exponent proportional to $c$,~\eqref{egw} is highly oscillatory. 


Equation~\eqref{egw} has zeros for real $u$. Let us focus on the first zero
\be \label{zeRo}
u_s = {\pi \ov b_2} = {1 \ov 2 \sqrt{3}} ,
\ee
which gives 
\bln
\lam (E_s)  &= - 2 \pi i \int_{J}^{E_s} dE' \,  \bar \rho_J (E') =   4 \pi i  c \int_{0}^{u_s} du u   \bar \rho_J (-u^2) \\
& = -i \sqrt{2} M_1 {12^{3 \ov 2} \ov c^{3 \ov 2}}  \le(1 - e^{-{i \pi \ov 12}} e^{{i \pi  c \ov 6}}\ri) \ .
\end{align}
From~\eqref{mueI}, we thus find that the corresponding multiverse instanton action is given by
\be\label{qpa}
I_{MI} = - \Re \lam (E_s)  =  \sqrt{2} M_1 \le({12 \ov c}\ri)^{3 \ov 2} \sin {\pi (c -\ha) \ov 6} \ .
\ee
We see that, owing to the highly oscillatory factor in~\eqref{egw}, the sign of the instanton action oscillates wildly with $c$. As discussed earlier in Sec.~\ref{sec:hyper-I}, self-consistency requires $I_{MI} > 0$. For those values of $c$ where~\eqref{qpa} is negative, we expect that $u_s$ does not correspond to a genuine saddle (e.g., the integration contour should not pass through it). 
 It remains an open question whether we should trust~\eqref{qpa} to represent a genuine multiverse instanton even for the values of $c$ where it is positive.

We note that the existence of zeros~\eqref{zeRo} in~\eqref{egw} depends crucially on the use of the primary density of states. It can be readily checked that these zeros disappear when considering the full density of states. In other words, the instantons are ``washed out'' by the inclusion of descendants, which smear out the chaotic structure of the primary density of states.

In the case of pure AdS$_3$ gravity, there are no light primaries other than the identity. The primary density of states is again given by~\eqref{kk1}, without the need to take the $E_L/c \to \infty$ limit. We can then obtain the density of states at fixed angular momentum $J$ and examine the possible existence of multiverse instantons. The subsequent analysis proceeds similarly to the discussion above and will not be repeated here.

 \subsection{Baby-universe propagator in pure AdS$_3$ gravity}

The wormhole amplitude connecting two copies of $T^2$ with modular parameters $\tau_1 , \tau_2$ can be found explicitly for pure gravity in AdS$_3$ and has the form~\cite{CotJen20}  
\bega
Z (\tau_1 , \tau_2) = {1 \ov 2 \pi^2} Z_0 (\tau_1) Z_0 (\tau_2) \sF (\tau_1, \tau_2),  \\
Z_0 (\tau) = {1 \ov \sqrt{\Im \tau} |\eta (\tau)|^2} , 
\quad \sF (\tau_1, \tau_2) =  \sum_{\ga \in \text{PSL} (2, \ZZ)} 
{\Im (\tau_1) \Im (\ga \tau_2) \ov |\tau_1 +\ga \tau_2|^2}   \ .
\end{gather} 

We define can define the amplitude for the primaries by stripping off the descendants 
\be
Z_{\rm prim}  (\tau_1 , \tau_2) = |\eta (\tau_1)|^2 |\eta (\tau_2)|^2 Z (\tau_1 , \tau_2) 
=  {1 \ov 2 \pi^2 \sqrt{\Im \tau_1}\sqrt{\Im \tau_2}} \sF (\tau_1, \tau_2) , 
\ee
and consider its Fourier modes in energy space (with $\tau_s = x_s + i {\b_s \ov 2 \pi}, s=1,2$) 
\bega
Z_{\rm prim}  (\tau_1, \tau_2) =\sum_{J_1, J_2=0}^\infty e^{-2 \pi i x_1 J_1 - 2 \pi i x_2 J_2}  F_{J_1, J_2} (\b_1, \b_2) , \\
g(J_1, z_1;  J_2 , z_2) = \int_0^\infty  {d\b_1 \ov \b_1} {d\b_2 \ov \b_2} \, e^{\b_1 z_1 + \b_2 z_2} \,F_{J_1, J_2} (\b_1, \b_2) \ .
\end{gather} 
Under the identification~\eqref{worm}, we have 
\be
g(J_1, z_1;  J_2 , z_2) = \Fil{X_{J_1} (z_1) X_{J_2} (z_2)}
\ee
where $X_J (z)$ is a Tier-II representation for the integrated resolvent projected to the subspace with a fixed angular momentum $J$, 
\be 
\bX_J (z) = {\rm Tr}_J \log (z- H)  \ .
\ee
To probe the spectral correlations we thus need to take $J_1 = J_2  = J$ and we will take $J > 0$. 

From~\cite{CotJen20} (see Sec. 4), $F_{J, J} (\b_1, \b_2)$ can be written as 
\bega 
 F_{J_1, J_2} (\b_1, \b_2) = e^{-(\b_1 + \b_2) J} \le(f_{DT} (\b_1, \b_2)  + 
 f_2 (J; \b_1, \b_2) \ri) \\
   \label{yeg2}
  f_2 (J; \b_1,\b_2)  ={\pi \ov \sqrt{\b_1 \b_2}}  \sum_{n=1}^\infty {1 \ov n^2 \sqrt{A}} S(J, J, n) 
  e^{- {\lam J \ov n^2  (1 + \sqrt{A})} } \\
  A = 1 + {4 \pi^2 \ov n^2 \b_1 \b_2} , \quad \lam =4 \pi^2  {\b_1 + \b_2 \ov \b_1 \b_2}  \ .
 \end{gather} 
 Here $S(J, J, n) $ is the Kloosterman sum, $f_{DT} (\b_1, \b_2)$ is the expression~\eqref{dtr},  and~\eqref{yeg2} can be obtained from (4.20)
 of~\cite{CotJen20}  through some manipulations.  We then find that 
 \bega 
 g(J; E_1, E_2) = g_{DT} (\hat E_1, \hat E_2) + \tilde g (J; \hat E_1, \hat E_2)  , \quad \hat E_s = E_s - J, \; s=1,2 ,  \\
 \tilde g (J; \hat E_1, \hat E_2)  =  \int_0^\infty  {d\b_1 \ov \b_1} {d\b_2 \ov \b_2} \, e^{\b_1 \hat E_1 + \b_2 \hat E_2} 
  f_2 (J; \b_1, \b_2),
  \label{geg1}
 \end{gather} 
where $g_{DT} (\hat E_1, \hat E_2)$ is given by~\eqref{dtr1}. While we cannot evaluate~\eqref{yeg2} or~\eqref{geg1} in a closed form, it can be shown that in the large $J$ limit, $\tilde g(J; \hat E_1, \hat E_2)  \sim O(J^{-\al})$ with $\al > 0$. 
Thus we find in the large $J$ limit, $g (J, E_1, E_2)$ reduces to the expression~\eqref{dtr1} of JT with $E_1, E_2$ replaced by $\hat E_1, \hat E_2$.

\section{Discussions}\label{sec:disc}

In this paper we explicitly illustrated the general framework proposed in~\cite{Liu25c} using the Gutzwiller representation as a paradigmatic example,  demonstrating  how many features characteristic of random matrix models can be derived within a single-system framework. 
Furthermore, we postulated the existence of a minimal Gutzwiller-like structure in the large-$N$ limit of holographic systems and explored the emergence of hyper-non-perturbative structures that give rise to double exponentials in $1/N^2$. On the gravity side, we showed how spacetime wormholes enable the construction of emergent hyper-non-perturbative objects, which correspondingly yield double exponentials in $G_N$. 

Below we highlight some immediate open questions (some of which were already mentioned in the main text):

\medskip
 $\bullet \;$  {\bf Gutzwiller representation for quantum many-body systems}
\medskip

It would clearly be desirable to derive a Gutzwiller-like trace formula operating directly in the fully many-body regime. As noted previously, the central challenge lies in identifying a macroscopic ``classical'' phase space in the limit~\eqref{manB} that is sufficiently large to accommodate time evolution on exponentially long timescales $t \sim e^{\mathcal{O}(\mathcal{N})}$. 

In standard large-$N$ gauge theories and matrix models, the large-$N$ limit is captured by a classical master field configuration space, whose classical coordinates are defined by single-trace expectation values via large-$N$ factorization. The algebraic structure of this phase space is, in turn, governed by free probability theory~\cite{VoiDyk92,Dou94,GopGro94}. However, this conventional master field space is too restricted: it is generated solely by the algebra of single-trace observables. Holographically, we may interpret it as the classical phase space of bulk string field theory. Consequently, it only parameterizes ``confined'' degrees of freedom. To probe the black hole sector and resolve microscopic spacings on the scale $\Delta E \sim e^{-\mathcal{O}(\mathcal{N})}$, we must go beyond single-trace observables, which necessitates expanding the effective classical configuration space to encompass ``deconfined'' degrees of freedom. Generalizing master field geometry and free probability to handle such deconfined, non-factorized sectors represents a highly valuable avenue for quantum gravity and many-body chaos.

Regardless of whether a Gutzwiller-like sum can be rigorously defined for a quantum many-body system, the existence of a general decomposition of the form~\eqref{Decom} requires much less---only that the large-$N$ limit need not be smooth. This motivates the existence of a canonical projection $\FF$ that extracts the smooth component from a microscopic large-$N$ description.\footnote{There may be alternative, system-specific methods for extracting the smooth component for certain observables. This non-uniqueness is conceptually akin to the freedom in choosing a coarse-graining procedure in classical thermodynamics. We emphasize, however, that $\FF$ should not be conflated with generic statistical coarse-graining; rather, it is a precise mathematical operation acting on the microscopic data.} The gravity dual provides strong motivation for such a canonical projection: the gravitational path integral furnishes a distinguished bulk prescription for extracting smooth macroscopic data from the underlying microscopic data of quantum gravity.

\medskip 
 $\bullet \;$  {\bf What is the Tier-II description in gravity?}
\medskip

The duality with the boundary CFT implies a corresponding three-tier description of the bulk gravitational system (recall Fig.~\ref{fig:3tiers}). In particular, even in the $G_N\to0$ limit, the gravity side should contain erratic contributions carrying ``microscopic'' information that is not captured by gravitational path integrals in the sense defined below~\eqref{gid}. Such contributions are necessary to restore factorization on the gravity side. In Sec.~\ref{sec:hmul}, we incorporated this Tier-II structure formally by associating the erratic parts of observables with vectors in the third-quantized multiverse Hilbert space $\sH_{\text{multiverse}}$. Their inner products are computed by the gravitational path integral, which thereby acts as a smooth filtering projection that produces macroscopic Tier-III quantities.

A natural question is whether Tier-II quantities on the gravity side can be characterized more directly. For a few-body quantum chaotic system, the erratic contribution in the Gutzwiller formula emerges from a semiclassical treatment of the underlying path integral. In searching for a gravitational analogue, however, it is important to recognize the distinctive character of the gravitational path integral and the aspects in which it differs fundamentally from the path integral of an ordinary quantum system.

In ordinary quantum mechanics or quantum field theory, the path integral is a purely ``mechanical'' object. The degrees of freedom are explicit: one tracks particle positions or fundamental field values. When evaluating the path integral $Z = \int \mathcal{D}x \, e^{-S[x]/\hbar}$ via saddle-point approximation, a saddle point represents merely a single classical trajectory. A single periodic orbit provides only a tiny fraction of the microscopic spectral data; building up a macroscopic thermodynamic quantity (such as the total density of states) requires summing over an infinite tower of these orbits. The standard saddles are thus strictly microscopic, containing no intrinsic thermodynamic information.

In stark contrast, the gravitational path integral is intrinsically thermodynamic and hydrodynamic. In the Euclidean gravitational path integral, $Z = \int \mathcal{D}g \, e^{-I_E[g]}$, we integrate over spacetime geometries subject to fixed boundary conditions (such as the length of a Euclidean time circle specifying the temperature). The dominant saddle point here is not a microscopic trajectory, but an entire macroscopic ``classical spacetime.'' Evaluating the classical Einstein--Hilbert action on a single saddle immediately yields macroscopic thermodynamic quantities, complete with the Bekenstein--Hawking entropy ($S = A/4G$) and phase transitions. 

Consequently, obtaining an explicit Tier-II description requires penetrating beyond these hydrodynamic-like degrees of freedom (such as the metric and low-energy matter fields). This reinforces the discussion of the previous bullet from the boundary perspective: to access Tier-II fine structures directly, one may need to explicitly uncover and parameterize the underlying ``deconfined'' configuration space. Explicit string theory calculations in solvable models, such as those in~\cite{Ebe20,Ebe21}, may serve as an instructive guide.

A deeper understanding of possible erratic boundary behavior is also crucial, as it may yield direct clues about the corresponding gravity description. In particular, elucidating erratic behavior within microcanonical ensembles~\cite{Moo98a,Moo98b,MilMoo99,DijMal00,Per26} and drawing connections to discussions of fortuity (see, e.g.,~\cite{ChaLin24}) should be highly fruitful. Alongside this, analyzing correlations in $N$ and exploring the Mellin-Laplace transform with respect to $N$~\cite{KudWit26} represent two other promising avenues. Furthermore, gaining a clearer picture of resurgence and the underlying analytic structure in the Borel plane could also prove insightful.

\medskip 
 $\bullet \;$  {\bf Universal features of OPE coefficients of heavy operators}
\medskip

In this paper, we have focused on the universal features associated with the density of states, the spectral form factor, and the  off-shell wormholes of~\cite{CotJen20,CotJen20b,CotJen21,CotJen22} (with the double cone~\cite{SaaShe18} emerging as a special limit), which play a crucial role as baby-universe propagators in the multiverse Hilbert space.

There are also on-shell wormholes with amplitudes scaling with $G_N$ as $e^{\mathcal{O}(1/G_N)}$, which have been argued in~\cite{Liu25c} to be connected to the erratic nature of matrix elements (or OPE coefficients) associated with heavy operators of dimension $\mathcal{O}(1/G_N)$. In particular, for a large class of wormholes in AdS$_3$, the discussion of~\cite{ChaCol22} (see also~\cite{ChaHar24,ChaPos26,GenHun25}) suggests deep connections between filtered correlations among the OPE coefficients of heavy operators and Liouville CFT structure constants. It would be desirable to develop an explicit decomposition of the form~\eqref{Decom} for such OPE coefficients (just as the Gutzwiller representation does for the resolvent and density of states), which amounts to formulating an ETH-like ansatz for the matrix elements of heavy operators (see~\cite{ColMal19,BeldeB20,Sas22} for examples of the smooth parts of such formulas). Crucially, the existence of bulk wormholes dictates that the erratic fluctuations of such matrix elements are not mere structureless noise, but rather encode deep, universal signatures of quantum chaos.

 \medskip 
 $\bullet \;$  {\bf A reinterpretation of JT gravity in a single system?}
\medskip

JT gravity can be \emph{defined} via the gravitational path integral, which can be shown to be dual to the \emph{perturbative} expansion of a double-scaled random matrix model~\cite{SaaShe19}. However, such a definition of JT gravity possesses ambiguities concerning the gravitational description of \emph{non-perturbative} contributions on the matrix model side, which correspond to double exponentials in $G_N$. The specific double-scaled random matrix model dual to JT gravity has a potential that is unbounded from below, and thus cannot serve as a rigorous, unambiguous UV completion of JT gravity on its own. One way to fix such ambiguities is to embed JT gravity into a higher-dimensional gravity theory with a boundary dual (an example is given in Sec.~\ref{sec:ads3}), such that its ``UV completion'' is warranted by the boundary CFT.

In light of the discussion in this paper, it is tempting to view the elegant duality between JT gravity and random matrix models not as a duality with an ensemble of boundary theories, but rather as an indication that the random matrix model description---and thus JT gravity itself---may simply capture the universal chaotic aspects of a \emph{single} quantum system. This would, in particular, suggest that there is a much larger set of universal quantities in quantum chaotic systems corresponding to the topological recursion relations of random matrix models, which are in turn connected to multi-boundary wormholes.

\medskip 
\noindent $\bullet \;$ {\bf Toward a universe effective theory for higher-dimensional gravity}
\medskip 

In Sec.~\ref{sec:gravity}, we highlighted the crucial role of the third-quantized Hilbert space $\sH_{\text{multiverse}}$ in capturing Tier-II fine structures and defining the hyper-non-perturbative objects responsible for double-exponential effects. A natural question then arises: is there a ``universe effective theory'' (to use the terminology of~\cite{PosVan22}) that systematically governs the transitions, splittings, and joinings of baby universes within this space? Closely tied to this is the question of whether we can formulate an effective theory capturing the ``interactions'' among baby-universe condensates (and/or anti-condensates)---a framework we might refer to as a {\it hyper-structure effective theory} (hyper-structure ET). 

For two-dimensional quantum gravity endowed with a target-space interpretation (i.e., where the two-dimensional world is viewed as a string worldsheet), these two effective theories have precise formulations in target space: the universe effective theory corresponds to non-perturbative closed string field theory, while the hyper-structure ET corresponds to open string field theory on D-branes (which serve as the hyper-non-perturbative objects). In the context of JT gravity, the corresponding universe effective theory is naturally provided by topological gravity~\cite{PosVan22,AltPos22,McNVaf20}. However, for general higher-dimensional gravity, which lacks a target-space description, the guiding symmetries and principles required to construct such effective theories remain elusive. We note that recent developments~\cite{McNWan26} exploring topological fluctuations and factorization in higher dimensions could provide relevant insights.

\vspace{0.2in}   \centerline{{\bf Acknowledgements}} \vspace{0.2in}
We would like to thank 
Alexander Altland, Justin Berman, Jordan Cotler, Hao Geng, Elliott Gesteau, Kristan Jensen, Marc Klinger, Jonah Kudler-Flam, Juan Maldacena,  Eric Perlmutter, Julian Sonner, Douglas Stanford, Erik Verlinde, Edward Witten, Jiuci Xu, and Zhenbin Yang for discussions. This work is supported by the Office of High Energy Physics of U.S. Department of Energy under grant Contract Number  DE-SC0012567 and DE-SC0020360 (MIT contract \# 578218), and was made possible through the support of grant \#63670 from the John Templeton Foundation.

\appendix

\section{Ramp in distributions of zeros of Riemann $\ze$-function without averaging}\label{app:Rie}

The Riemann Hypothesis conjectures that the nontrivial zeros of the Riemann zeta function $\zeta(s)$ all lie at positions $s = \frac{1}{2} + i t_n$, where $t_n$  are real numbers. There has long been speculation (tracing back to the Hilbert-P\'olya conjecture) that these $t_n$ 
may correspond to the energy eigenvalues of some quantum system. In fact, pair correlations of $t_n$ exhibit a
ramp~\cite{Mon73,Odl87}, suggesting that the corresponding quantum system is classically chaotic~\cite{Ber86,Ber88,BerKea99}. 

Under this scenario, the density of the Riemann zeros corresponds exactly to the density of states of a quantum chaotic system. While the absence of a known Hamiltonian precludes a direct Gutzwiller trace formula, the Riemann-Weil explicit formula serves as a striking mathematical analogue, where prime numbers play the exact role of primitive periodic orbits.
Just as in the quantum chaotic case, the density of zeros can be formally decomposed into a smooth Weyl-like background and an erratic, highly oscillatory piece summed over primes. Berry~\cite{Ber86} reinterpreted the observation of the ramp by Montgomery~\cite{Mon73} as the exact analogue of a semiclassical diagonal approximation for quantum chaotic systems. 

Here, we revisit this classic calculation to further reinterpret it as a smooth filter projection $\Fil{\cdot}$, demonstrating that in this number-theoretic analogue, spectral correlations of Riemann zeros can similarly arise from the product of the erratic part of the density.

Specifically, the Riemann-Weil explicit formula expresses the density of zeros as (see e.g.~\cite{BerKea99})
\be
\rho(t) = \sum_n \delta(t - t_n) = \bar{\rho}(t) + \rho^{\text{osc}}(t)
\ee
where $\bar{\rho}(t)$ is the smooth part 
\be
\bar{\rho}(t) = \frac{1}{2\pi}\log\frac{t}{2\pi}
\ee
which grows logarithmically. The oscillatory part $\rho^{\text{osc}}(t)$ is expressed as a sum over primes $p$,
\be
\rho^{\text{osc}}(t) = -\frac{1}{\pi}\sum_p \sum_{k=1}^{\infty} \frac{\log p}{p^{k/2}}\cos(kt\log p) \ .
\ee
We again identify $\bar \rho$ and $\rho^{\text{osc}}$ with $\rho^{\rm (sm)}$ and $\rho^{\text{err}}$, respectively. 
The analogy with the Gutzwiller formula goes as follows: the energy $E$ corresponds to $t$, the sum over primitive periodic orbits corresponds to the sum over primes $p$, the action of a periodic orbit $S_a (E)/\hbar$ 
corresponds to $t \log p$, and the multiple traversals of periodic orbits (which were suppressed in~\eqref{dosD1}) correspond to the sum over $k$.\footnote{As discussed earlier, higher traversals yield negligible corrections. Similarly, in the $\ze$-function case, we will see below that only $k=1$ is relevant for the ramp.}

Now consider the product of oscillatory parts, 
\be\label{eyn}
\rho^{\text{osc}}\!\left(t+\tfrac{\omega}{2}\right)\rho^{\text{osc}}\!\left(t-\tfrac{\omega}{2}\right) = \frac{1}{\pi^2}\sum_{p,p'}\sum_{k,k'} \frac{\log p\,\log p'}{p^{k/2}\,p'^{k'/2}}\cos\!\left(k(t+\tfrac{\omega}{2})\log p\right)\cos\!\left(k'(t-\tfrac{\omega}{2})\log p'\right) \ .
\ee
The product of cosines in~\eqref{eyn} splits into two terms 
\bega
\frac{1}{2}\cos\!\Big((k\log p + k'\log p')\,t + \tfrac{\omega}{2}(k\log p - k'\log p')\Big) \\
+ \frac{1}{2}\cos\!\Big((k\log p - k'\log p')\,t + \tfrac{\omega}{2}(k\log p + k'\log p')\Big) \ .
\end{gather} 
The first term is always highly oscillatory.
The rapidly oscillating component of the second term cancels when $k\log p = k'\log p'$, which forces the diagonal pairing $p = p', \; k = k'$. Only these paired terms survive the smooth filtering operation, and thus
\be
I \equiv \Fil{\rho^{\text{osc}}\rho^{\text{osc}}}= \frac{1}{2\pi^2}\sum_p\sum_{k=1}^{\infty}\frac{(\log p)^2}{p^k}\cos(k\omega\log p)
= \frac{1}{2\pi^2} \text{Re}  \sum_p \sum_{k=1}^{\infty} { (\log p)^2 \ov p^{k (1-i \om )}} 
 \ .
\label{ywn}
\ee
 The expression~\eqref{ywn} is singular as $\om \to 0$ because the sum over $p$ diverges for $k=1$ at $\om =0$ (the 
sum $\sum_p\sum_{k=2}^\infty  \frac{(\log p)^2}{p^{ks}}$ is convergent for $s=1$). 

To find the small-$\om$ behavior, recall the logarithmic derivative of $\zeta$-function\footnote{Note the the Euler product formula for the Riemann zeta function, $\ze(s)= \prod_{p} {1 \ov 1-p^{-s}}$.},
\be\label{ywn1}
-\frac{\zeta'}{\zeta}(s) = \sum_p\sum_{k=1}^\infty \frac{\log p}{p^{ks}} \quad \to \quad 
\p_s \le(\frac{\zeta'}{\zeta}(s)\ri) = \sum_p\sum_{k=1}^\infty  \frac{k(\log p)^2}{p^{ks}} \ .
\ee
$\zeta$-function is singular at $s=1$, so is~\eqref{ywn1}. Furthermore,  the singular behavior at $s=1$ in~\eqref{ywn1} solely comes from the $k=1$ as $\sum_p\sum_{k=2}^\infty  \frac{k(\log p)^2}{p^{ks}}$
is also convergent. 

Comparing~\eqref{ywn} and~\eqref{ywn1} for $k=1$, we thus conclude that 
\be
I \approx {\rm Re} \, \p_s \le(\frac{\zeta'}{\zeta}(s)\ri)\biggr|_{s = 1-i \om}  = -{1 \ov 2 \pi^2 \om^2} , \quad \om \to 0,
\ee
where we have used the Laurent expansion $\zeta(s) \approx \frac{1}{s-1} + O(1)$ near the pole $s=1$. We have thus precisely recovered the universal linear ramp purely through filtering.

\linespread{1.5}

\bibliographystyle{jhep}
\bibliography{all}

\end{document}